\documentclass[...,alpha-refs]{wiley-article}

\usepackage{siunitx}
\usepackage{hyperref}
\usepackage{adjustbox}
\usepackage{float}
\usepackage[final]{changes}
\usepackage{orcidlink}

\renewcommand{\Authfont}{\fontsize{10pt}{14pt}\selectfont\bfseries\raggedright}

\newcommand{\tabref}[1]{[#1]}
\newcounter{box}
\newcounter{savedfigure}
\renewcommand{\thebox}{\arabic{box}} 

\newmdenv[
  linewidth=4pt,
  roundcorner=0pt,
  backgroundcolor=white,
  linecolor=black!20,
  innertopmargin=8pt,
  innerbottommargin=8pt,
  innerleftmargin=8pt,
  innerrightmargin=8pt,
  skipabove=\baselineskip,
  skipbelow=\baselineskip,
  frametitlefont=\bfseries\large,
  frametitlealignment=\raggedright
]{wileyboxTitledMD}

\newenvironment{wileyboxTitled}[1][]{%
  \refstepcounter{box}%
  \setcounter{savedfigure}{\value{figure}}%
  \setcounter{figure}{0}%
  \begin{wileyboxTitledMD}[#1]%
}{%
  \end{wileyboxTitledMD}%
  \setcounter{figure}{\value{savedfigure}}%
}

\papertype{Review}

\title{Future of Edge AI in biodiversity monitoring}

\renewcommand{\Authfont}{\fontsize{11pt}{14pt}\selectfont\bfseries\raggedright}
\author[1,2\authfn{1}]{Aude Vuilliomenet}
\author[2]{Kate E. Jones}
\author[1]{Duncan Wilson}

\affil[1]{The Bartlett Centre for Advanced Spatial Analysis, Faculty of the Built Environment, University College London, London, W1T 4TJ, United Kingdom}
\affil[2]{Centre for Biodiversity and Environment Research, Department of Genetics, Evolution and Environment, University College London, London, WC1E 6BT, United Kingdom}

\corraddress{Aude Vuilliomenet}
\corremail{aude.vuilliomenet.18@ucl.ac.uk}

\fundinginfo{EPSRC, Grant/Award Number: EP/R513143/1 and EP/T517793/1}
\runningauthor{Vuilliomenet et al.}

\begin{document}

    \begin{frontmatter}
        \maketitle
        \begin{abstract}
\textbf{Abstract}

\begin{enumerate}
    \item  
    Many ecological decisions are slowed by the gap between collecting and analysing biodiversity information. 
    Edge computing moves data processing closer to the sensor, with edge artificial intelligence (AI) using AI models for data processing, reducing reliance on data transfer and continuous connectivity. 
    In principle, this shifts biodiversity monitoring from passive data logging towards autonomous, responsive sensing systems. 
    In practice, however, the adoption of edge AI in ecological research remains fragmented, with key design trade-offs, performance constraints, and implementation challenges rarely reported in a systematic way.
    \item
    Here, we analyse 82 studies published between 2017 and 2025 that apply edge computing to biodiversity monitoring across acoustic, vision-based, tracking, and multi-modal systems.
    We synthesise hardware platforms, AI model optimisation, and wireless communication to critically assess how design choices impact ecological inference, deployment longevity, and operational feasibility.    
    \item 
    Annual publications rise moderately from 3 in 2017 to 19 in 2025. 
    We identify four system types represents how edge AI has been implemented across acoustic, vision, and tracking applications.
    These include low-power microcontrollers (MCUs) for single-taxon or rare-event detection, single-board computers (SBCs) supporting multi-species classification and real-time alerts, and retrospective cloud processing.
    Each system type represents context-dependent compromises among power consumption, computational capability, and communication requirements.
    \item 
    Our analysis reveals the evolution of edge computing systems from proof-of-concept to robust, scalable tools.
    We argue that edge computing offers opportunities for responsive biodiversity management, but realising this potential requires increased collaboration between ecologists, engineers, and data scientists to align model development and system design with ecological questions, field constraints, and ethical considerations.
\end{enumerate}



\end{abstract}
    \end{frontmatter}
    
    \section{Introduction}
\label{sec:introduction}

Biodiversity underpins the health and resilience of ecosystems on which human societies depend \citep{pigot_macroecological_2025, costanza_value_1997}. 
Nevertheless, global assessment reports continue to highlight the degradation, depletion, and disruption of the biosphere at rates faster and larger than ever \citep{sakschewski_planetary_2025}.
Unlike climate change, for which there is consensus on tracking straightforward metrics such as atmospheric carbon concentrations \citep{lee_ipcc_2023}, the change in biodiversity is inherently complex and multifaceted \citep{ipbes_global_2019}. 
Species respond to anthropogenic pressures through adaptation, range shifts, or decline toward extinction, however, significant knowledge gaps persist regarding species distributions, abundances, and ecological interactions \citep{otto_adaptation_2018}.
Addressing these knowledge gaps requires rethinking how biodiversity is monitored and measured \citep{dasgupta_economics_2024}.
The goal is to transition from sporadic and localised surveys to continuous, scalable, and autonomous systems that deliver accurate, timely ecological insights for effective conservation and sustainable ecosystem management \citep{king_implementation_2025}.

Traditionally, biodiversity assessments have relied on in-person field surveys, which are time consuming, labour intensive, and hard to replicate across large spatial and temporal scales.
The development of automated sensing technologies has increased observation capacity beyond direct human presence, enabling long-term data collection \citep{allan_futurecasting_2018}.
These technologies facilitate the acquisition of billions of high-resolution data points, resulting in a "data deluge" that far exceeds the capacity of manual analysis \citep{porter_staying_2012, farley_situating_2018}.
Modern monitoring systems support two complementary, but analytically distinct approaches: environmental sensing, which captures abiotic conditions, and biodiversity monitoring, which targets biotic factors \citep{lahoz-monfort_comprehensive_2021}. 
Abiotic parameters such as temperature, humidity, soil moisture, and chemical concentrations, provide contextual information on habitat conditions and enable early warning of environmental disturbances such as wildfires or pollution \citep{caron_ai_2025}.
In contrast, biotic monitoring extract species-level information using four primary data modalities: acoustic systems record species vocalisations and soundscapes \citep{browning_passive_2017}; vision-based approaches record morphological characteristics \citep{glover-kapfer_camera-trapping_2019}; movement technologies track trajectories and physiological states; and, eDNA sampling collects genetic material shed by organisms into the environment \citep{beng_applications_2020}.
While both approaches contribute to ecosystem understanding, biotic data streams are high-dimensional and unstructured, requiring efficient autonomous processing pipelines to deliver timely, actionable ecological insights \citep{tuia_perspectives_2022}.

Advances in artificial intelligence (AI), including machine learning (ML) and deep learning (DL), have enabled substantial progress in species identification 
and the detection of ecological patterns across acoustic, visual, movement, and environmental datasets \citep{christin_applications_2019}. 
Such progress has been driven by collaborative efforts between experts in ecology and AI, and enabled the development of models tailored to ecological applications, capable of extracting relevant insights from large, heterogeneous, and unbalanced datasets, leading to the proliferation of AI models spanning multiple taxa and various sensing modalities \citep{borowiec_deep_2022}.
Examples of popular AI models used by the ecological community include vision-based detectors such as MegaDetector and SpeciesNet, identifying animals, people and vehicles in camera trap images \citep{beery_efficient_2019, gadot_crop_2024}, but also acoustic-based classifiers such as BirdNET, identifying north American and European bird species \citep{kahl_birdnet_2021}, BatDetect2 identifying UK bat species \citep{mac_aodha_towards_2022}, and SurfPerch a foundational model to distinguish coral reef and marine sounds \citep{williams_using_2025}.
These pretrained models reduce the analytical bottleneck once imposed by manual processing.

Current ecological data processing workflows remain dependent on centralised infrastructure, whereby data collected over months are uploaded to high-performance computing clusters, processed in batches, and only after considerable delay translated into ecological insights. 
This sequential "collect-upload-analyse" workflow limits proactive responses, constraining the rapid interventions that are imperative in certain conservation contexts.
Such interventions include preventing collisions between cetaceans and ships \citep{paoletti_seadetect_2023}, containing the spread of invasive species \citep{wood_real-time_2024}, mitigating human-wildlife conflicts \citep{abrahms_climate_2023}, and combating poaching and illegal activities \citep{lynam_rising_2025}. 
As conservation challenges intensify, the demand for tools delivering timely and accurate information will continue to grow.

To overcome temporal and logistical constraints, data can be processed at or near the point of collection, an approach known as edge computing \citep{ding_roadmap_2022}.
This idea evolved alongside the rapid expansion of the Internet of Things (IoT), a global network of interconnected objects capable of sensing, processing, and communicating data \citep{pan_future_2018}.
Edge computing devices are often part of IoT networks, and when combined with wireless sensor networks (WSN) such as Wi-Fi, cellular, long-range wide-area network (LoRaWAN), or satellite, enable decentralised and real-time monitoring systems.
In this context, edge computing refers to the distributed processing of data across IoT sensing nodes, with edge AI indicating that the data are processed using an AI-based model \citep{shi_edge_2016}. 
Pioneering studies include Shazam for Bats, which used Wi-Fi-connected acoustic sensors to detect and classify bat calls \citep{gallacher_shazam_2021}, 
and AI-enabled camera traps, which process wildlife images locally and send alerts through the Iridium satellite network \citep{whytock_real-time_2023}.
The convergence of edge computing, AI, and IoT technologies has attracted growing academic interest, reflected in initiatives such as the Global Biodiversity Observing System (GBiOS) \citep{gonzalez_global_2023} and visionary frameworks like "Nature 4.0" \citep{zeuss_nature_2024}.

\begin{figure}[hbbt]
\includegraphics[width=0.98\textwidth]{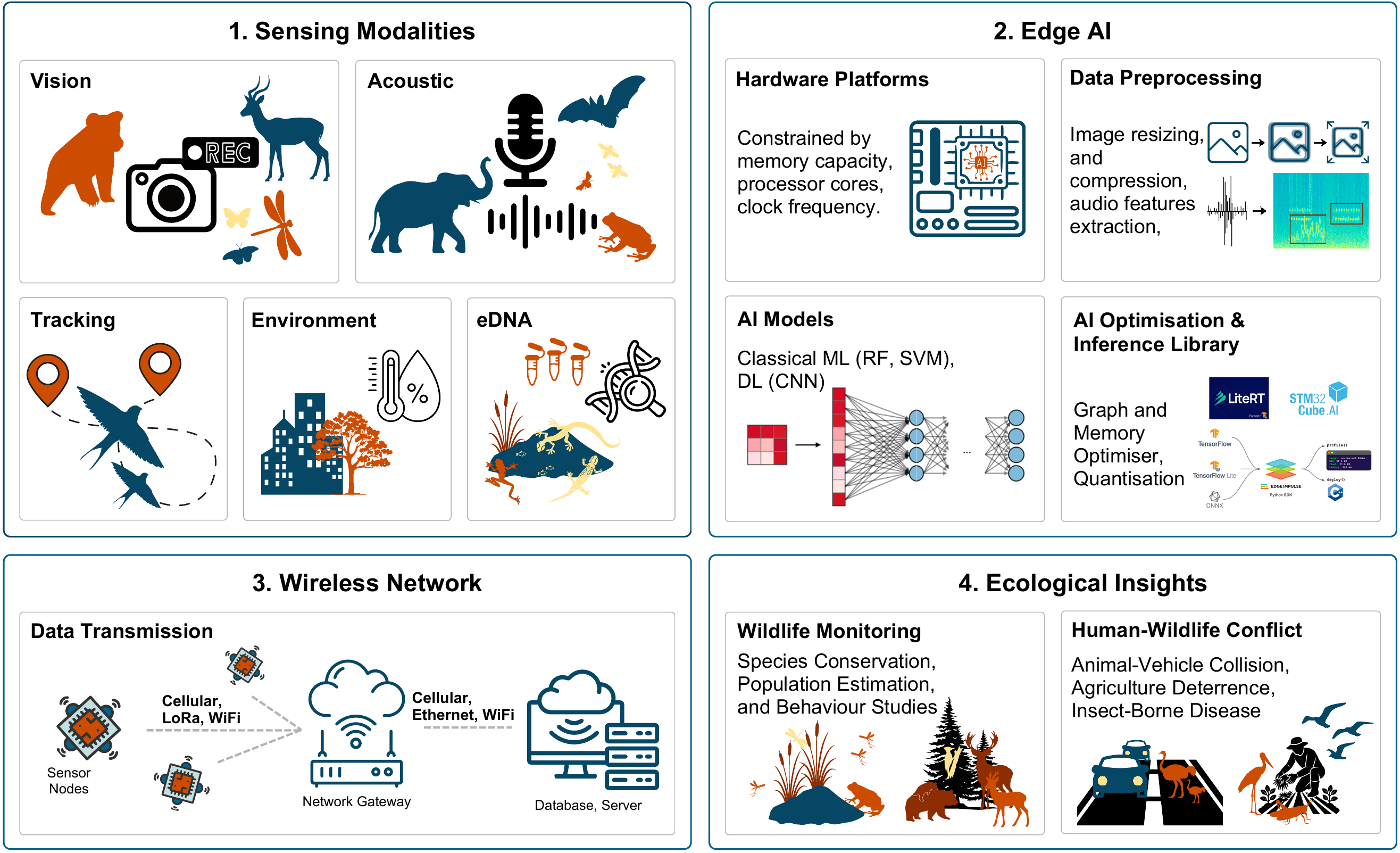}
\centering
\caption{
    \textbf{Conceptual overview of edge computing systems in biodiversity monitoring.}
    Edge computing integrates four key components to enable autonomous, in-situ data processing and decision-making. 
    (1) Biodiversity data are captured through various sensing modalities including vision-based camera systems, acoustic recorders, tracking technologies, environmental sensors, and environmental DNA. 
    (2) AI inference occurs on resource-constrained hardware platforms, where data undergo preprocessing before being analysed by classical ML algorithms or DL models. Model optimisation techniques, including quantisation and graph optimisation, combined with specialised inference libraries, enable efficient on-device computation despite limited memory capacity, processor cores, and clock frequency. 
    (3) Wireless networks support selective data transmission from sensor nodes to network gateways via cellular, LoRa, or WiFi before final storage in remote infrastructure. 
    (4) Ecological insights derived from edge computing systems include wildlife monitoring and human-wildlife conflict management.}
\label{figure_section1:edgecomputing_4_biodiversity}
\end{figure}

Despite the promise of edge AI, its development within ecological research remains fragmented and largely disconnected from the constraints of real-world deployment. 
Most studies prioritise AI model development under assumptions of cloud-based computational resources \citep{pollock_harnessing_2025}, a pattern reflected across recent reviews of AI applications in biological conservation \citep{reynolds_potential_2025}, ML for animal movement and behaviour \citep{tuia_perspectives_2022}, and low-cost field computing platforms such as Raspberry Pi \citep{jolles_broad-scale_2021}.
Few studies address the edge computing pipeline as an integrated system, spanning sensor selection, embedded hardware design, on-device model optimisation, wireless connectivity, and ecological interpretation of outputs \citep{yu_edge_2024} (Figure \ref{figure_section1:edgecomputing_4_biodiversity}).
As a result, ecologists adopting on-device AI are often required to draw on computing literature to identify appropriate solutions for model compression, embedded platforms, and network protocols \citep{zhou_edge_2019, wang_empowering_2025}.
If fully autonomous, real-time biodiversity monitoring is to be realised, a consolidated, cross-disciplinary synthesis is needed.

Here, we synthesise the current knowledge and implementations of on-device AI in biodiversity monitoring, focusing primarily on edge computing systems capturing morphological or behavioural data for taxonomic identification (Appendix \ref{appendix:review_methods}).
Through the analysis of existing case studies and field deployments that integrate on-device AI inference with data transmission capabilities, we identify challenges and opportunities of current edge computing solutions (i.e., embedded platforms, AI model optimisations, and wireless networks) in ecological applications.

    \section{Edge Computing Basics}
\label{sec:edge_computing_draft}

\subsection{What is edge computing and edge AI?}
Successful on-device AI deployment depends on the coordinated design of three tightly coupled components: hardware platforms for data collection and processing; AI models for on-device inference; and, wireless networks for data transmission (Figure \ref{figure_section1:edgecomputing_4_biodiversity}). 
Edge hardware spans a wide range from microcontrollers to embedded computers in unmanned aerial vehicles, but all operate under strict constraints on memory, processing speed, and energy availability.
These constraints directly limit the complexity of AI models that can be deployed, including model size, architecture, and computational demand \citep{david_tensorflow_2021}.
Wireless sensor networks (WSNs) enable remote data transmission and system updates, though the energy cost of communication often exceeds that of computation, making energy-efficient networks essential for battery-powered deployments. 
As a result, edge computing systems must be designed holistically: choices about hardware, AI model architecture, and network configuration are inseparable, and optimising one component inevitably shapes the performance, feasibility, and energy budget of the others \citep{iodice_edge_2025}.

Processing data at the point of collection offers advantages compared to centralised processing in high-performance computer clusters \citep{zhou_edge_2019}.  
The benefits of edge AI are commonly framed in terms of Bandwidth, Latency, Economic cost, Reliability, and Privacy (BLERP) \citep{bier_whats_2020}.
Processing data locally significantly reduces the volume of information requiring transmission, thereby lowering demand for bandwidth and decreasing both transfer costs and device energy consumption. 
Moreover, local computation enables immediate responses and enhances reliability, as devices can function autonomously without continuous network connectivity.
Furthermore, on-device processing strengthens privacy and security by ensuring sensitive information is analysed and filtered locally rather than transmitted to external servers.
These operational advantages have driven interest in embedding increasingly complex AI algorithms in highly resource-constrained edge devices located in remote and adverse environments \citep{merenda_edge_2020}. 
The following subsections explore the multi-faceted characteristics defining hardware platforms, edge AI models, and WSNs, highlighting key considerations for the successful implementation of edge computing systems.

\subsection{Hardware} \label{subsection:hardware}
\begin{figure}[hbbt]
\centering
\includegraphics[width=1.0\textwidth]{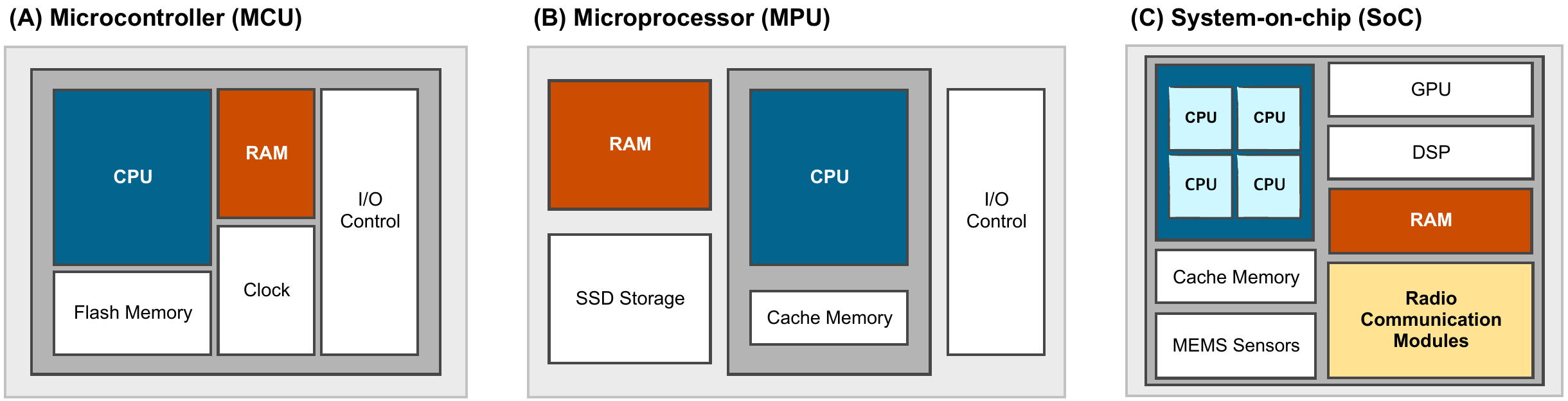}
\centering
\caption{
    \textbf{Different categories of hardware chips} 
    (A) Microcontroller units (MCUs) integrate all core elements of a computer system, which are a central processing unit (CPU), a volatile memory (i.e. RAM), and programmable input and output (I/O) peripherals into one single chip.
    (B) Microprocessor units (MPUs) integrate only the CPU on a chip. They are often mounted on a printed circuit board (PCB) and interact with other computing elements (i.e., RAM, I/O Control, SSD Storage) via conductive tracks.
    (C) Systems-on-chip (SoCs) integrate multiple CPUs along with additional processing chips (i.e. graphic processing unit (GPU), digital signal processing (DSP), radio communication modules, and MEMS sensors (e.g. accelerator, pressure, microphones) in one compact chip.
}
\label{figure_section2:hardware_types}
\end{figure}

The first component of an edge computing system is the physical hardware supporting the execution of the edge application.
This hardware consists of compact integrated circuits designed to perform sensing, computation and control.
Three broad categories of chips are commonly used in edge systems: microcontroller units (MCUs), microprocessor units (MPUs), and system-on-chips (SoCs) (Figure \ref{figure_section2:hardware_types}). 
Their capability is highly dependent on their microarchitecture, including their number of processor cores and their memory capacity, dictating the number of operations that can be executed per second (i.e. clock frequency), and the way fractional numbers are computed. 
Understanding these architectural characteristics is essential to evaluate the efficiency of on-device AI implementations \citep{david_tensorflow_2021}.

MCUs are widely available, low-cost, and energy-efficient, making them a natural choice for long-term field deployments despite limited computational resources \citep{warden_tinyml_2020}. 
MCUs offer low power consumption, usually in the milliwatt (mW) range, limited volatile memory in the kilobyte (kB) range, and only solid-state storage (i.e. flash memory), where the computer programme and AI model need to be written and stored. 
Computer boards integrating MCUs include, for instance, the AudioMoth, Arduino devices (e.g., Nano and MKR), ESP32, and STM32 (Appendix \ref{appendix:hardware_table}). 
The simplicity and energy efficiency of MCUs make them well-suited for long-term monitoring deployments where power availability is severely constrained. 


In contrast, MPUs and SoCs support multitasking environments ranging from real-time lightweight operating system (OS) like FreeRTOS (i.e. RTOS kernel) to fully-featured OS like Linux (i.e. OS distribution). 
OS support allows MPUs and SoCs to manage complex data pipelines and execute edge AI inferences efficiently, often utilising on-chip vector instructions or hardware accelerators to handle high-level model complexity. 
An example of an MPU is the Broadcom BCM2711 at the core of the Raspberry Pi 4, while an example of a highly integrated SoC is the ESP32 series from Espressif. 
While MPUs often rely on external components mounted alongside them on a printed circuit board (PCB), such as RAM, radio modules, or specialised accelerators (e.g. NPU / TPU), to form a Single-Board Computer (SBC), SoCs like the ESP32 integrate many of these elements, including Wi-Fi and Bluetooth radios, directly into a single silicon package.
This high level of integration provides the extreme compactness required for wearable and small-scale IoT devices. 
Common SBCs used in edge AI include the Raspberry Pi 5, Google Coral Dev Board, and the NVIDIA Jetson series (Appendix \ref{appendix:hardware_table}).

\begin{figure}[hbbt]
\includegraphics[width=0.75\textwidth]{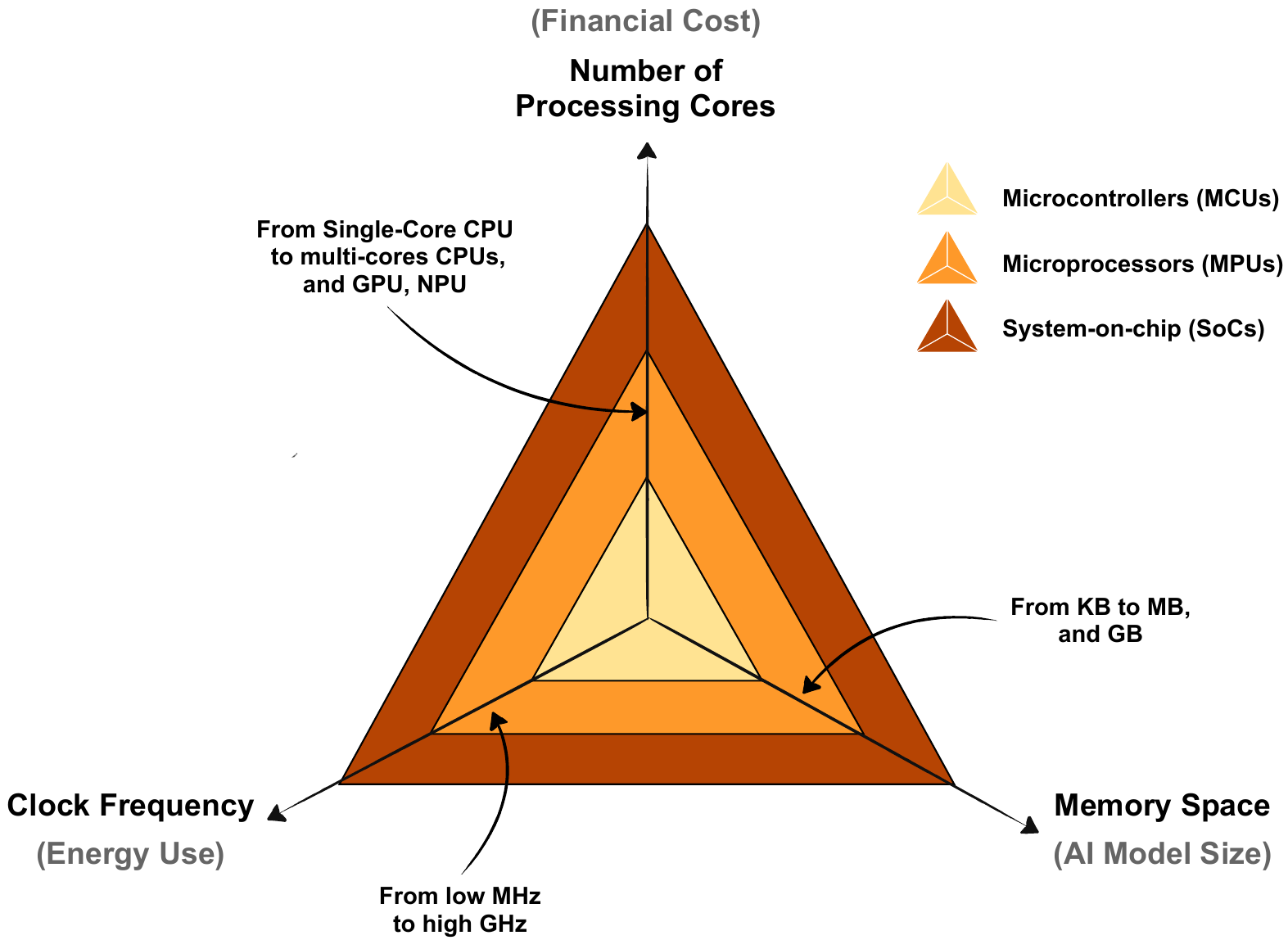}
\centering
\caption{
    \textbf{Hardware Trade-Offs in edge computing.}
    Platform selection balances three primary considerations: AI model complexity, energy consumption, and device cost. 
    Triangle sizes illustrate these trade-offs. MCUs (small yellow) typically feature single core, around 100 kB memory, and operate at frequencies near 100 MHz, while MPUs and SoCs (medium orange and large brown) offer multiple CPUs cores, gigabytes of memory, and operate at frequencies above 1 or 2 GHz, with SoCs incorporating specialised processors (GPU, NPU, TPU).
    Quantitative metrics for specific hardware platforms are provided in Appendix (\ref{appendix:hardware_table}).
}
\label{figure_section2:hardware_tradeoffs}
\end{figure}

Selecting appropriate hardware requires balancing three competing priorities: computational resources which constrain AI model size and complexity; power consumption, which determines deployment duration; and device cost, which influences the achievable spatial scale of monitoring (Figure \ref{figure_section2:hardware_tradeoffs}).
MCUs enable year-long autonomous operation at minimal cost (e.g., AudioMoth at around £50) but typically support only lightweight AI models, such as keyword spotting \citep{warden_tinyml_2020}, whereas SBCs offer greater computational capacity for complex computer vision or audio analysis at the expense of higher power budgets and costs (£40–200+) \citep{jolles_broad-scale_2021}.
Understanding these trade-offs is critical to ensure hardware selection matches deployment constraints.
However, the hardware landscape is evolving rapidly, with ongoing improvements in processing capability, energy efficiency, and specialised AI accelerators blurring the boundaries between resource-constrained and complex AI applications \citep{banbury_edge_2023}.
For instance, the progression from AudioMoth (single-core, 16 KB memory, 32 MHz) to Arduino Portenta H7 (dual-core, 8 MB memory, 480 MHz) demonstrates expanding MCU capabilities, whilst the evolution of RPi devices from 3 to 4B, and 5, and the emergence of AI accelerators (e.g., Google Coral's TPU, RPi AI HAT) are enabling more sophisticated on-device inference at manageable power consumption.
Ultimately, the architectural features of each hardware platform determine which AI models are feasible to deploy, whilst AI model design and software optimisation techniques influence computational requirements, establishing a direct relationship between hardware capabilities and software design \citep{iodice_edge_2025}.

\subsection{Software and AI Models} \label{subsection:software}
The second core component of an edge computing system is the software that runs on the hardware, most notably AI models that perform on-device inference.
Within ecological research, AI models based on machine learning (ML) and deep learning (DL) approaches support tasks such as species detection, classification, localisation, and behavioural inference \citep{pichler_machine_2022, borowiec_deep_2022}.
While most ecological AI models are developed and evaluated on desktop computers or cloud infrastructure with effectively unconstrained resources, edge computing shifts AI execution to embedded devices.
At the most resource-constrained end of this spectrum lies Tiny Machine Learning (TinyML), a subfield of on-device AI focused on deploying ML models on ultra-low-power MCUs, with only kilobytes of memory and milliwatt power budgets \citep{warden_tinyml_2020}. 
TinyML has driven the development of aggressive model compression techniques and specialised inference frameworks that enable AI deployment on hardware platforms previously considered too limited.

AI inference represents one of the most computationally demanding operations within edge computing systems \citep{wang_empowering_2025}.
Inference applies learned model parameters to incoming sensor data, executing millions of mathematical operations, each consuming time, memory, and energy
(Box \ref{box:ai_overview}). 
On resource-constrained devices, inference operations can dominate memory usage and energy consumption, creating inherent tension between the development of highly accurate models and the physical constraints of embedded devices \citep{reddi_generative_2025}.
For instance, MegaDetector v5, a widely used object detector for camera trap images, contains 121 million parameters based on the large YOLOv5x architecture, achieving 96\% precision and 73\% recall but proving incompatible with resource-constrained edge devices. 
A recent and compact reimplementation, MegaDetector v6-compact (MDv6-c) reduces the model to 22 million parameters, by adopting the more efficient YOLOv9-compact architecture, while MDv6-c has a modest reduction in precision from 96\% to 92\%, it gains improving detection recall from 73\% to 85\%, demonstrating the optimisation trade-offs necessary to improving inference speed and memory efficiency for edge devices \citep{hernandez_pytorch-wildlife_2024}. 
Consequently, edge AI deployments require not only accurate model training but also implementation and evaluation of optimisation strategies essential for efficient execution under strict resource constraints.

Model optimisation techniques aim to reduce memory access, storage requirements, and arithmetic complexity, enabling faster and more energy-efficient inference while maintaining acceptable predictive performance.
One of the most widely adopted technique is to narrow the numerical precision used to store and process model parameters, a technique known as quantisation (Box \ref{box:ai_overview}) \citep{gupta_deep_2015, jacob_quantization_2017}.
For instance, the Bird@Edge system applied FP16 quantisation, reducing inference time by 10 ms per spectrogram (i.e. from 64 ms to 54 ms) on the NVIDIA Jetson Nano board whilst maintaining 95.2\% mean average \citep{hochst_birdedge_2022}.
Similarly, the burrowing owl vocalization TinyML detection model, TinyOwlNet, employed INT8 quantisation to enable deployment on a STM32 MCU (i.e. STM32H747I-DISCO), decreasing model size by a factor of three, allowing the resulting 11 KB model to be deployed within the constraints of the 512 KB flash memory, whilst maintaining average accuracy within 5\% of the full-precision baseline \citep{lawrence_tinyml_2025}.
These examples demonstrate that carefully implemented quantisation strategies can achieve accuracy comparable to full-precision models whilst substantially reducing memory requirements and improving inference speed, making deployment on resource-constrained edge devices practical for real-time ecological monitoring applications.

\clearpage
\begin{wileyboxTitled}[frametitle={BOX \thebox: Overview of AI Models and AI Inference}]
\label{box:ai_overview}

Artificial neural networks are computational models inspired by biological neural systems, composed of layers of interconnected processing units called neurons. 
Each neuron applies a weighted transformation to incoming values, adds a bias term, and passes the result through a non-linear activation function, enabling the network to learn complex relationships between inputs and outputs.
Machine learning (ML) encompasses a broad set of algorithms that learn patterns from data, including decision trees, random forest (RF), support vector machines (SVM), and neural networks, whilst deep learning (DL) refers specifically to neural networks with multiple hidden layers that automatically extract hierarchical feature representations from raw data.
Deep neural networks models typically comprises an \textbf{input layer} that receives raw data (e.g., audio spectrograms, image pixels), one or more \textbf{hidden layers} that extract increasingly abstract features, and an \textbf{output layer} that generates predictions such as species classifications (Figure \ref{figure:box_ai_overview}.A).
\\ 
\\
Neural networks typically store parameters as 32-bit floating-point numbers (FP32), which can represent $2^{32}$ (4,294,967,296) unique values.
Quantisation converts these parameters to lower-precision formats, most commonly 8-bit integers (INT8), which can only take $2^{8}$ (256) different unique values, or 16-bit floating point (FP16, $2^{16}$ or 65,536), thus reducing the range of possible values to represent any numbers.
This process reduces the range of possible values to represent numbers as well as the memory requirements by approximately 75\% and 50\% for INT8 and FP16, respectively, thus accelerating computation (Figure \ref{figure:box_ai_overview}.B).

\begin{center}
\includegraphics[width=1.0\textwidth]{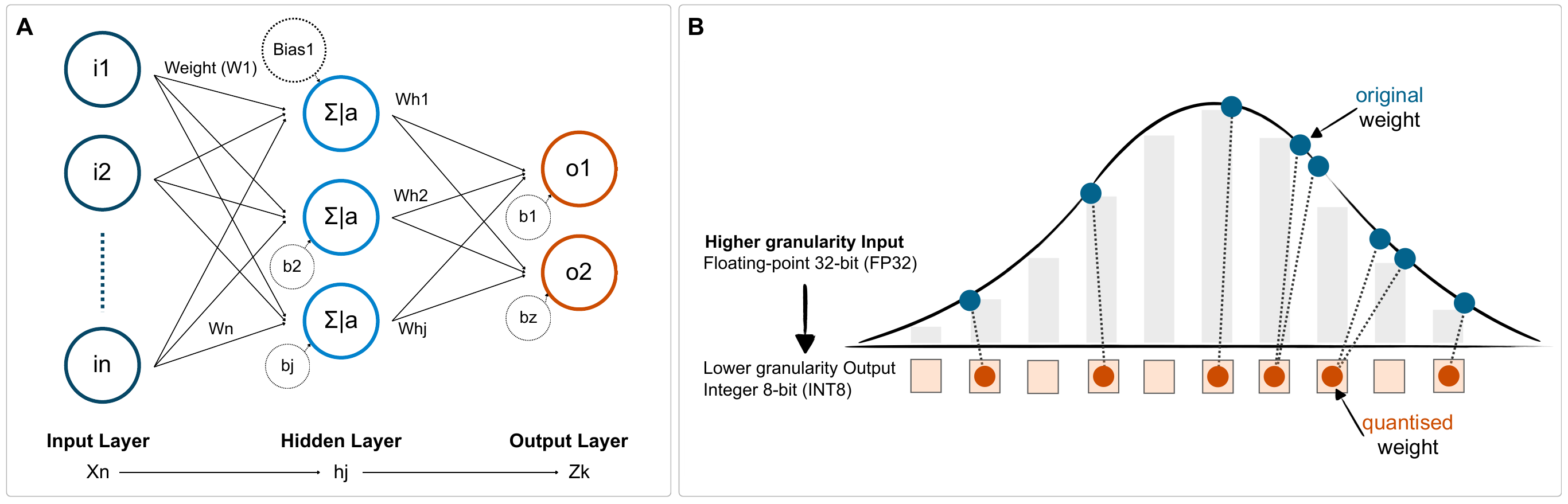}
\captionof{figure}{
    \textbf{Overview of neural networks and quantisation.} 
    (A) Structure of an artificial neural network, illustrating information flow from input to output through weighted connections and bias terms.
    (B) Effect of quantisation. The distribution of FP32 values (blue original weight) is mapped to 256 discrete INT8 bins (orange quantised weight). This trades numerical precision for reduced memory and faster computation.
}
\label{figure:box_ai_overview}
\end{center}
\end{wileyboxTitled}

Complementary optimisation techniques include pruning, which removes parameters that contribute minimally to prediction accuracy \citep{lecun_optimal_1989, han_learning_2015}, and knowledge distillation, which transfers knowledge from a large "teacher" model to a smaller "student" model through specialised training \citep{hinton_distilling_2015}.
The WrenNet model exemplifies the combined application of multiple optimisation strategies, integrating INT8 quantisation and knowledge distillation using the popular bird sound classifier, BirdNET-Analyzer, as teacher model, resulting in a compact student model achieving 70.1\% accuracy across 70 alpine bird species, and representing a 16-fold improvement in energy and time per inference over the BirdNET (INT8 quantised) on RPi 3B+ \citep{ciapponi_enabling_2025}.
Additionally, efficient model architectures such as MobileNet \citep{howard_mobilenets_2017} designed specifically for resource-constrained environments provide foundations for further optimisation.
Determining optimal combinations of optimisation strategies requires identifying computational bottlenecks while balancing ecological consideration, ensuring that the technical trade-offs between model complexity and hardware performance do not compromise reliable predictions for species of conservation concern or result in unmanageable false positive rates.

For successful deployment of AI models on edge devices, AI inference runtime frameworks are a necessary step.
While AI models are typically developed using ML libraries such as PyTorch or TensorFlow on desktop or cloud computers with unlimited resources, deployment on embedded hardware requires substantial model transformation and optimisation.
Lightweight inference runtimes including LiteRT and its MCU variant, specifically targeting TinyML deployments, LiteRT for Microcontrollers (formerly TensorFlow Lite and TensorFlow Lite Micro), PyTorch ExecuTorch, and the Open Neural Network Exchange (ONNX) Runtime, provide the tools needed to convert trained models into compact representations ready for execution.
These frameworks apply graph simplification, operator fusion, and numerical optimisation to reduce computational and memory overhead. 
The resulting portable model formats decouple model development from deployment, allowing researchers to train models in familiar environments before executing optimised versions across heterogeneous hardware platforms \citep{david_tensorflow_2021}.
As edge devices range from bare-metal MCUs to Linux-based SBCs, the choice of inference framework has direct implications for system robustness, portability, and long-term maintainability \citep{theocharides_tinymlefficient_2025}.
Consequently, understanding this software ecosystem is essential for selecting deployment workflows that balance ecological monitoring objectives with practical hardware constraints.

Despite these successes, deploying compact and optimised models remains challenging, as workflows often involve multiple conversion, optimisation, and validation steps that require careful scrutiny \citep{lawrence_tinyml_2025}.
End-to-end platforms such as Edge Impulse seek to address these barriers by providing integrated pipeline for data collection, feature extraction, model training, and deployment through a web-based graphical interface. 
The platform suite of tools enables users with varying technical expertise to develop and test edge AI deployment on diverse targets devices, with the platform Edge Optimised Neural (EON) compiler directly translating models into C/C++ source code, enabling rapid deployment of efficient models across microcontrollers and single-board computers. 
The growing adoption of Edge Impulse underscores that successful edge AI applications depend not only on model performance, but also on accessible toolchains that manage optimisation, compilation, and hardware integration \citep{banbury_edge_2023}. 
By lowering technical barriers and standardising deployment workflows, inference software toolkits  play a central role in translating complex AI models into compact and efficient formats for embedded devices.

\subsection{Wireless Communication Technologies}
One final important aspect of edge computing system is connectivity, which links edge devices to remote infrastructure and users.
Wireless communication networks (WSNs) allow devices to transmit data, receive configuration updates, and enable near-real-time ecological monitoring workflows \citep{davison_automated_2025}.
The choice of wireless communication influences the energy consumption of edge devices, as data transmission consumes substantially more energy than sensing or on-device processing, particularly when transmitting uncompressed sensor data like audio, images, or video \citep{karic_send_2025}.
Consequently, edge computing deployments typically adopt one of two strategies: sending raw sensor data to remote servers for processing, characteristic of traditional Internet of Things (IoT) devices, or transmitting only processed AI outputs and metadata, reducing bandwidth requirements by orders of magnitude.
Evaluating the trade-offs inherent to different communication technology is essential to ensure selected connectivity align with monitoring objectives, deployment environment, and device energy constraints.

WSNs vary in transmission range, data throughput, power consumption, and infrastructure requirements (Table \ref{table:table2.0_wireless_networks}), creating distinct application niches.
Short-range, high-bandwidth technologies such as Wi-Fi (IEEE 802.11) and Bluetooth Low Energy (BLE) operate at 2.4 to 5 GHz, providing megabits per second (Mbps) data rates over tens of metres, suitable for deployments near permanent monitoring stations, where raw data can be periodically offloaded for detailed analysis.
In contrast, long-range, low-power wide-area networks (LPWAN) such as LoRaWAN and cellular IoT standards (LTE-M, NB-IoT) operate at sub-gigahertz frequencies, achieving kilometre-scale ranges whilst consuming only milliwatts, though with limited data rates in kilobits per second (kbps) and payload sizes (under 250 bytes).
Cellular networks (4G LTE, 5G) provide a compromise between range, throughput, and infrastructure availability, but impose subscription costs and elevated power consumption that may constrain long-term deployments. 
Satellite connectivity (e.g., Iridium, Globalstar) extends coverage to remote locations but comes with high transmission costs and requires careful message scheduling \citep{whytock_real-time_2023}.
Finally, telecommunications regulations governing radio spectrum use and site-specific conditions mean that communication technologies cannot be selected in isolation, but must be evaluated alongside the physical realities of deployment sites.

\newcolumntype{L}{>{\RaggedRight\arraybackslash}X}

\begin{table}[!htbp]
    \caption[Overview of wireless communication technologies.]
    {\textbf{Overview of wireless communication technologies.}
    Comparison of wireless protocols for sensor networks, showing trade-offs between transmission distance (range), data throughput (data rate) from kilobits per second (kbps) to gigabits per second (Gbps), operating frequency, and power consumption, with examples of compatible hardware platforms.
    Short-range protocols (Wi-Fi, BLE) provide high throughput for local data transfer, while long-range options (LoRaWAN, Cellular, Satellite) extend coverage to remote sites at reduced throughput.}
\label{table:table2.0_wireless_networks}
\begin{threeparttable}
\small
\setlength\tabcolsep{4pt} 
\setlength{\arrayrulewidth}{0.1pt} 

\begin{tabularx}{\textwidth}{
    >{\hsize=0.8\hsize}L     
    >{\hsize=1.0\hsize}L     
    >{\hsize=0.7\hsize}L     
    >{\hsize=1.0\hsize}L     
    >{\hsize=1.6\hsize}L     
    >{\hsize=0.7\hsize}L     
    >{\hsize=1.2\hsize}L     
}
\headrow
\thead{Protocol} & 
\thead{Standard} & 
\thead{Range} & 
\thead{Data Rate} & 
\thead{Frequency} & 
\thead{Power} & 
\thead{Hardware} 
\\ \hline

Wi-Fi & IEEE 802.11 & 10-100 m & Up to 6.0 Gbps & 2.4 GHz / 5 GHz & High & RPi, Jetson Nano \\

BLE & IEEE 802.15.1 & 1-50 m & 1-2 Mbps / 125-500 kbps & 2.4 GHz & Low & STM32, ESP32 \\

Zigbee & IEEE 802.15.4 & 10–100 m & 20–250 kbps & 2.4 GHz & Very Low & 
\\

LoRaWAN & LoRa Alliance & 1–15 km & 0.3–50 kbps & EU: 433 / 863-870 MHz, US: 902-928 MHz, \newline AU: 915-928 MHz & Very Low & Arduino MKR1310, STM32WLE5 \\

Cellular \newline (4G LTE, 5G, NB-IoT) & 3GPP & 1–20 km & 200 kbps – 1 Mbps & 800-2600 MHz & Moderate & STM32 + Quectel BG95 / RPi + SIM7600 HAT \\

Satellite (Iridium) & Proprietary & Global & ~2.4 kbps & 1616-1626.5 MHz & High & RockBLOCK \newline modules \\

\hline
\end{tabularx}
\end{threeparttable}
\end{table}

\subsection{Practical Deployment Considerations}
Apart from computation and connectivity, the environmental context in which an edge computing system is deployed imposes practical constraints that strongly influence its performance and longevity.
The dominant constraint regards energy consumption. 
Like other passive monitoring biodiversity systems, edge devices rely on batteries or small solar panels, making careful energy budgeting essential \citep{miquel_energy-efficient_2023}.
Hardware selection, sensor duty cycles, inference frequency, and data transmission all contribute to cumulative power demand. 
Power management strategies fall into three main categories: event-triggered capture using external factors like motion, heat, or vibration, a method often used in camera trapping surveys \citep{lahoz-monfort_comprehensive_2021}; time-interval activation, typical for acoustic monitoring \citep{browning_passive_2017}, location tracking \citep{gauld_characterisation_2023} or environmental sensing  \citep{campagnaro_monitoring_2025}; and adapting sampling based on prior detections \citep{millar_terracorder_2024}.
Battery chemistry and capacity further influence deployment longevity, with cold temperatures accelerating batteries discharge and elevated temperatures posing overheating risks. 
Solar panels must also account for seasonal variation and extended periods of low irradiance \citep{callebaut_art_2021}, requiring careful capacity planning to ensure year-round operation.
 
Beyond power considerations, physical robustness is equally critical. 
Edge devices must withstand temperature fluctuations, humidity, precipitation, dust, and interference from animals. 
Enclosure design must therefore balance environmental protection with sensing performance \citep{rhinehart_acoustic_2020, glover-kapfer_camera-trapping_2019}.
Ingress Protection (IP) ratings provide a standardised benchmark, with IP65 generally adequate for terrestrial deployments, while IP67 is recommended for flood-prone environments.
However, increased sealing introduces trade-offs. 
In bioacoustic, sealed housings may attenuate sound, while in vision-based systems, they may exacerbate heat build-up and condensation. 
Mitigation strategies include acoustic membranes, passive ventilation, heat sinks, and desiccants.
Finally, enclosure design must also consider site-specific installation constraints, including mounting methods, wind exposure, and the risk of vandalism.

Effective edge computing deployment requires navigating interconnected design decisions spanning hardware, AI model optimisation, connectivity, energy provisioning, and physical protection. These trade-offs should be guided by clearly defined scientific and operational objectives, starting with the monitoring requirements, such as target taxa, spatial coverage, temporal resolution, and acceptable detection latency, alongside explicit constraints related to budget, site accessibility, and maintenance frequency.

The following section reviews how existing ecological studies have addressed these challenges, synthesising current practice and highlighting emerging opportunities for edge computing in biodiversity monitoring.
    \section{Edge computing applications in ecology}
\label{sec:edge_computing_applications}

\subsection{Edge Computing Review Scope and Analysis Framework}
We conducted a targeted systematic review to identify literature at the intersection of edge computing and biodiversity monitoring, with a specific focus on systems performing on-device or near-sensor AI processing at taxonomic level. 
The search strategy was purposeful rather than fully systematic, designed to capture methodological developments in a rapidly evolving field instead of cataloguing all existing work.
We combined computing terms (e.g., "edge computing", "edge AI", "TinyML", "IoT") with ecological keywords (i.e., "biodiversity", "wildlife", and "ecology") and searched across the ACM Digital Library, arXiv, bioRxiv, IEEE Xplore, and Web of Science databases in October 2025 (Appendix \ref{appendix:review_methods}).
After reviewing abstracts, we included publications describing biodiversity monitoring systems that either execute AI models on resource-constrained hardware, or employed wireless connectivity to support remote inference workflows.
We included studies from laboratory prototypes to long-term field experiments, enabling comparison of systems design trade-offs across deployment contexts.

We focus our scope on sensing systems capturing morphological or behavioural data, being primarily vision-based (e.g., camera traps, video cameras, imaging sensors), acoustic-based, and tracking-based. 
This choice is motivated by the maturity in development of pre-trained AI models for automated taxonomic identification.
We excluded several categories of relevant biodiversity monitoring technologies as operating mostly out-of-the edge computing paradigms, remaining at early development stages, or having been reviewed elsewhere (Appendix \ref{appendix:review_methods}).
We also excluded bio-logging systems without on-device AI or connectivity, theoretical frameworks proposing but not implementing edge systems, and duplicate publications describing earlier iterations of the same system.
In total, we retained 82 publications from 2017 to 2025 for analysis.

\begin{figure}[!h]
\includegraphics[width=0.98\textwidth]{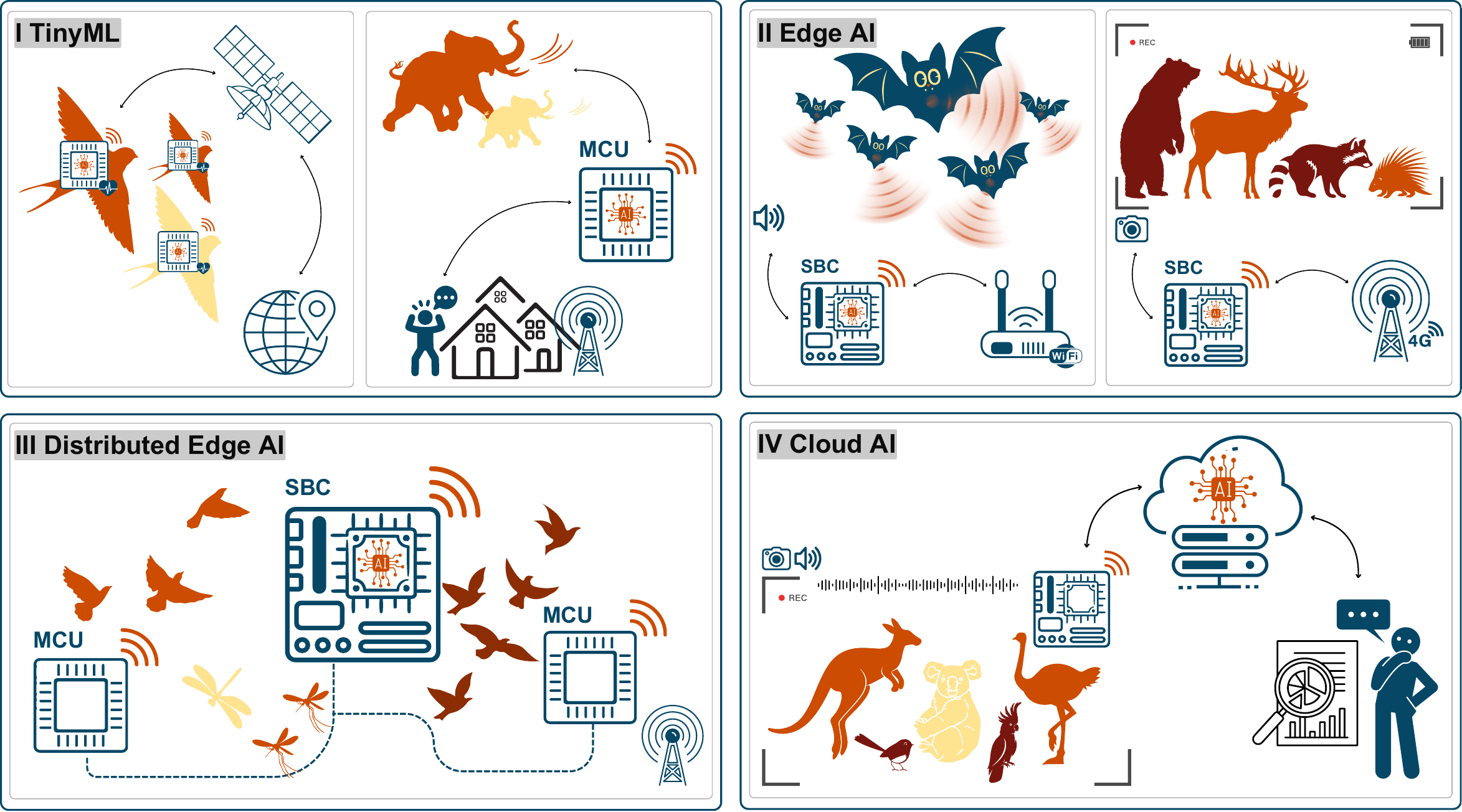}
\centering
\caption{
    \textbf{Conceptual overview of edge AI systems types for biodiversity monitoring.}
    \textbf{I TinyML} systems (or Edge AI on MCUs) run quantised models on microcontrollers (MCUs) for specific events or species detections, transmitting AI inference results via low-power networks such as LoRaWAN. Common for acoustic and movement applications.
    \textbf{II Edge AI} systems (or Edge AI on SBCs) execute computationally intensive models on single-board computers (SBCs) for real-time multi-species classifications, using wide range of connectivity. Common in vision-based and acoustic applications.
    \textbf{III Distributed Edge AI} coordinates multiple sensing nodes with a centralised gateway, allowing spatially distributed data capture while aggregating or coordinating inference locally. Well suited to landscape-scale monitoring.
    \textbf{IV Cloud AI} streams raw data to remote servers, trading transmission energy and inference latency for model sophistication. Common when no sufficiently accurate pre-trained models are available, or for retrospective long-term analysis.
}
\label{figure_section3:edgeai_types}
\end{figure}

Edge computing systems for ecological applications can be categorised into four distinct architectural types, each defined by hardware limitations, power consumption, and connectivity needs (Table \ref{table:table3.0_edgeai_system_types}).
Rather than treating hardware platforms, AI models, and networking technologies as independent components, these typology highlight edge computing as a highly interconnected design space, in which trade-offs between data extend, temporal resolution, energy autonomy, and spatial coverage need to be negotiated in relation to ecological priorities.
This classification provides a unifying framework for understanding how engineering choices constrain biodiversity monitoring applications, addressing distinct scenarios across acoustic, vision, movement, and few other sensing modalities (Figure \ref{figure_section3:edgeai_types}).
\newline

\newcolumntype{L}{>{\RaggedRight\arraybackslash}X}

\begin{table}[hbtp]
\caption{
    \textbf{Edge Computing System Types: Technical Details.}
    Comparison of four architectural configurations distinguished by hardware platforms, AI model complexity, wireless connectivity, and data transfer strategies, revealing essential trade-offs  that define which ecological monitoring questions each system can feasibly address.
}
  
\label{table:table3.0_edgeai_system_types}
\begin{threeparttable}
\footnotesize 
\setlength\tabcolsep{3pt}
\setlength{\arrayrulewidth}{0.1pt}
\begin{tabularx}{\textwidth}{
    >{\hsize=0.1\hsize}L     
    >{\hsize=0.75\hsize}L     
    >{\hsize=1.10\hsize}L     
    >{\hsize=1.15\hsize}L     
    >{\hsize=0.75\hsize}L     
    >{\hsize=1.0\hsize}L     
    >{\hsize=1.25\hsize}L     
    >{\hsize=1.20\hsize}L     
    >{\hsize=0.70\hsize}L     
}
\headrow
\thead{} & 
\thead{System} & 
\thead{Hardware} & 
\thead{AI Models} & 
\thead{WSN} &
\thead{Transfer} & 
\thead{Applications} & 
\thead{Limitations} & 
\thead{Ref.\tnote{*}} 
\\ \hline

\textbf{I} &
\textbf{TinyML} On-device AI, Edge AI on MCUs & 
Cortex-M CPU \newline (e.g. STM32, Arduino MKR, Nano 33 BLE), EFM32 Gecko, ESP32 & 
Quantised CNN (INT8), Custom Lightweight via TFLite micro, CMSIS-NN, \newline Edge Impulse & 
LoRaWAN, Cellular & 
Classification results. \newline Event-driven, \newline opportunistic (near gateway) or time-driven & 
Real-time \newline detection \newline (e.g. illegal logging, invasive species). Human-wildlife conflict. Acoustic and tracking. & 
Small, Tiny AI models, require optimisation, sensitive to model drift & 
\tabref{4, 5, 6, 19, 27, 28, 30, 34, 41, 50, 52, 56, 58, 63, 65, 74}
\\

\textbf{II} &
\textbf{Edge AI} On-device AI, Edge AI on SBCs & 
Cortex-A CPU (e.g. RPi 4), Nvidia Jetson, Google Coral & 
YOLO Models not always \newline quantised. Frameworks TFLite, TensorRT, ONNX & 
Cellular 4G-LTE, WiFi. & 
Crop images w. bounding-box, classification results. \newline Event- or time-driven &
Real-time species classifications. Possible expert validation. \newline Acoustic and vision. & 
High energy consumption limits deployment duration. Rely on available power, connectivity infrastructure & 
\tabref{12, 17, 26, 32, 33, 39, 40, 42, 44, 45, 68, 77, 78, 80}
\\

\textbf{III} &
\textbf{Distributed Edge AI} &
Mixed. \newline MCUs nodes \newline (e.g. STM32, ESP32) + \newline SBC gateway (e.g. RPi 4) & 
Mixed. Lightweight (see Type I) or complex (see Type II) & 
LoRaWAN, BLE, \newline Cellular, WiFi,  \newline Satellite. &
Event- or time-driven &
Landscape-scale surveys, spatially explicit monitoring (e.g. soundscape ecology). \newline Multi-modal. & 
Network maintenance. Careful scheduling of data transmission & 
\tabref{8, 23, 25, 37, 46, 59}
\\

\textbf{IV} &
\textbf{Cloud AI} Remote, central inference &
MCUs, SBCs, retrofitted trail cameras (data collection only). & 
N/A (cloud inference) & 
Cellular \newline 4G-LTE, WiFi, Satellite (Iridium). &
Raw audio, images or videos. \newline Continuous stream or time-driven &
High-resolution images, ultrasonic audio, multi-species classifications. Retrospective analysis. &
Data transmission dominates energy budget, latency precludes real-time use, privacy concerns with cloud storage &
\tabref{1, 2, 21, 29, 31, 51, 55, 61, 67, 69, 71, 76, 79}
\\

\hline
\end{tabularx}
\begin{tablenotes}
\small
\item * Reference numbers correspond to the publication list in Appendix \ref{appendix:list_of_publications}. 
Full details of each publication, including abstracts and extracted data, are available in the Supplementary Materials (see \hyperref[sec:data_availability]{Data Availability Statement}).
\end{tablenotes}
\end{threeparttable}
\end{table}

\subsection{Overview of the reviewed publications}
\begin{figure}[hbtp]
\includegraphics[width=0.94\textwidth]{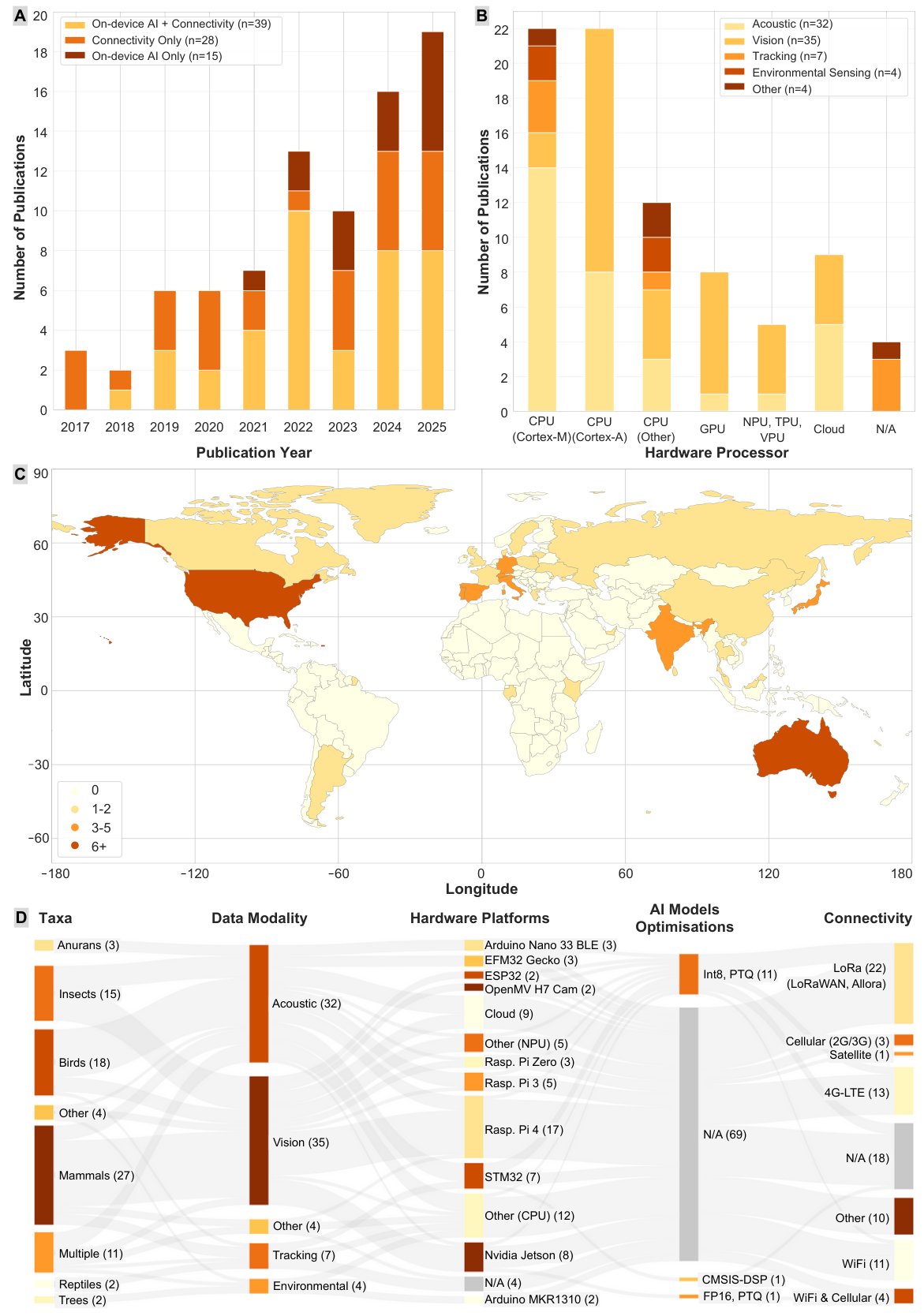}
\centering
\caption{
    \textbf{Publications Analysis Overview}
    Studies using edge computing systems for biodiversity monitoring are summarised by (A) year of publication, (B) processor characteristics and data modalities, 
    (C) country, and (D) ecological and technical categories including monitored taxa, data types, hardware platforms, AI models optimisation techniques, and supported connectivity. Results based on data extracted from 82 articles (See Supplementary Materials \hyperref[sec:data_availability]{Data Availability Statement}).
    }
\label{figure_section3:results_overview}
\end{figure}

The number of publications shows a modest increase over time, with annual publications rising from 3 in 2017 to 19 in 2025 (Figure \ref{figure_section3:results_overview}A).
Systems integrating both on-device AI (i.e. regardless of edge AI system types) and connectivity show a continuous increase, representing nearly half of all studies, reflecting the complementarity between on-device processing and networked communication.
A slight lag exists between the maturity of wireless sensor networks (WSNs) and on-device AI.
While WSNs for environmental sensing and habitat monitoring date back to the mid-2000s \citep{mainwaring_wireless_2002} and expanded to animal tracking and acoustic sensing \citep{alippi_lightweight_2017, elias_wheres_2017, sethi_robust_2018}, research into the deployment of ML and DL algorithms on resource-constrained edge devices gained momentum from 2021 onwards \citep{disabato_birdsong_2021, sabbella_always-tinyml_2022}.
This delayed, yet consistent upward trend reflects the growing maturity and reduced barriers of edge AI deployment,
and the growth of community platforms supporting knowledge sharing and practical advice \citep{wildlabs_2024_2024, speaker_global_2022}.

The retrieved publications span wide range of taxa, sensing modalities, and deployment contexts.
Applications include acoustic monitoring of birds, bats, insects, amphibians, and large mammals \citep{sethi_safe_2020, gallacher_shazam_2021, mainetti_acoustic_2023, monedero_cyber-physical_2021, stahli_development_2022}, vision-based monitoring of large mammals and insects \citep{ibraheam_accurate_2023, ma_smart_2022, elias_wheres_2017, sittinger_insect_2024}, and tracking systems for reptiles, birds, and mammals \citep{de_luca_design_2025, luder_anitrack_2025, alippi_lightweight_2017, babu_perovskite_2024}.
These applications are implemented across different hardware platforms, ranging from ultra-low-power MCUs to GPU-accelerated SBCs, and across communication networks spanning LoRaWAN, cellular, WiFi, and satellite links (Figure \ref{figure_section3:results_overview}B, D).
A missing piece of this review regards publications discussing systems for marine ecosystems.
While examples exist in the literature, including surveillance systems to detect cetaceans and prevent collisions with boats \citep{sanguineti_real-time_2021}, or bomb fishing events \citep{showen_locating_2018}, implementations remain rare, as most marine  monitoring systems follow typical biologging approaches \citep{lamont_hydromoth_2022}.

Geographically, research outputs originate mainly from Europe, North America, and Australia (Figure \ref{figure_section3:results_overview} C).
Although this pattern reflects longstanding disparities in ecological research capacity \citep{speaker_global_2022}, collaborations between institutions and conservation organisations have enabled deployments in biodiversity-rich regions, including Southeast Asian rainforests \citep{sethi_robust_2018}, Africa savannahs \cite{whytock_real-time_2023}, and Latin America \citep{xprize_rainforest_xprize_2024}.

\subsection{Edge Computing Architectural Types}
\subsubsection{Type I: TinyML (Edge AI on MCUs)}
The most power-constrained on-device AI systems are built around ultra-low-power MCUs, most commonly Arm Cortex-M platforms such as STM32, Arduino Nano 33 BLE, or EFM32 Gecko, and are deployed across acoustic, vibrational, and animal-borne sensing contexts. 
These systems run highly optimised INT8 quantised models or classical ML algorithms, trading model complexity and taxonomic resolution for extreme energy efficiency, thus referred to as TinyML systems (Table \ref{table:table3.0_edgeai_system_types} and Figure \ref{figure_section3:edgeai_types}). 
Such design choices enable long-term deployment in remote environments with limited maintenance access, but constrain the ecological questions that can be addressed, favouring targeted detection tasks and binary presence–absence inference for focal species or rare events.

Most edge acoustic applications adopt \textit{type I} architecture. 
Representative examples include detecting chainsaw activity to combat illegal logging \citep{andreadis_monitoring_2021}, identifying disease-vector mosquitoes from wingbeat frequencies \citep{kimutai_low-cost_2023}, recognising wood-boring insects \citep{mainetti_acoustic_2023}, and inferring honeybee queen presence from hive acoustics \citep{de_simone_advancing_2024}. 
Across these studies, target taxa or events produce acoustically constrained signals with distinctive spectral–temporal structure, allowing reliable detection using compact models. \textit{Type I} systems are used for on-animal tracking where strict mass limits constrain hardware choice. 
The adoption of LPWAN protocols in place of GSM has substantially increased deployment lifetimes and spatial coverage \citep{alippi_lightweight_2017, gauld_characterisation_2023, wild_multi-species_2023}, enabling deployment durations from several weeks on battery power to multiple years when supplemented with solar energy harvesting.
Examples include solar-augmented collars enabling near-continuous tracking of bison \citep{babu_perovskite_2024}, centimetre-scale localisation of Aldabra giant tortoises \citep{luder_anitrack_2025}, and Sigfox-connected tags deployed across 30 species from bats to  elephants \citep{wild_multi-species_2023}. 
At present, most \textit{type I} systems transmit raw sensor data for off-device analysis with the deployment of AI models directly on tags remaining limited.
However, recent work propose inferring behavioural states from accelerometer and gyroscope data using lightweight neural networks on iguana \citep{de_luca_design_2025}, representing a potential transition from location tracking to animal-centric sensing, thus expanding ecological insights derived from bio-logging.

To sustain extended deployments, \textit{Type I} systems prioritise power efficiency over data richness. 
This is achieved through energy-triggered pipelines that activate recording or inference only when predefined acoustic thresholds are exceeded \citep{prince_deploying_2019, huang_tinychirp_2024}, combined with the use of low-power connectivity protocols such as LoRaWAN. 
These design choices favour the transmission of lightweight payloads (e.g. alerts, classification scores, GPS coordinates) rather than raw sensor data, thereby limiting downstream expert validation from original recordings.
Ecological inference in type I systems is therefore shaped by the interaction between triggering logic, species vocal behaviour, and site-specific noise conditions, collectively influencing what events are recorded and unobserved.

\subsubsection{Type II: Edge AI (Edge AI on SBCs)}
In contrast to MCU-based systems, \textit{type II} architectures perform on-device inference using more powerful SBCs, such as the Raspberry Pi 4, or neural accelerator platforms including NVIDIA Jetson and Google Coral TPU. These systems illustrate typical edge AI applications (Table \ref{table:table3.0_edgeai_system_types} and Figure \ref{figure_section3:edgeai_types}).
These devices support on-device inference of computationally intensive DL models for both acoustic and visual data.
\textit{Type II} is the most prevalent architecture among the reviewed studies, enabling broader ecological investigations such as community-level surveys using multi-species classifiers and low-latency expert validation workflows. 
However, the increased computational capacity of SBCs entails substantially higher power consumption, typically restricting battery-powered deployments to days or weeks unless supported by mains electricity or solar infrastructure.

In acoustic applications, \textit{type II} systems commonly deploy multi-species convolutional neural networks, exemplified by BirdNET for avian community monitoring \citep{kahl_birdnet_2021} and BatDetect2 for ultrasonic bat call recognition \citep{mac_aodha_towards_2022}. 
The computational capabilities of SBCs enable model deployment without extensive optimisation, as demonstrated by the acoupi system \citep{vuilliomenet_acoupi_2025}. 
Moreover, \textit{type II} systems support the integration of multiple sensors, such as environmental units that record temperature, humidity, soil moisture, and light levels, as illustrated by the BioSense platform \citep{contina_biosense_2024}.

In vision-based applications, \textit{type II} architectures support real-time object detection using models such as You Only Look Once (YOLO), enabling applications including human–wildlife conflict mitigation and road collision prevention without reliance on network connectivity.
For instance, the cassowary detection system provides real-time alerts to motorists \citep{li_endangered_2024}, while the bear prevention system on the Tibetan Plateau processes thermal imagery and activates deterrent mechanisms with millisecond latency \citep{chen_intelligent_2025}. 
Connectivity options such as WiFi and cellular networks are commonly used to transmit selected audio segments or cropped images, as demonstrated by systems including AviEar \citep{verma_aviear_2024} and VespAI \citep{oshea-wheller_vespai_2024}.
AviEar adopts probability-based uplink filtering to transmit only segments containing detected bird vocalisations, optimising cellular data consumption, while the VespAI system sends captured images of the invasive hornet \textit{Vespa velutina} via a local WiFi connection to a paired computer, both enabling expert validation.

The primary limitation of \textit{type II} architectures is energy consumption.
Power draw typically ranges from hundreds of milliamps to several watts, constraining deployments to locations where power infrastructure is accessible, such as ecological field stations \citep{contina_biosense_2024, arshad_where_2020} or urban landscapes \citep{gallacher_shazam_2021}.

\subsubsection{Type III: Distributed Edge AI} 
Whereas \textit{types I} and \textit{II} operate as standalone units, \textit{type III} architectures coordinate multiple lightweight peripheral sensing nodes, typically ESP32 or STM32 MCUs, with a local gateway that aggregates their outputs (Table \ref{table:table3.0_edgeai_system_types} and and Figure \ref{figure_section3:edgeai_types}). 
Peripheral nodes perform coarse-grained inference or signal capture before transmitting compact representations or raw data streams to a higher-capability central processor for processing or further analysis.
\textit{Type III} systems are used to address landscape-scale ecological questions, such as soundscapes and habitat occupancy, by enabling broad spatial coverage while maintaining low power consumption at the node level. 

Illustrative examples include the acoustic system Bird@Edge, which deploys ESP32-based microphone nodes that stream audio over WiFi to a central Jetson Nano for species classification \citep{hochst_birdedge_2022}. 
Similarly, the vision-based Elemantra system that coordinates distributed infrared and seismic sensing nodes with a central Raspberry Pi to enable early detection of elephants and automated alert transmission \citep{bandara_elemantra_2025}.

While \textit{type III} architectures support spatially explicit inference across multiple sites, they introduce additional system-level complexity. 
Communication reliability is influenced by vegetation density, topography, and node–gateway distance, and network constraints can result in variable latency or data loss, as reported for satellite-linked camera trap systems under dense canopy \citep{whytock_real-time_2023}. 
Effective deployment, therefore, requires careful network design, temporal synchronisation, and robust node failure management.

\subsubsection{Type IV: Cloud AI}
Reversing the edge AI paradigm entirely, \textit{type IV} systems separate sensing from inference, transferring raw data to cloud or remote servers for processing (Table \ref{table:table3.0_edgeai_system_types} and Figure \ref{figure_section3:edgeai_types}). 
This approach removes computational constraints, accommodating complex classification tasks and supporting retrospective ecological analysis, which is particularly useful when sufficiently accurate AI models are unavailable and expert validation workflows are required to establish ground-truth data.
Hardware includes commercial camera traps equipped with connectivity modules, Arduino MKR MCUs, and Raspberry Pi devices configured for data acquisition. 
Data are transmitted via cellular, WiFi, or satellite networks, often using batch upload schedules. 

An example is the autonomous eco-acoustic network deployed in Borneo’s tropical rainforests, where solar-powered Raspberry Pi units continuously record audio and transmit files over a mobile network to a central server for indexing, public access, and large-scale soundscape analysis \citep{sethi_safe_2020}. 
Similarly, motion-triggered camera systems for koala monitoring upload short video clips to cloud infrastructure, allowing researchers to generate expert-labelled data to incrementally improve model accuracy \citep{trevathan_computer_2025}.
Beyond acoustic and vision, environmental monitoring systems often adopt \textit{type IV} architecture, transmitting raw sensor readings to cloud platforms for threshold alerts or time-series analysis.
Examples include LoRaWAN-connected systems for urban tree health monitoring \citep{zhao_lorawan-based_2021, zakaria_iot_2020} and water-quality sensing platforms such as SENSWICH deployed in Venice Lagoon \citet{campagnaro_monitoring_2025}.

\textit{Type IV} systems prioritise data completeness and analytical flexibility over real-time responsiveness, making them well suited to research contexts requiring iterative model development.
These systems require significant power and incur substantial connectivity costs as video clips and high-resolution images require transmission, thereby limiting their scalability and popularity. 
\section{Discussion}
\label{sec:discussion}

\subsection{From passive biodiversity monitoring to active ecological knowledge.}
Biodiversity monitoring and the generation of ecological insights are being transformed by major advances in computing technologies, from novel sensing devices to sophisticated AI methods, and extensive data repositories
\citep{sutherland_horizon_2026, tuia_towards_2026, king_implementation_2025}.
Edge AI systems contribute to this transformation by moving inference near the point of data collection, altering not only the efficiency of processing, but when and how ecological insights become actionable.
These systems offer opportunities to interact dynamically with ecological processes as they unfold, shifting monitoring away from the traditional "collect-upload-analyse" paradigm toward timely, proactive actions \citep{davison_automated_2025}.
This distinction, between passive data accumulation and systems sending real-time alerts and performing adaptive sampling, reconfigures ecological sensing, with implications for how monitoring devices are designed, evaluated, and interpreted.
Our synthesis of 82 publications reveals that edge AI systems have matured beyond proof-of-concept prototypes to operational deployments spanning acoustic, visions, and movement modalities.
By categorising systems into four architectural types (Table \ref{table:table3.0_edgeai_system_types} and Figure \ref{figure_section3:edgeai_types}), we show that no single configuration is universally appropriate.
Instead, deployment must be guided by ecological objectives, including spatial coverage, taxonomic resolution, temporal granularity, and intervention requirements, before translation into context-specific hardware and deployment architectures.
While the implications discussed here should be interpreted within the scope of the review (Appendix \ref{appendix:review_methods}), this synthesis underscores that edge AI systems are ecological instruments, not universal technical solutions.

The transition from retrospective analysis to responsive sensing is essential to adapt conservation practice under rising environmental change, land-use conversion, and climate extremes \citep{dietze_iterative_2018}.
Advancements in both model architectures and sensing hardware could enable edge AI systems to infer richer ecological signals than species presence-absence.
A first opportunity is to extend current classification systems to behavioural and physiological assessments.
On-device models could analyse movement patterns, vocalisations, or morphological features to infer individual condition, enabling population health monitoring at spatial and temporal scales previously unattainable \citep{sanakoyeu_transferring_2020, mathis_deep_2020, zuffi_three-d_2019}.
Computer vision models are being developed to estimate injury, body condition, or reproductive status from video \citep{muramatsu_wildpose_2025}. 
Similarly, bioacoustic models are trained to distinguish socially meaningful vocalisations, such as feeding buzzes, alarm calls, and mating songs, providing windows into social dynamics, foraging success, and predation risk \citep{stowell_computational_2022}. 
These advances remain constrained by imbalanced datasets and limited behavioural or physiological labels in ecological training data. 
Whilst emerging approaches such as transfer learning, few-shot learning, and synthetic data augmentation show promise in mitigating these limitations \citep{reynolds_potential_2025}, targeted field data collection by ecologists remains essential, underscoring the interdependence of technological development and domain expertise \citep{tuia_perspectives_2022}.

A second advance concerns the integration of multimodal sensing. 
Current edge AI systems typically rely on a single data modality, limiting interpretation of broader ecosystem interactions \citep{stephenson_measuring_2022, christin_applications_2019}.
Edge platforms, however, offer opportunities to fuse environmental, acoustic, visual, and chemical sensors locally \citep{wagele_towards_2022}, an approach well-suited to \textit{type III} distributed architectures, where fusion tasks are offloaded to gateway devices.
Multimodal models could be trained to exploit relationships among species occurrences, environmental conditions, and soundscape profiles, improving classification confidence and reducing false positives \citep{miao_new_2025, pollock_harnessing_2025}. 
Processing multiple high-bandwidth data streams, however, remains computationally demanding, requiring optimisation of sensing schedules and fusion strategies to maintain acceptable energy budgets.
Yet, emerging hardware innovations, including affordable AI accelerators such as the RPi AI HATs (HAT+ and HAT+ 2), as well as better power management hardware exemplified by the PV-Pi system \citep{ditria_pv_2025}, expand the feasibility of such approaches.

Finally, a third approach regards the potential to integrate edge-processed outputs directly into operational biodiversity infrastructures.
Online dashboards and decision-support interfaces have already allowed practitioners to visualise trends and receive alerts in near real-time \citep{gallacher_shazam_2021, sethi_safe_2020}, although they often require technical expertise and infrastructure maintenance that many ecological teams lack \citep{beardsley_addressing_2018}.
More sustainably, integrating edge observations into existing infrastructures such as Movebank, GBIF, and LifeWatch could consolidate regional monitoring efforts and enable interpretation of local observations in a global context.
While animal tracking data flows routinely via Movebank \citep{kays_movebank_2022}, comparable pipelines for acoustic or camera-trap inferences remain rare, with BirdWeather \citep{clark_birdweather_2021} representing a notable exception.
Extending these digital would help reveal large-scale patterns of change invisible to isolated studies \citep{cowans_improving_2026}.
Taken together, these developments reinforce that edge AI systems extend the potential of ecological conservation, yet effective system design requires convergence between hardware innovation, software accessibility, and field implementation.

\subsection{Edge AI systems as a set of ecological–technical trade-offs, not a single solution}
Edge AI systems are best understood as negotiated compromises between ecological objectives and technical constraint. 
Design choices related to hardware, model complexity, energy availability, and connectivity directly shape what ecological questions can be addressed.
The distinction between \textit{type I} TinyML system, which enable long-duration monitoring to the expense of taxonomic resolution, and \textit{type II} SBC architecture, which offer richer outputs at the cost of higher power consumption, illustrates a core tension between model complexity and energy autonomy. 
This tension is further exemplified by \textit{Type III} distributed architectures, which coordinates multiple sensing nodes, whilst introducing network maintenance complexity, and \textit{type IV} cloud AI system,s which trade inference latency and transmission costs in exchange for unlimited compute. 
Ultimately, these diverse system types demonstrate that their capabilities are defined by how successfully ecological goals are translated into viable technical specifications.

The practical implementation of edge AI systems is further governed by the maturity and accessibility of AI deployment toolchains.
\textit{Type I} TinyML systems often require quantisation and low-level optimisation, yet accessible deployment tools remain limited beyond platforms such as Edge Impulse \citep{banbury_edge_2023}.
In contrast, \textit{type II, III} and \textit{IV} systems can deploy larger models without extensive compression, and benefit from mature ecosystems for data annotation and model training, including low-code platforms \citep{martinez_balvanera_whombat_2025, kellenberger_aide_2020}.
Model-sharing repositories such as HuggingFace reduce barriers to access pre-trained models, but ecology-specific models that are both well-documented and edge-optimised remain scarce \citep{horwitz_we_2025}.
As a result, ecologists seeking to deploy edge AI often face steep technical barriers, frequently requiring manual implementation of optimisation techniques that demand considerable machine learning expertise \citep{warden_tinyml_2020}.

Large pre-trained and foundation models offer the most promising route to reducing the model development burden.
Models such as BirdNET \citep{kahl_birdnet_2021}, DeepFaune \citep{rigoudy_deepfaune_2023}, MegaDetector \citep{pocock_developing_2015}, and Perch \citep{merrienboer_perch_2026} demonstrate good baseline performance on global datasets and can be fine-tuned to local species assemblages, reducing annotation requirements and computational training costs \citep{dumoulin_search_2025, dussert_zero-shot_2025}.
However, deploying foundation models across \textit{types I, II}, and \textit{III} systems remains challenging, typically necessitating extensive model compression and optimisation \citep{giorgetti_transitioning_2025}.  
As optimisation toolchains and edge hardware continue to advance, deployment of increasingly complex models is likely to accelerate.
Realising this potential requires adherence to FAIR principles \citep{wilkinson_fair_2016}, including transparent reporting of training data, performance across taxa and environmental conditions, and use-case limitations.

A further under-discussed dimension of edge AI deployment concerns the irreversible decisions that on-device systems make about data retention.
Edge AI systems inherently determine what constitutes signal versus noise and whether data are retained, transmitted, or discarded. 
Data treated as noise in one study may hold ecological value in another, particularly for long-term or retrospective analyses \citep{darras_worldwide_2025}. 
There is, therefore, an imperative for researchers to clearly report model confidence thresholds governing on-device filtering, data retention strategies, and device duty cycles.
Balancing comprehensive data preservation, which incurs storage and computation costs often acceptable in academic contexts, with aggressive on-device filtering, which may be necessary for conservation organisations with limited resources, remains an unresolved tension in the design and use of edge AI systems.
Further investigation is needed to understand how much data should be retained to enable both immediate monitoring objectives and the long-term scientific goals and values archived sensor data.

\subsection{Towards reproducible and responsible deployment.}
A persistent limitation across the reviewed literature is the inconsistent reporting of deployment and performance metrics, a gap that undermines reproducibility and system comparison.
Our analysis reveals that only half of publications report quantitative energy consumption, fewer than one-third describe duty-cycle configurations, only 16\% specify AI model size and 40\% report inference time.
Moreover, model drift and long-term performance degradation remain systematically under-studied, with few surveys reporting classification accuracy over deployment duration or documenting how seasonal environmental changes affect detection rates. 
Laboratory validation on benchmark datasets provides initial performance estimates but may bear little relation to field conditions characterised by variable weather, shifting ambient noise, sensor fouling, and calibration drift.
To address this reporting fragmentation, we propose a set of minimum reporting standards to support reproducible field deployment of edge AI systems (Table \ref{table:table4.0_reporting_standard}).


\newcolumntype{L}{>{\RaggedRight\arraybackslash}X}

\begin{table}[hbtp]
\caption[Minimum reporting standards for edge AI systems in biodiversity monitoring.]
{
    \textbf{Minimum reporting standards for edge AI systems in biodiversity monitoring.}
    Proposition of five domains to improve the documentation of edge AI systems, and enable reproducibility and cross-study comparison.
}   

\label{table:table4.0_reporting_standard}
\begin{threeparttable}
\setlength\tabcolsep{4pt}
\setlength{\arrayrulewidth}{0.1pt}
\begin{tabularx}{\textwidth}{
    >{\hsize=0.7\hsize}L     
    >{\hsize=1.5\hsize}L     
    >{\hsize=1.0\hsize}L     
    >{\hsize=0.8\hsize}L     
}

\headrow
\thead{Reporting Domain} & 
\thead{Minimum Information} & 
\thead{Rationale} & 
\thead{Examples}
\\ \hline

\textbf{Bill of materials} &
Comprehensive list of hardware \newline components, including part numbers, descriptions, quantities, and costs. &
Facilitate reproducibility and cost estimation for future deployments. &
\citet{contina_biosense_2024}
\\

\textbf{Energy budget} &
Energy profile across operational state (i.e., idle, sensing, processing, transmission); Battery chemistry and capacity; Solar panel specifications. &
Estimate deployment \newline duration. &
\citet{hochst_birdedge_2022, disabato_birdsong_2021}
\\

\textbf{AI model \newline specification} &
Model architecture, version, and size; \newline Parameter count; Quantisation format (e.g., INT8, FP16); Inference runtime framework; Inference time per sample on target hardware. &
Assessment of model \newline performance vs. efficiency across hardware; 
\newline Help clarify resource \newline requirements. &
\citet{bjerge_towards_2024, mahbub_bat2web_2024}
\\

\textbf{Duty cycle \newline configuration} &
Wake-up policy (e.g., event-triggered, time-interval, adaptive sampling);
\newline Sampling rate (e.g., continuous, periodic, adaptive); Transmission interval \newline (e.g., real-time, batch, opportunistic). &
Define energy consumption and data resolution; 
Clarify proportion of ecological activity captured. &
\citet{knyva_iot_2023, de_simone_advancing_2024}
\\

\textbf{In-situ detection performance} &
Precision, recall, and confidence \newline thresholds under field conditions; \newline 
classification accuracy over deployment duration; 
documentation of model drift. &
Laboratory benchmarks \newline vs. field performance;
&
\citet{trevathan_computer_2025, li_endangered_2024}
\\

\textbf{Data management} &
Specification of data storage and retention;
Confidence thresholds governing on-device filtering; 
Data transmission strategy;
Raw data archival policy &
On-device filtering is \newline irreversible. &
\citet{vuilliomenet_acoupi_2025}
\\ \hline

\end{tabularx}
\end{threeparttable}
\end{table}

These standards focus on aspects of system design that influence ecological inference, including energy budgets, AI model specification, duty cycle configuration, in situ detection performance, and data retention policies.
While tools such as the AudioMoth deployment calculator set a standard for battery-operated recorders \citep{prince_deploying_2019}, equivalent frameworks that incorporate AI inference frequency, wireless overhead, and energy harvesting remain underdeveloped. 
Most current software tools provide limited support for configuring edge AI device parameters for sensing, inference, and data transmission, a shortcoming that the acoupi system  addresses \citep{vuilliomenet_acoupi_2025}. 
Nevertheless, establishing community reporting standards covering bills of materials, energy profiles, device settings, and AI metrics, would improve comparability and accelerate learning across deployments \citep{speaker_global_2022}.

As deployments scale, the environmental footprint of the monitoring technologies requires greater attention.
Device manufacture, alongside computation and storage of large ecological datasets, contributes to greenhouse gas emissions and resource use \citep{zikulnig_life_2025}.
Long-term deployments raise concerns around electronic waste, material toxicity, and degradation, including battery leakage and fragmentation of polymer housings \citep{bathaei_environmental_2025}. 
While biodegradable electronics based on cellulose substrates are promising, these remain constrained by durability and performance limitations \citep{aliyana_review_2023, teixeira_review_2023}.
More immediate solutions could come from improved enclosure using biopolymer composites and cellulose nanofibres \citep{madsen_materials_2025}.
Evaluating the sustainability of edge AI systems requires considering both operational longevity and end-of-life impacts.

Ethical governance is equally important as edge computing systems become more prevalent and sophisticated.
Edge AI systems should be designed to minimise the risk of privacy encroachment, ensuring that inadvertently captured sensitive data, such as human conversations or imagery, is directly filtered out and deleted \citep{sandbrook_principles_2021}.
Balancing the imperative for open data sharing to advance scientific research with the need to protect sensitive information from misuse presents ongoing challenges that requires sustained collaboration among local communities stewarding ecosystems, researchers advancing knowledge and governmental agencies regulating conservation \cite{jennings_governance_2025, pritchard_data_2022}.
Grounding technological development in ecological questions, underpinned by transparent reporting and ethical practice, is crucial to move the field beyond fragmented experimentation into a robust component of biodiversity monitoring at scale.
Encouragingly, community platforms and initiatives (e.g., WILDLABS, PyTorch Wildlife, XPrize Rainforest) increasingly support cross-domain collaboration and should be celebrated for advancing the field forward and contributing to the dissemination of best practices.
    \section{Concluding Remarks}
\label{sec:conclusion}
Edge computing is rapidly transforming how biodiversity data are collected, processed, and acted upon, enabling a shift from retrospective analysis to autonomous, preemptive ecological monitoring. 
Our synthesis of 82 publications demonstrates that edge AI systems have advanced from proof-of-concept prototypes to operational field deployments across acoustic, vision-based, and movement modalities.
We identified four distinct architectural types, from TinyML and Edge AI to Distributed Edge AI and Cloud AI, each representing trade-offs among computational capacity, energy autonomy, inference latency, and spatial coverage.
Field implementations of each system type demonstrate the feasibility of responsive biodiversity monitoring, supporting applications such as human–wildlife conflict mitigation and species behavioural studies, yet no single type is universally optimal.
Realising the full potential of edge AI systems  requires sustained interdisciplinary collaboration between ecologists, engineers, and local stakeholders, alongside transparent reporting practices and modular system designs.
Grounding technological development in ecological questions is essential for moving the field beyond fragmented experimentation into a robust component of biodiversity monitoring capable of delivering  observation at the spatial and temporal scales conservation demands.

    \clearpage
    \section*{authors contributions}
Aude Vuilliomenet: Conceptualisation (lead); investigation (lead); visualisation (lead); writing—original draft (lead). Kate E. Jones: Resources (equal); supervision (equal); writing—review and editing (equal) Duncan Wilson: Resources (equal); supervision (equal); writing—review and editing (equal).
    \section*{acknowledgements}
This work was financially supported by URKI EPSRC (EP/R513143/1 and EP/T517793/1). 
    \section*{conflict of interest}
The authors declare that they have no conflicts of interest.
    \section*{data availability statement}
\label{sec:data_availability}

Supplementary table with complete reference list used to create Figure \ref{figure_section3:results_overview}, along with their abstracts, is available as Supporting Materials to this article as well as on the following publicly interactive \href{https://docs.google.com/spreadsheets/d/1VsmnpAhPw3fcmjqI5Tb2d8ppendoLl2FGnyKLpVg04A/edit?usp=sharing}{Google Spreadsheet}.
    \section*{ORCID}
\textit{Aude Vuilliomenet} \orcidlinkf{0000-0003-1517-9296} \\
\textit{Kate E. Jones }\orcidlinkf{0000-0001-5231-3293} \\
\textit{Duncan Wilson} \orcidlinkf{0000-0001-6041-8044} \\
    
    \bibliography{bibliography/references_edgeai}

\begin{thebibliography}{147}
\expandafter\ifx\csname natexlab\endcsname\relax\def\natexlab#1{#1}\fi
\expandafter\ifx\csname url\endcsname\relax
  \def\url#1{\texttt{#1}}\fi
\expandafter\ifx\csname urlprefix\endcsname\relax\def\urlprefix{URL: }\fi

\bibitem[{Abrahms et~al.(2023)Abrahms, Carter, Clark-Wolf, Gaynor, Johansson,
  McInturff, Nisi, Rafiq and West}]{abrahms_climate_2023}
Abrahms, B., Carter, N.~H., Clark-Wolf, T.~J., Gaynor, K.~M., Johansson, E.,
  McInturff, A., Nisi, A.~C., Rafiq, K. and West, L. (2023) Climate change as a
  global amplifier of human--wildlife conflict.
\newblock \textit{Nature Climate Change}, \textbf{13}, 224--234.
\newblock \urlprefix\url{https://www.nature.com/articles/s41558-023-01608-5}.

\bibitem[{Alippi et~al.(2017)Alippi, Ambrosini, Longoni, Cogliati and
  Roveri}]{alippi_lightweight_2017}
Alippi, C., Ambrosini, R., Longoni, V., Cogliati, D. and Roveri, M. (2017) A
  lightweight and energy-efficient {Internet}-of-birds tracking system.
\newblock In \textit{2017 {IEEE} {International} {Conference} on {Pervasive}
  {Computing} and {Communications} ({PerCom})}, 160--169.
\newblock \urlprefix\url{https://ieeexplore.ieee.org/document/7917862}.

\bibitem[{Aliyana and Stylios(2023)}]{aliyana_review_2023}
Aliyana, A.~K. and Stylios, G. (2023) A {Review} on the {Progress} in
  {Core}-{Spun} {Yarns} ({CSYs}) {Based} {Textile} {TENGs} for {Real}-{Time}
  {Energy} {Generation}, {Capture} and {Sensing}.
\newblock \textit{Advanced Science}, \textbf{10}, 2304232.
\newblock
  \urlprefix\url{https://onlinelibrary.wiley.com/doi/abs/10.1002/advs.202304232}.
\newblock \_eprint:
  https://advanced.onlinelibrary.wiley.com/doi/pdf/10.1002/advs.202304232.

\bibitem[{Allan et~al.(2018)Allan, Nimmo, Ierodiaconou, VanDerWal, Koh and
  Ritchie}]{allan_futurecasting_2018}
Allan, B.~M., Nimmo, D.~G., Ierodiaconou, D., VanDerWal, J., Koh, L.~P. and
  Ritchie, E.~G. (2018) Futurecasting ecological research: the rise of
  technoecology.
\newblock \textit{Ecosphere}, \textbf{9}, e02163.
\newblock
  \urlprefix\url{https://onlinelibrary.wiley.com/doi/abs/10.1002/ecs2.2163}.
\newblock \_eprint:
  https://esajournals.onlinelibrary.wiley.com/doi/pdf/10.1002/ecs2.2163.

\bibitem[{Andreadis et~al.(2021)Andreadis, Giambene and
  Zambon}]{andreadis_monitoring_2021}
Andreadis, A., Giambene, G. and Zambon, R. (2021) Monitoring {Illegal} {Tree}
  {Cutting} through {Ultra}-{Low}-{Power} {Smart} {IoT} {Devices}.
\newblock \textit{Sensors}, \textbf{21}, 7593.
\newblock \urlprefix\url{https://www.mdpi.com/1424-8220/21/22/7593}.

\bibitem[{Arshad et~al.(2020)Arshad, Barthelemy, Pilton and
  Perez}]{arshad_where_2020}
Arshad, B., Barthelemy, J., Pilton, E. and Perez, P. (2020) Where is my
  {Deer}?-{Wildlife} {Tracking} {And} {Counting} via {Edge} {Computing} {And}
  {Deep} {Learning}.
\newblock In \textit{2020 {IEEE} {SENSORS}}, 1--4.
\newblock \urlprefix\url{https://ieeexplore.ieee.org/document/9278802/}.

\bibitem[{Babu et~al.(2024)Babu, Pineda, Bizan, Wojak, Wierzowiecki,
  Gerv{\'a}sio, Szklarz, Castriotta and Carlo}]{babu_perovskite_2024}
Babu, V., Pineda, R.~F., Bizan, M., Wojak, A., Wierzowiecki, S., Gerv{\'a}sio,
  J., Szklarz, J., Castriotta, L.~A. and Carlo, A.~D. (2024) Perovskite {Solar}
  {Module} {Enabled} {IoT} {Asset} {Tracking} for {Wildlife} {Conservation}.
\newblock \textit{IEEE Journal of Photovoltaics}, \textbf{14}, 337--343.
\newblock \urlprefix\url{https://ieeexplore.ieee.org/document/10430371/}.

\bibitem[{Banbury et~al.(2023)Banbury, Janapa~Reddi, Hymel, Elium, Tischler,
  Situnayake, Ward, Moreau, Plunkett, Kelcey, Baaijens, Grande, Maslov, Beavis,
  Jongboom and Quaye}]{banbury_edge_2023}
Banbury, C., Janapa~Reddi, V., Hymel, S., Elium, A., Tischler, D., Situnayake,
  D., Ward, C., Moreau, L., Plunkett, J., Kelcey, M., Baaijens, M., Grande, A.,
  Maslov, D., Beavis, A., Jongboom, J. and Quaye, J. (2023) Edge {Impulse}:
  {An} {MLOps} {Platform} for {Tiny} {Machine} {Learning}.
\newblock \textit{Proceedings of Machine Learning and Systems}, \textbf{5},
  254--268.
\newblock
  \urlprefix\url{https://proceedings.mlsys.org/paper_files/paper/2023/hash/49fe55f5e9574714dda575bfb2177662-Abstract-mlsys2023.html}.

\bibitem[{Bandara and Bandara(2025)}]{bandara_elemantra_2025}
Bandara, N.~S. and Bandara, D.~P. (2025) Elemantra: {An} {End}-to-{End}
  {Framework} {Empowered} with {Edge} {AI} to {Tackle} {Human}-{Elephant}
  {Conflict}.
\newblock In \textit{2025 {IEEE} {Applied} {Sensing} {Conference} ({APSCON})},
  1--4. Hyderabad, India: IEEE.
\newblock \urlprefix\url{https://ieeexplore.ieee.org/document/11144360/}.

\bibitem[{Bathaei et~al.(2025)Bathaei, Bathaei, Liao, Yazdanmehr, Sethi,
  Nikolayev, Cardoso and Boutry}]{bathaei_environmental_2025}
Bathaei, M.~J., Bathaei, Y., Liao, Z., Yazdanmehr, M., Sethi, S.~S., Nikolayev,
  D., Cardoso, F.~A. and Boutry, C.~M. (2025) Environmental and {Ecological}
  {Monitoring} with {Biodegradable} {Technologies}.
\newblock \textit{Advanced Science}, e11452.
\newblock
  \urlprefix\url{https://advanced.onlinelibrary.wiley.com/doi/10.1002/advs.202511452}.

\bibitem[{Beardsley et~al.(2018)Beardsley, Gropp and
  Verdier}]{beardsley_addressing_2018}
Beardsley, T.~M., Gropp, R.~E. and Verdier, J.~M. (2018) Addressing
  {Biological} {Informatics} {Workforce} {Needs}: {A} {Report} from the {AIBS}
  {Council}.
\newblock \textit{BioScience}, \textbf{68}, 847--853.
\newblock \urlprefix\url{https://doi.org/10.1093/biosci/biy116}.

\bibitem[{Beery et~al.(2019)Beery, Morris and Yang}]{beery_efficient_2019}
Beery, S., Morris, D. and Yang, S. (2019) Efficient {Pipeline} for {Camera}
  {Trap} {Image} {Review}.
\newblock \urlprefix\url{http://arxiv.org/abs/1907.06772}.
\newblock ArXiv:1907.06772 [cs].

\bibitem[{Beng and Corlett(2020)}]{beng_applications_2020}
Beng, K.~C. and Corlett, R.~T. (2020) Applications of environmental {DNA}
  ({eDNA}) in ecology and conservation: opportunities, challenges and
  prospects.
\newblock \textit{Biodiversity and Conservation}, \textbf{29}, 2089--2121.
\newblock \urlprefix\url{https://doi.org/10.1007/s10531-020-01980-0}.

\bibitem[{Bier(2020)}]{bier_whats_2020}
Bier, J. (2020) What's {Driving} {AI} and {Vision} to the {Edge}.
\newblock
  \urlprefix\url{https://www.eetasia.com/whats-driving-ai-and-vision-to-the-edge/}.

\bibitem[{Bjerge et~al.(2024)Bjerge, Karstoft and
  H{\o}ye}]{bjerge_towards_2024}
Bjerge, K., Karstoft, H. and H{\o}ye, T.~T. (2024) Towards edge processing of
  images from insect camera traps.
\newblock \textit{bioRxiv}, 2024.07.01.601488.
\newblock
  \urlprefix\url{https://www.biorxiv.org/content/10.1101/2024.07.01.601488v2}.

\bibitem[{Borowiec et~al.(2022)Borowiec, Dikow, Frandsen, McKeeken, Valentini
  and White}]{borowiec_deep_2022}
Borowiec, M.~L., Dikow, R.~B., Frandsen, P.~B., McKeeken, A., Valentini, G. and
  White, A.~E. (2022) Deep learning as a tool for ecology and evolution.
\newblock \textit{Methods in Ecology and Evolution}, \textbf{13}, 1640--1660.
\newblock
  \urlprefix\url{https://onlinelibrary.wiley.com/doi/abs/10.1111/2041-210X.13901}.
\newblock \_eprint:
  https://besjournals.onlinelibrary.wiley.com/doi/pdf/10.1111/2041-210X.13901.

\bibitem[{Browning et~al.(2017)Browning, Gibb, Glover-Kapfer and
  Jones}]{browning_passive_2017}
Browning, E., Gibb, R., Glover-Kapfer, P. and Jones, K.~E. (2017) Passive
  acoustic monitoring in ecology and conservation.
\newblock \textit{{WWF} {Conservation} {Technology}}, WWF-UK, Woking, United
  Kingdom.
\newblock
  \urlprefix\url{https://www.wwf.org.uk/sites/default/files/2019-04/Acousticmonitoring-WWF-guidelines.pdf}.

\bibitem[{Callebaut et~al.(2021)Callebaut, Leenders, Van~Mulders, Ottoy,
  De~Strycker and Van~der Perre}]{callebaut_art_2021}
Callebaut, G., Leenders, G., Van~Mulders, J., Ottoy, G., De~Strycker, L. and
  Van~der Perre, L. (2021) The {Art} of {Designing} {Remote} {IoT}
  {Devices}---{Technologies} and {Strategies} for a {Long} {Battery} {Life}.
\newblock \textit{Sensors}, \textbf{21}, 913.
\newblock \urlprefix\url{https://www.mdpi.com/1424-8220/21/3/913}.
\newblock Number: 3.

\bibitem[{Calvin et~al.(2023)Calvin, Dasgupta, Krinner, Mukherji, Thorne,
  Trisos, Romero, Aldunce, Barrett, Blanco, Cheung, Connors, Denton,
  Diongue-Niang, Dodman, Garschagen, Geden, Hayward, Jones, Jotzo, Krug, Lasco,
  Lee, Masson-Delmotte, Meinshausen, Mintenbeck, Mokssit, Otto, Pathak, Pirani,
  Poloczanska, P{\"o}rtner, Revi, Roberts, Roy, Ruane, Skea, Shukla, Slade,
  Slangen, Sokona, S{\"o}rensson, Tignor, Van~Vuuren, Wei, Winkler, Zhai,
  Zommers, Hourcade, Johnson, Pachauri, Simpson, Singh, Thomas, Totin, Arias,
  Bustamante, Elgizouli, Flato, Howden, M{\'e}ndez-Vallejo, Pereira,
  Pichs-Madruga, Rose, Saheb, S{\'a}nchez~Rodr{\'i}guez, {\"U}rge-Vorsatz,
  Xiao, Yassaa, Alegr{\'i}a, Armour, Bednar-Friedl, Blok, Ciss{\'e}, Dentener,
  Eriksen, Fischer, Garner, Guivarch, Haasnoot, Hansen, Hauser, Hawkins,
  Hermans, Kopp, Leprince-Ringuet, Lewis, Ley, Ludden, Niamir, Nicholls, Some,
  Szopa, Trewin, Van Der~Wijst, Winter, Witting, Birt, Ha, Romero, Kim, Haites,
  Jung, Stavins, Birt, Ha, Orendain, Ignon, Park, Park, Reisinger, Cammaramo,
  Fischlin, Fuglestvedt, Hansen, Ludden, Masson-Delmotte, Matthews, Mintenbeck,
  Pirani, Poloczanska, Leprince-Ringuet and P{\'e}an}]{lee_ipcc_2023}
Calvin, K., Dasgupta, D., Krinner, G., Mukherji, A., Thorne, P.~W., Trisos, C.,
  Romero, J., Aldunce, P., Barrett, K., Blanco, G., Cheung, W.~W., Connors, S.,
  Denton, F., Diongue-Niang, A., Dodman, D., Garschagen, M., Geden, O.,
  Hayward, B., Jones, C., Jotzo, F. et~al. (2023) {IPCC}, 2023: {Climate}
  {Change} 2023: {Synthesis} {Report}. {Contribution} of {Working} {Groups}
  {I}, {II} and {III} to the {Sixth} {Assessment} {Report} of the
  {Intergovernmental} {Panel} on {Climate} {Change} [{Core} {Writing} {Team},
  {H}. {Lee} and {J}. {Romero} (eds.)]. {IPCC}, {Geneva}, {Switzerland}.
\newblock \textit{Tech. rep.}, Intergovernmental Panel on Climate Change
  (IPCC).
\newblock \urlprefix\url{https://www.ipcc.ch/report/ar6/syr/}.
\newblock Edition: First.

\bibitem[{Campagnaro et~al.(2025)Campagnaro, Ghalkhani, Tumiati, Marin,
  Dal~Grande, Pozzebon, De~Battisti, Francescon and
  Zorzi}]{campagnaro_monitoring_2025}
Campagnaro, F., Ghalkhani, M., Tumiati, R., Marin, F., Dal~Grande, M.,
  Pozzebon, A., De~Battisti, D., Francescon, R. and Zorzi, M. (2025) Monitoring
  the {Venice} {Lagoon}: {An} {IoT} {Cloud}-{Based} {Sensor} {Network}
  {Approach}.
\newblock \textit{IEEE Journal of Oceanic Engineering}, \textbf{50}, 570--582.

\bibitem[{Caron et~al.(2025)Caron, Noura, Nakache, Guyeux and
  Aynes}]{caron_ai_2025}
Caron, N., Noura, H.~N., Nakache, L., Guyeux, C. and Aynes, B. (2025) {AI} for
  {Wildfire} {Management}: {From} {Prediction} to {Detection}, {Simulation},
  and {Impact} {Analysis}---{Bridging} {Lab} {Metrics} and {Real}-{World}
  {Validation}.
\newblock \textit{AI}, \textbf{6}.
\newblock \urlprefix\url{https://www.mdpi.com/2673-2688/6/10/253}.

\bibitem[{Chen et~al.(2025)Chen, Fei, Du, Yi, Li and
  Kupfer}]{chen_intelligent_2025}
Chen, P., Fei, T., Du, Y., Yi, J., Li, Y. and Kupfer, J.~A. (2025) Intelligent
  {Bear} {Prevention} {System} {Based} on {Computer} {Vision}: {An} {Approach}
  to {Reduce} {Human}-{Bear} {Conflicts} in the {Tibetan} {Plateau} {Area},
  {China}.
\newblock \urlprefix\url{http://arxiv.org/abs/2503.23178}.
\newblock ArXiv:2503.23178 [cs].

\bibitem[{Christin et~al.(2019)Christin, Hervet and
  Lecomte}]{christin_applications_2019}
Christin, S., Hervet, {\'E}. and Lecomte, N. (2019) Applications for deep
  learning in ecology.
\newblock \textit{Methods in Ecology and Evolution}, \textbf{10}, 1632--1644.
\newblock
  \urlprefix\url{https://onlinelibrary.wiley.com/doi/abs/10.1111/2041-210X.13256}.
\newblock \_eprint:
  https://onlinelibrary.wiley.com/doi/pdf/10.1111/2041-210X.13256.

\bibitem[{Ciapponi et~al.(2025)Ciapponi, Mannini, Scanferla, Anderle and
  Farella}]{ciapponi_enabling_2025}
Ciapponi, S., Mannini, L., Scanferla, J., Anderle, M. and Farella, E. (2025)
  Enabling {Multi}-{Species} {Bird} {Classification} on {Low}-{Power}
  {Bioacoustic} {Loggers}.
\newblock \urlprefix\url{http://arxiv.org/abs/2509.20103}.
\newblock ArXiv:2509.20103 [cs].

\bibitem[{Clark(2021)}]{clark_birdweather_2021}
Clark, T. (2021) {BirdWeather} - a living library of bird vocalizations.
\newblock \urlprefix\url{https://www.birdweather.com/}.

\bibitem[{Contina et~al.(2024)Contina, Abelson, Allison, Stokes, Sanchez,
  Hernandez, Kepple, Tran, Kazen, Brown, Powell and
  Keitt}]{contina_biosense_2024}
Contina, A., Abelson, E., Allison, B., Stokes, B., Sanchez, K.~F., Hernandez,
  H.~M., Kepple, A.~M., Tran, Q., Kazen, I., Brown, K.~A., Powell, J.~H. and
  Keitt, T.~H. (2024) {BioSense}: {An} automated sensing node for organismal
  and environmental biology.
\newblock \textit{HardwareX}, \textbf{20}, e00584.
\newblock
  \urlprefix\url{https://linkinghub.elsevier.com/retrieve/pii/S2468067224000786}.

\bibitem[{Costanza et~al.(1997)Costanza, D'Arge, De~Groot, Farber, Grasso,
  Hannon, Limburg, Naeem, O'Neill, Paruelo, Raskin, Sutton and Van
  Den~Belt}]{costanza_value_1997}
Costanza, R., D'Arge, R., De~Groot, R., Farber, S., Grasso, M., Hannon, B.,
  Limburg, K., Naeem, S., O'Neill, R.~V., Paruelo, J., Raskin, R.~G., Sutton,
  P. and Van Den~Belt, M. (1997) The value of the world's ecosystem services
  and natural capital.
\newblock \textit{Nature}, \textbf{387}, 253--260.

\bibitem[{Cowans et~al.(2026)Cowans, Lambin, Hare and
  Sutherland}]{cowans_improving_2026}
Cowans, A., Lambin, X., Hare, D. and Sutherland, C. (2026) Improving the
  integration of artificial intelligence into existing ecological inference
  workflows.
\newblock \textit{Methods in Ecology and Evolution}, \textbf{17}, 228--237.
\newblock
  \urlprefix\url{https://onlinelibrary.wiley.com/doi/abs/10.1111/2041-210X.14485}.
\newblock \_eprint:
  https://besjournals.onlinelibrary.wiley.com/doi/pdf/10.1111/2041-210X.14485.

\bibitem[{Darras et~al.(2025)Darras, Rountree, Van~Wilgenburg, Cord, Pitz,
  Chen, Dong, Rocquencourt, Desjonqu{\`e}res, Diaz, Lin, Turco, Emmerson,
  Bradfer-Lawrence, Gasc, Marley, Salton, Schill{\'e}, Wensveen, Wu,
  Acero-Murcia, Acevedo-Charry, Adam, Aguzzi, Akoglu, Amorim, Anders,
  Andr{\'e}, Antonelli, Do~Nascimento, Appel, Archer, Astaras, Atemasov,
  Atkinson, Attia, Baltag, Barbaro, Basan, Batist, Baumgarten, Bayle~Sempere,
  Bellisario, David, Berger-Tal, Bertucci, Betts, Bhalla, Bicudo, Bolgan,
  Bombaci, Bota, Boullhesen, Briers, Buchan, Budka, Burchard, Buscaino,
  Calvente, Campos-Cerqueira, Gon{\c{c}}alves, Ceraulo, Cerezo-Araujo,
  Cerw{\'e}n, Chaskda, Chistopolova, Clark, Cox, Cretois, Czarnecki, da~Silva,
  da~Silva, De~Clippele, de~la Haye, de~Oliveira~Tissiani, de~Zwaan, Degano,
  Deichmann, del Rio, Devenish, D{\'i}az-Delgado, Diniz, Oliveira-J{\'u}nior,
  Dorigo, Dr{\"o}ge, Duarte, Duarte, Dunleavy, Dziak, Elise, Enari, Enari,
  Erbs, Eriksson, Ert{\"o}r-Akyazi, Ferrari, Ferreira, Fleishman, Fonseca,
  Freitas, Friedman, Froidevaux, Gogoleva, Gonzaga, Correa, Goodale, Gottesman,
  Grass, Greenhalgh, Gregoire, Hach{\'e}, Hagge, Halliday, Hammer,
  Hanf-Dressler, Haupert, Haver, Heath, Hending, Hernandez-Blanco, Higgs,
  Hiller, Huang, Hutchinson, Hyacinthe, Ieronymidou, Iniunam, Jackson, Jacot,
  Jahn, Juanes, Kanes, Kenchington, Kepfer-Rojas, Kitzes, Kusuminda, Lehnardt,
  Lei, Leitman, Le{\'o}n, Li, Lima-Santos, Lloyd, Looby, L{\'o}pez-Baucells,
  L{\'o}pez-Bosch, Louth-Robins, Maeda, Malige, Mammides, Marcacci, Markolf,
  Marques, Martin, Martin, Martin, McArthur, McKown, McLeod, M{\'e}doc,
  Metcalf, Meyer, Mikusinski, Miller, Monteiro, Mooney, Moreira, Sugai, Morris,
  M{\"u}ller, Mu{\~n}oz-Duque, Murchy, Nagelkerken, Mas, Nouioua, Ocampo-Ariza,
  Olden, Oppel, Osiecka, Papale, Parsons, Pashkevich, Patris, Marques,
  P{\'e}rez-Granados, Piatti, Pichorim, Pine, Pinheiro, Pradervand, Quinn,
  Quintella, Radford, Raick, Rainho, Ramalho, Ramesh, R{\'e}taux, Reynolds,
  Riede, Rimmer, R{\'i}os, Rocha, Rocha, Roe, Ross, Rosten, Ryan,
  Salustio-Gomes, Samarra, Samartzis, Santos, Sattler, Scharffenberg, Schoeman,
  Schuchmann, Sebasti{\'a}n-Gonz{\'a}lez, Seibold, Sethi, Shabangu, Shaw, Shen,
  Singer, {\v{S}}irovi{\'c}, Slater, Spriel, Stanley, Sueur, da~Cunha~Tavares,
  Thomisch, Thorn, Tong, Torrent, Traba, Tremblay, Trevelin, Tseng, Tuanmu,
  Valverde, Vernasco, Vieira, da~Paz, Ward, Watson, Weldy, Wiel, Willie, Wood,
  Xu, Zhou, Li, Sousa-Lima and Wanger}]{darras_worldwide_2025}
Darras, K. F.~A., Rountree, R.~A., Van~Wilgenburg, S.~L., Cord, A.~F., Pitz,
  F., Chen, Y., Dong, L., Rocquencourt, A., Desjonqu{\`e}res, C., Diaz, P.~M.,
  Lin, T.-H., Turco, T., Emmerson, L., Bradfer-Lawrence, T., Gasc, A., Marley,
  S., Salton, M., Schill{\'e}, L., Wensveen, P.~J., Wu, S.-H. et~al. (2025)
  Worldwide {Soundscapes}: {A} {Synthesis} of {Passive} {Acoustic} {Monitoring}
  {Across} {Realms}.
\newblock \textit{Global Ecology and Biogeography}, \textbf{34}, e70021.
\newblock
  \urlprefix\url{https://onlinelibrary.wiley.com/doi/abs/10.1111/geb.70021}.
\newblock \_eprint: https://onlinelibrary.wiley.com/doi/pdf/10.1111/geb.70021.

\bibitem[{Dasgupta(2024)}]{dasgupta_economics_2024}
Dasgupta, P. (2024) \textit{The {Economics} of {Biodiversity}: {The} {Dasgupta}
  {Review}}.
\newblock Cambridge University Press, 1 edn.
\newblock
  \urlprefix\url{https://www.cambridge.org/core/product/identifier/9781009494359/type/book}.

\bibitem[{David et~al.(2021)David, Duke, Jain, Janapa~Reddi, Jeffries, Li,
  Kreeger, Nappier, Natraj, Wang, Warden and Rhodes}]{david_tensorflow_2021}
David, R., Duke, J., Jain, A., Janapa~Reddi, V., Jeffries, N., Li, J., Kreeger,
  N., Nappier, I., Natraj, M., Wang, T., Warden, P. and Rhodes, R. (2021)
  {TensorFlow} {Lite} {Micro}: {Embedded} {Machine} {Learning} for {TinyML}
  {Systems}.
\newblock \textit{Proceedings of Machine Learning and Systems}, \textbf{3},
  800--811.
\newblock
  \urlprefix\url{https://proceedings.mlsys.org/paper_files/paper/2021/hash/6c44dc73014d66ba49b28d483a8f8b0d-Abstract.html}.

\bibitem[{Davison et~al.(2025)Davison, De~Koning, Taubert and
  Schakel}]{davison_automated_2025}
Davison, A.~M., De~Koning, K., Taubert, F. and Schakel, J.-K. (2025) Automated
  near real-time monitoring in ecology: {Status} quo and ways forward.
\newblock \textit{Ecological Informatics}, \textbf{89}, 103157.
\newblock
  \urlprefix\url{https://linkinghub.elsevier.com/retrieve/pii/S1574954125001669}.

\bibitem[{De~Luca et~al.(2025)De~Luca, Loreti, Bracciale, Colosimo, Gentile,
  Mastrangeli, Gerber, Haakonsson, Waters, Allegra, Capuano, Natale and
  Catini}]{de_luca_design_2025}
De~Luca, M., Loreti, P., Bracciale, L., Colosimo, G., Gentile, G., Mastrangeli,
  F., Gerber, G.~P., Haakonsson, J., Waters, G., Allegra, V., Capuano, R.,
  Natale, C.~D. and Catini, A. (2025) Design of a {LoRa}-{Based} {Multisensor}
  {Device} for the {Internet} of {Animals}.
\newblock \textit{IEEE Internet of Things Journal}, \textbf{12}, 31588--31600.
\newblock \urlprefix\url{https://ieeexplore.ieee.org/document/11027532/}.

\bibitem[{De~Simone et~al.(2024)De~Simone, Barbisan, Turvani and
  Riente}]{de_simone_advancing_2024}
De~Simone, A., Barbisan, L., Turvani, G. and Riente, F. (2024) Advancing
  {Beekeeping}: {IoT} and {TinyML} for {Queen} {Bee} {Monitoring} {Using}
  {Audio} {Signals}.
\newblock \textit{IEEE Transactions on Instrumentation and Measurement},
  \textbf{73}, 1--9.

\bibitem[{Dietze et~al.(2018)Dietze, Fox, Beck-Johnson, Betancourt, Hooten,
  Jarnevich, Keitt, Kenney, Laney, Larsen, Loescher, Lunch, Pijanowski,
  Randerson, Read, Tredennick, Vargas, Weathers and
  White}]{dietze_iterative_2018}
Dietze, M.~C., Fox, A., Beck-Johnson, L.~M., Betancourt, J.~L., Hooten, M.~B.,
  Jarnevich, C.~S., Keitt, T.~H., Kenney, M.~A., Laney, C.~M., Larsen, L.~G.,
  Loescher, H.~W., Lunch, C.~K., Pijanowski, B.~C., Randerson, J.~T., Read,
  E.~K., Tredennick, A.~T., Vargas, R., Weathers, K.~C. and White, E.~P. (2018)
  Iterative near-term ecological forecasting: {Needs}, opportunities, and
  challenges.
\newblock \textit{Proceedings of the National Academy of Sciences},
  \textbf{115}, 1424--1432.
\newblock
  \urlprefix\url{https://www.pnas.org/doi/full/10.1073/pnas.1710231115}.

\bibitem[{Ding et~al.(2022)Ding, Peltonen, Meuser, Aral, Becker, Dustdar,
  Hiessl, Kranzlm{\"u}ller, Liyanage, Maghsudi, Mohan, Ott, Rellermeyer,
  Schulte, Schulzrinne, Solmaz, Tarkoma, Varghese and Wolf}]{ding_roadmap_2022}
Ding, A.~Y., Peltonen, E., Meuser, T., Aral, A., Becker, C., Dustdar, S.,
  Hiessl, T., Kranzlm{\"u}ller, D., Liyanage, M., Maghsudi, S., Mohan, N., Ott,
  J., Rellermeyer, J.~S., Schulte, S., Schulzrinne, H., Solmaz, G., Tarkoma,
  S., Varghese, B. and Wolf, L. (2022) Roadmap for edge {AI}: a {Dagstuhl}
  perspective.
\newblock \textit{ACM SIGCOMM Computer Communication Review}, \textbf{52},
  28--33.
\newblock \urlprefix\url{https://dl.acm.org/doi/10.1145/3523230.3523235}.

\bibitem[{Disabato et~al.(2021)Disabato, Canonaco, Flikkema, Roveri and
  Alippi}]{disabato_birdsong_2021}
Disabato, S., Canonaco, G., Flikkema, P.~G., Roveri, M. and Alippi, C. (2021)
  Birdsong {Detection} at the {Edge} with {Deep} {Learning}.
\newblock In \textit{2021 {IEEE} {International} {Conference} on {Smart}
  {Computing} ({SMARTCOMP})}, 9--16.
\newblock Journal Abbreviation: 2021 IEEE International Conference on Smart
  Computing (SMARTCOMP).

\bibitem[{Ditria(2025)}]{ditria_pv_2025}
Ditria, L. (2025) {PV} {PI} - {Power} your {Raspberry} {PI} with the {Sun}!
\newblock
  \urlprefix\url{https://www.kickstarter.com/projects/pvpi/pv-pi-power-your-raspberry-pi-with-the-sun}.

\bibitem[{Dumoulin et~al.(2025)Dumoulin, Stretcu, Hamer, Harrell, Laber,
  Larochelle, Merri{\"e}nboer, Navine, Hart, Williams, Lamont, Razak, Team,
  Brodie, Doohan, Eichinski, Roe, Schwarzkopf and
  Denton}]{dumoulin_search_2025}
Dumoulin, V., Stretcu, O., Hamer, J., Harrell, L., Laber, R., Larochelle, H.,
  Merri{\"e}nboer, B.~v., Navine, A., Hart, P., Williams, B., Lamont, T. A.~C.,
  Razak, T.~B., Team, M. C.~R., Brodie, S., Doohan, B., Eichinski, P., Roe, P.,
  Schwarzkopf, L. and Denton, T. (2025) The {Search} for {Squawk}: {Agile}
  {Modeling} in {Bioacoustics}.
\newblock \urlprefix\url{http://arxiv.org/abs/2505.03071}.
\newblock ArXiv:2505.03071 [eess].

\bibitem[{Dussert et~al.(2025)Dussert, Miele, Van~Reeth, Delestrade, Dray and
  Chamaill{\'e}-Jammes}]{dussert_zero-shot_2025}
Dussert, G., Miele, V., Van~Reeth, C., Delestrade, A., Dray, S. and
  Chamaill{\'e}-Jammes, S. (2025) Zero-shot animal behaviour classification
  with vision-language foundation models.
\newblock \textit{Methods in Ecology and Evolution}, \textbf{16}, 1460--1472.
\newblock
  \urlprefix\url{https://onlinelibrary.wiley.com/doi/abs/10.1111/2041-210X.70059}.
\newblock \_eprint:
  https://besjournals.onlinelibrary.wiley.com/doi/pdf/10.1111/2041-210X.70059.

\bibitem[{Elias et~al.(2017)Elias, Golubovic, Krintz and
  Wolski}]{elias_wheres_2017}
Elias, A.~R., Golubovic, N., Krintz, C. and Wolski, R. (2017) Where's {The}
  {Bear}? {Automating} {Wildlife} {Image} {Processing} {Using} {IoT} and {Edge}
  {Cloud} {Systems}.
\newblock In \textit{Proceedings of the {Second} {International} {Conference}
  on {Internet}-of-{Things} {Design} and {Implementation}}, {IoTDI} '17,
  247--258. New York, NY, USA: Association for Computing Machinery.
\newblock \urlprefix\url{https://dl.acm.org/doi/10.1145/3054977.3054986}.

\bibitem[{Farley et~al.(2018)Farley, Dawson, Goring and
  Williams}]{farley_situating_2018}
Farley, S.~S., Dawson, A., Goring, S.~J. and Williams, J.~W. (2018) Situating
  {Ecology} as a {Big}-{Data} {Science}: {Current} {Advances}, {Challenges},
  and {Solutions}.
\newblock \textit{BioScience}, \textbf{68}, 563--576.
\newblock
  \urlprefix\url{https://academic.oup.com/bioscience/article/68/8/563/5049569}.

\bibitem[{Gadot et~al.(2024)Gadot, Istrate, Kim, Morris, Beery, Birch and
  Ahumada}]{gadot_crop_2024}
Gadot, T., Istrate, {\c{S}}., Kim, H., Morris, D., Beery, S., Birch, T. and
  Ahumada, J. (2024) To crop or not to crop: {Comparing} whole-image and
  cropped classification on a large dataset of camera trap images.
\newblock \textit{IET Computer Vision}, \textbf{18}, 1193--1208.
\newblock
  \urlprefix\url{https://onlinelibrary.wiley.com/doi/abs/10.1049/cvi2.12318}.
\newblock \_eprint:
  https://ietresearch.onlinelibrary.wiley.com/doi/pdf/10.1049/cvi2.12318.

\bibitem[{Gallacher et~al.(2021)Gallacher, Wilson, Fairbrass, Turmukhambetov,
  Firman, Kreitmayer, Mac~Aodha, Brostow and Jones}]{gallacher_shazam_2021}
Gallacher, S., Wilson, D., Fairbrass, A., Turmukhambetov, D., Firman, M.,
  Kreitmayer, S., Mac~Aodha, O., Brostow, G. and Jones, K. (2021) Shazam for
  bats: {Internet} of {Things} for continuous real-time biodiversity
  monitoring.
\newblock \textit{IET Smart Cities}, \textbf{3}, 171--183.
\newblock
  \urlprefix\url{https://onlinelibrary.wiley.com/doi/abs/10.1049/smc2.12016}.
\newblock \_eprint: https://onlinelibrary.wiley.com/doi/pdf/10.1049/smc2.12016.

\bibitem[{Gauld et~al.(2023)Gauld, Atkinson, Silva, Senn and
  Franco}]{gauld_characterisation_2023}
Gauld, J., Atkinson, P.~W., Silva, J.~P., Senn, A. and Franco, A. M.~A. (2023)
  Characterisation of a new lightweight {LoRaWAN} {GPS} bio-logger and
  deployment on griffon vultures {Gyps} fulvus.
\newblock \textit{Animal Biotelemetry}, \textbf{11}, 17.
\newblock \urlprefix\url{https://doi.org/10.1186/s40317-023-00329-y}.

\bibitem[{Giorgetti and Pau(2025)}]{giorgetti_transitioning_2025}
Giorgetti, G. and Pau, D.~P. (2025) Transitioning from {TinyML} to {Edge}
  {GenAI}: {A} {Review}.
\newblock \textit{Big Data and Cognitive Computing}, \textbf{9}, 61.
\newblock \urlprefix\url{https://www.mdpi.com/2504-2289/9/3/61}.
\newblock Number: 3.

\bibitem[{Glover-Kapfer et~al.(2019)Glover-Kapfer, Soto-Navarro and
  Wearn}]{glover-kapfer_camera-trapping_2019}
Glover-Kapfer, P., Soto-Navarro, C.~A. and Wearn, O.~R. (2019) Camera-trapping
  version 3.0: current constraints and future priorities for development.
\newblock \textit{Remote Sensing in Ecology and Conservation}, \textbf{5},
  209--223.
\newblock
  \urlprefix\url{https://onlinelibrary.wiley.com/doi/abs/10.1002/rse2.106}.
\newblock \_eprint:
  https://zslpublications.onlinelibrary.wiley.com/doi/pdf/10.1002/rse2.106.

\bibitem[{Gonzalez et~al.(2023)Gonzalez, Vihervaara, Balvanera, Bates,
  Bayraktarov, Bellingham, Bruder, Campbell, Catchen, Cavender-Bares, Chase,
  Coops, Costello, Cz{\'u}cz, Delavaud, Dornelas, Dubois, Duffy, Eggermont,
  Fernandez, Fernandez, Ferrier, Geller, Gill, Gravel, Guerra, Guralnick,
  Harfoot, Hirsch, Hoban, Hughes, Hugo, Hunter, Isbell, Jetz, Juergens,
  Kissling, Krug, Kullberg, Le~Bras, Leung, Londo{\~n}o-Murcia, Lord, Loreau,
  Luers, Ma, MacDonald, Maes, McGeoch, Mihoub, Millette, Molnar, Montes, Mori,
  Muller-Karger, Muraoka, Nakaoka, Navarro, Newbold, Niamir, Obura, O'Connor,
  Paganini, Pelletier, Pereira, Poisot, Pollock, Purvis, Radulovici, Rocchini,
  Roeoesli, Schaepman, Schaepman-Strub, Schmeller, Schmiedel, Schneider,
  Shakya, Skidmore, Skowno, Takeuchi, Tuanmu, Turak, Turner, Urban,
  Urbina-Cardona, Valbuena, Van~de Putte, van Havre, Wingate, Wright and
  Torrelio}]{gonzalez_global_2023}
Gonzalez, A., Vihervaara, P., Balvanera, P., Bates, A.~E., Bayraktarov, E.,
  Bellingham, P.~J., Bruder, A., Campbell, J., Catchen, M.~D., Cavender-Bares,
  J., Chase, J., Coops, N., Costello, M.~J., Cz{\'u}cz, B., Delavaud, A.,
  Dornelas, M., Dubois, G., Duffy, E.~J., Eggermont, H., Fernandez, M. et~al.
  (2023) A global biodiversity observing system to unite monitoring and guide
  action.
\newblock \textit{Nature Ecology \& Evolution}, \textbf{7}, 1947--1952.
\newblock \urlprefix\url{https://www.nature.com/articles/s41559-023-02171-0}.

\bibitem[{Gupta et~al.(2015)Gupta, Agrawal, Gopalakrishnan and
  Narayanan}]{gupta_deep_2015}
Gupta, S., Agrawal, A., Gopalakrishnan, K. and Narayanan, P. (2015) Deep
  learning with limited numerical precision.
\newblock In \textit{Proceedings of the 32nd {International} {Conference} on
  {International} {Conference} on {Machine} {Learning} - {Volume} 37},
  {ICML}'15, 1737--1746. Lille, France: JMLR.org.

\bibitem[{Hammad et~al.(2025)Hammad, Wu, Ghafar-Zadeh and
  Magierowski}]{hammad_nanopore-aware_2025}
Hammad, K., Wu, Z., Ghafar-Zadeh, E. and Magierowski, S. (2025)
  Nanopore-{Aware} {Embedded} {Detection} for {Mobile} {DNA} {Sequencing}: {A}
  {Viterbi}--{HMM} {Design} {Versus} {Deep} {Learning} {Approaches}.
\newblock \textit{Biosensors}, \textbf{15}.
\newblock \urlprefix\url{https://www.mdpi.com/2079-6374/15/9/569}.

\bibitem[{Han et~al.(2015)Han, Pool, Tran and Dally}]{han_learning_2015}
Han, S., Pool, J., Tran, J. and Dally, W.~J. (2015) Learning both {Weights} and
  {Connections} for {Efficient} {Neural} {Networks}.
\newblock \urlprefix\url{http://arxiv.org/abs/1506.02626}.
\newblock ArXiv:1506.02626 [cs].

\bibitem[{Hernandez et~al.(2024)Hernandez, Miao, Vargas, Beery, Dodhia,
  Arbelaez and Ferres}]{hernandez_pytorch-wildlife_2024}
Hernandez, A., Miao, Z., Vargas, L., Beery, S., Dodhia, R., Arbelaez, P. and
  Ferres, J. M.~L. (2024) Pytorch-{Wildlife}: {A} {Collaborative} {Deep}
  {Learning} {Framework} for {Conservation}.
\newblock \urlprefix\url{http://arxiv.org/abs/2405.12930}.
\newblock ArXiv:2405.12930 [cs].

\bibitem[{Hinton et~al.(2015)Hinton, Vinyals and Dean}]{hinton_distilling_2015}
Hinton, G., Vinyals, O. and Dean, J. (2015) Distilling the {Knowledge} in a
  {Neural} {Network}.
\newblock \urlprefix\url{http://arxiv.org/abs/1503.02531}.
\newblock ArXiv:1503.02531 [stat].

\bibitem[{H{\"o}chst et~al.(2022)H{\"o}chst, Bellafkir, Lampe, Vogelbacher,
  M{\"u}hling, Schneider, Lindner, R{\"o}sner, Schabo, Farwig and
  Freisleben}]{hochst_birdedge_2022}
H{\"o}chst, J., Bellafkir, H., Lampe, P., Vogelbacher, M., M{\"u}hling, M.,
  Schneider, D., Lindner, K., R{\"o}sner, S., Schabo, D.~G., Farwig, N. and
  Freisleben, B. (2022) Bird@{Edge}: {Bird} {Species} {Recognition}
  at~the~{Edge}.
\newblock In \textit{Networked {Systems}} (eds. M.-A. Koulali and M.~Mezini),
  69--86. Cham: Springer International Publishing.

\bibitem[{Horwitz et~al.(2025)Horwitz, Kurer, Kahana, Amar and
  Hoshen}]{horwitz_we_2025}
Horwitz, E., Kurer, N., Kahana, J., Amar, L. and Hoshen, Y. (2025) We {Should}
  {Chart} an {Atlas} of {All} the {World}'s {Models}.
\newblock \urlprefix\url{http://arxiv.org/abs/2503.10633}.
\newblock ArXiv:2503.10633 [cs].

\bibitem[{Howard et~al.(2017)Howard, Zhu, Chen, Kalenichenko, Wang, Weyand,
  Andreetto and Adam}]{howard_mobilenets_2017}
Howard, A.~G., Zhu, M., Chen, B., Kalenichenko, D., Wang, W., Weyand, T.,
  Andreetto, M. and Adam, H. (2017) {MobileNets}: {Efficient} {Convolutional}
  {Neural} {Networks} for {Mobile} {Vision} {Applications}.
\newblock \urlprefix\url{http://arxiv.org/abs/1704.04861}.
\newblock ArXiv:1704.04861 [cs].

\bibitem[{Huang et~al.(2024)Huang, Tousnakhoff, Kozyr, Rehausen, Bie{\ss}mann,
  Lachlan, Adjih and Baccelli}]{huang_tinychirp_2024}
Huang, Z., Tousnakhoff, A., Kozyr, P., Rehausen, R., Bie{\ss}mann, F., Lachlan,
  R., Adjih, C. and Baccelli, E. (2024) {TinyChirp}: {Bird} {Song}
  {Recognition} {Using} {TinyML} {Models} on {Low}-power {Wireless} {Acoustic}
  {Sensors}.
\newblock \urlprefix\url{http://arxiv.org/abs/2407.21453}.
\newblock ArXiv:2407.21453 [cs].

\bibitem[{Ibraheam et~al.(2023)Ibraheam, Li and
  Gebali}]{ibraheam_accurate_2023}
Ibraheam, M., Li, K.~F. and Gebali, F. (2023) An {Accurate} and {Fast} {Animal}
  {Species} {Detection} {System} for {Embedded} {Devices}.
\newblock \textit{IEEE Access}, \textbf{11}, 23462--23473.
\newblock \urlprefix\url{https://ieeexplore.ieee.org/document/10058200/}.

\bibitem[{Iodice and Furtner(2025)}]{iodice_edge_2025}
Iodice, G.~M. and Furtner, W. (2025) Edge {AI}---{An} {Industry} {View}.
\newblock \textit{IEEE Design \& Test}, \textbf{42}, 27--34.
\newblock \urlprefix\url{https://ieeexplore.ieee.org/document/11023847}.

\bibitem[{IPBES(2019)}]{ipbes_global_2019}
IPBES (2019) Global assessment report on biodiversity and ecosystem services of
  the {Intergovernmental} {Science}-{Policy} {Platform} on {Biodiversity} and
  {Ecosystem} {Services}.
\newblock \textit{Tech. rep.}, Zenodo, Bonn, Germany.
\newblock \urlprefix\url{https://zenodo.org/doi/10.5281/zenodo.3831673}.
\newblock Version Number: 1.

\bibitem[{Jacob et~al.(2017)Jacob, Kligys, Chen, Zhu, Tang, Howard, Adam and
  Kalenichenko}]{jacob_quantization_2017}
Jacob, B., Kligys, S., Chen, B., Zhu, M., Tang, M., Howard, A., Adam, H. and
  Kalenichenko, D. (2017) Quantization and {Training} of {Neural} {Networks}
  for {Efficient} {Integer}-{Arithmetic}-{Only} {Inference}.
\newblock \urlprefix\url{http://arxiv.org/abs/1712.05877}.
\newblock ArXiv:1712.05877 [cs].

\bibitem[{Jennings et~al.(2025)Jennings, Jones, Taitingfong, Martinez,
  David-Chavez, Alegado, Tofighi-Niaki, Maldonado, Thomas, Dye, Weber,
  Spellman, Ketchum, Duerr, Johnson, Balch and
  Carroll}]{jennings_governance_2025}
Jennings, L., Jones, K., Taitingfong, R., Martinez, A., David-Chavez, D.,
  Alegado, R.~A., Tofighi-Niaki, A., Maldonado, J., Thomas, B., Dye, D., Weber,
  J., Spellman, K.~V., Ketchum, S., Duerr, R., Johnson, N., Balch, J. and
  Carroll, S.~R. (2025) Governance of {Indigenous} data in open earth systems
  science.
\newblock \textit{Nature Communications}, \textbf{16}, 572.
\newblock \urlprefix\url{https://www.nature.com/articles/s41467-024-53480-2}.

\bibitem[{Jolles(2021)}]{jolles_broad-scale_2021}
Jolles, J.~W. (2021) Broad-scale applications of the {Raspberry} {Pi}: {A}
  review and guide for biologists.
\newblock \textit{Methods in Ecology and Evolution}, \textbf{12}, 1562--1579.
\newblock
  \urlprefix\url{https://onlinelibrary.wiley.com/doi/abs/10.1111/2041-210X.13652}.
\newblock \_eprint:
  https://onlinelibrary.wiley.com/doi/pdf/10.1111/2041-210X.13652.

\bibitem[{Kahl et~al.(2021)Kahl, Wood, Eibl and Klinck}]{kahl_birdnet_2021}
Kahl, S., Wood, C., Eibl, M. and Klinck, H. (2021) {BirdNET}: {A} deep learning
  solution for avian diversity monitoring.
\newblock \textit{ECOLOGICAL INFORMATICS}, \textbf{61}.

\bibitem[{Karic et~al.(2025)Karic, Herrmann, Stenkamp, Scharf, Gieseke and
  Schwering}]{karic_send_2025}
Karic, B., Herrmann, N., Stenkamp, J., Scharf, P., Gieseke, F. and Schwering,
  A. (2025) Send {Less}, {Save} {More}: {Energy}-{Efficiency} {Benchmark} of
  {Embedded} {CNN} {Inference} vs. {Data} {Transmission} in {IoT}.
\newblock \urlprefix\url{http://arxiv.org/abs/2510.24829}.
\newblock ArXiv:2510.24829 [cs].

\bibitem[{Kays et~al.(2022)Kays, Davidson, Berger, Bohrer, Fiedler, Flack,
  Hirt, Hahn, Gauggel, Russell, K{\"o}lzsch, Lohr, Partecke, Quetting, Safi,
  Scharf, Schneider, Lang, Schaeuffelhut, Landwehr, Storhas, van Schalkwyk,
  Vinciguerra, Weinzierl and Wikelski}]{kays_movebank_2022}
Kays, R., Davidson, S.~C., Berger, M., Bohrer, G., Fiedler, W., Flack, A.,
  Hirt, J., Hahn, C., Gauggel, D., Russell, B., K{\"o}lzsch, A., Lohr, A.,
  Partecke, J., Quetting, M., Safi, K., Scharf, A., Schneider, G., Lang, I.,
  Schaeuffelhut, F., Landwehr, M. et~al. (2022) The {Movebank} system for
  studying global animal movement and demography.
\newblock \textit{Methods in Ecology and Evolution}, \textbf{13}, 419--431.
\newblock
  \urlprefix\url{https://onlinelibrary.wiley.com/doi/abs/10.1111/2041-210X.13767}.
\newblock \_eprint:
  https://besjournals.onlinelibrary.wiley.com/doi/pdf/10.1111/2041-210X.13767.

\bibitem[{Kellenberger et~al.(2020)Kellenberger, Tuia and
  Morris}]{kellenberger_aide_2020}
Kellenberger, B., Tuia, D. and Morris, D. (2020) {AIDE}: {Accelerating}
  image-based ecological surveys with interactive machine learning.
\newblock \textit{Methods in Ecology and Evolution}, \textbf{11}, 1716--1727.
\newblock
  \urlprefix\url{https://onlinelibrary.wiley.com/doi/abs/10.1111/2041-210X.13489}.
\newblock \_eprint:
  https://besjournals.onlinelibrary.wiley.com/doi/pdf/10.1111/2041-210X.13489.

\bibitem[{Kimutai and F{\"o}rster(2023)}]{kimutai_low-cost_2023}
Kimutai, G. and F{\"o}rster, A. (2023) A low-cost {TinyML} model for {Mosquito}
  {Detection} in {Resource}-{Constrained} {Environments}.
\newblock In \textit{Proceedings of the 2023 {ACM} {Conference} on
  {Information} {Technology} for {Social} {Good}}, {GoodIT} '23, 23--30. New
  York, NY, USA: Association for Computing Machinery.
\newblock \urlprefix\url{https://dl.acm.org/doi/10.1145/3582515.3609514}.

\bibitem[{King and Halpern(2025)}]{king_implementation_2025}
King, R.~A. and Halpern, B.~S. (2025) Implementation of automated biodiversity
  monitoring lags behind its potential.
\newblock \textit{Environmental Research Letters}, \textbf{20}, 064022.
\newblock \urlprefix\url{https://doi.org/10.1088/1748-9326/add02d}.

\bibitem[{Knyva et~al.(2023)Knyva, Gailius, Bal{\v{c}}i{\=u}nas,
  Prata{\v{s}}ius, Kuzas and Kukanauskait{\.e}}]{knyva_iot_2023}
Knyva, M., Gailius, D., Bal{\v{c}}i{\=u}nas, G., Prata{\v{s}}ius, D., Kuzas, P.
  and Kukanauskait{\.e}, A. (2023) {IoT} {Sensor} {Network} for {Wild}-{Animal}
  {Detection} near {Roads}.
\newblock \textit{Sensors}, \textbf{23}, 8929.
\newblock \urlprefix\url{https://www.mdpi.com/1424-8220/23/21/8929}.
\newblock Number: 21.

\bibitem[{Lahoz-Monfort and Magrath(2021)}]{lahoz-monfort_comprehensive_2021}
Lahoz-Monfort, J.~J. and Magrath, M. J.~L. (2021) A {Comprehensive} {Overview}
  of {Technologies} for {Species} and {Habitat} {Monitoring} and
  {Conservation}.
\newblock \textit{BioScience}, \textbf{71}, 1038--1062.
\newblock \urlprefix\url{https://doi.org/10.1093/biosci/biab073}.

\bibitem[{Lamont et~al.(2022)Lamont, Chapuis, Williams, Dines, Gridley,
  Frainer, Fearey, Maulana, Prasetya, Jompa, Smith and
  Simpson}]{lamont_hydromoth_2022}
Lamont, T. A.~C., Chapuis, L., Williams, B., Dines, S., Gridley, T., Frainer,
  G., Fearey, J., Maulana, P.~B., Prasetya, M.~E., Jompa, J., Smith, D.~J. and
  Simpson, S.~D. (2022) {HydroMoth}: {Testing} a prototype low-cost acoustic
  recorder for aquatic environments.
\newblock \textit{Remote Sensing in Ecology and Conservation}, \textbf{8},
  362--378.
\newblock
  \urlprefix\url{https://onlinelibrary.wiley.com/doi/abs/10.1002/rse2.249}.
\newblock \_eprint:
  https://zslpublications.onlinelibrary.wiley.com/doi/pdf/10.1002/rse2.249.

\bibitem[{Lawrence et~al.(2025)Lawrence, Lal and Shen}]{lawrence_tinyml_2025}
Lawrence, Z., Lal, A. and Shen, M. (2025) {TinyML} : {Acoustic} {Burrowing}
  {Owl} {Vocalization} {Detection} on {STM32}.
\newblock \urlprefix\url{https://api.semanticscholar.org/CorpusID:280037754}.

\bibitem[{LeCun et~al.(1989)LeCun, Denker and Solla}]{lecun_optimal_1989}
LeCun, Y., Denker, J. and Solla, S. (1989) Optimal {Brain} {Damage}.
\newblock In \textit{Advances in {Neural} {Information} {Processing}
  {Systems}}, vol.~2. Morgan-Kaufmann.
\newblock
  \urlprefix\url{https://proceedings.neurips.cc/paper/1989/hash/6c9882bbac1c7093bd25041881277658-Abstract.html}.

\bibitem[{Li et~al.(2024)Li, Shan, Perez, Luo and Worrall}]{li_endangered_2024}
Li, K., Shan, M., Perez, S.~B., Luo, K. and Worrall, S. (2024) Endangered
  {Alert}: {A} {Field}-{Validated} {Self}-{Training} {Scheme} for {Detecting}
  and {Protecting} {Threatened} {Wildlife} on {Roads} and {Roadsides}.
\newblock \urlprefix\url{http://arxiv.org/abs/2412.12222}.
\newblock ArXiv:2412.12222 [cs].

\bibitem[{Luder et~al.(2025)Luder, Schulthess, Cortesi, Davis and
  Magno}]{luder_anitrack_2025}
Luder, V., Schulthess, L., Cortesi, S., Davis, L.~R. and Magno, M. (2025)
  {AniTrack}: {A} {Power}-{Efficient}, {Time}-{Slotted} and {Robust} {UWB}
  {Localization} {System} for {Animal} {Tracking} in a {Controlled} {Setting}.
\newblock \urlprefix\url{http://arxiv.org/abs/2506.00216}.
\newblock ArXiv:2506.00216 [cs].

\bibitem[{Lynam et~al.(2025)Lynam, Cronin, Wich, Steward, Howe, Kolla,
  Markovina, Torrico, Reyes, Sophalrachana, Stevens, Schmidt and
  Cox}]{lynam_rising_2025}
Lynam, A.~J., Cronin, D.~T., Wich, S.~A., Steward, J., Howe, A., Kolla, N.,
  Markovina, M., Torrico, O., Reyes, V., Sophalrachana, K., Stevens, X.,
  Schmidt, E. and Cox, H. (2025) The rising tide of conservation technology:
  empowering the fight against poaching and unsustainable wildlife harvest.
\newblock \textit{Frontiers in Ecology and Evolution}, \textbf{13}.
\newblock
  \urlprefix\url{https://www.frontiersin.org/journals/ecology-and-evolution/articles/10.3389/fevo.2025.1527976/full}.

\bibitem[{Ma(2022)}]{ma_smart_2022}
Ma, A. (2022) Smart {Wildlife} {Sentinel} ({SWS}): {Preventing}
  {Wildlife}-{Vehicle} {Collisions} and {Monitoring} {Road} {Ecology} with
  {Embedded} {IoT} {Systems} and {Machine} {Learning}.
\newblock In \textit{2022 {IEEE} {MIT} {Undergraduate} {Research} {Technology}
  {Conference} ({URTC})}, 1--4.
\newblock \urlprefix\url{https://ieeexplore.ieee.org/document/10002174}.

\bibitem[{Mac~Aodha et~al.(2022)Mac~Aodha, Mart{\'i}nez~Balvanera, Damstra,
  Cooke, Eichinski, Browning, Barataud, Boughey, Coles, Giacomini, Mac
  Swiney~G., Obrist, Parsons, Sattler and Jones}]{mac_aodha_towards_2022}
Mac~Aodha, O., Mart{\'i}nez~Balvanera, S., Damstra, E., Cooke, M., Eichinski,
  P., Browning, E., Barataud, M., Boughey, K., Coles, R., Giacomini, G., Mac
  Swiney~G., M.~C., Obrist, M.~K., Parsons, S., Sattler, T. and Jones, K.~E.
  (2022) Towards a {General} {Approach} for {Bat} {Echolocation} {Detection}
  and {Classification}.
\newblock
  \urlprefix\url{https://www.biorxiv.org/content/10.1101/2022.12.14.520490v1}.
\newblock Pages: 2022.12.14.520490 Section: New Results.

\bibitem[{Madsen et~al.(2025)Madsen, Flavin and Rogers}]{madsen_materials_2025}
Madsen, K.~E., Flavin, M.~T. and Rogers, J.~A. (2025) Materials advances for
  distributed environmental sensor networks at scale.
\newblock \textit{Nature Reviews Materials}, 1--24.
\newblock \urlprefix\url{https://www.nature.com/articles/s41578-025-00838-7}.

\bibitem[{Magierowski et~al.(2025)Magierowski, Wu, Beyene and
  Hammad}]{magierowski_sequencing_2025}
Magierowski, S., Wu, Z., Beyene, A. and Hammad, K. (2025) Sequencing on
  {Silicon}: {AI} {SoC} {Design} for {Mobile} {Genomics} at the {Edge}.
\newblock \urlprefix\url{http://arxiv.org/abs/2510.09339}.
\newblock ArXiv:2510.09339 [cs] version: 1.

\bibitem[{Mahbub et~al.(2024)Mahbub, Bhagwagar, Chand, Zualkernan, Judas and
  Dghaym}]{mahbub_bat2web_2024}
Mahbub, T., Bhagwagar, A., Chand, P., Zualkernan, I., Judas, J. and Dghaym, D.
  (2024) {Bat2Web}: {A} {Framework} for {Real}-{Time} {Classification} of {Bat}
  {Species} {Echolocation} {Signals} {Using} {Audio} {Sensor} {Data}.
\newblock \textit{Sensors}, \textbf{24}, 2899.
\newblock \urlprefix\url{https://www.mdpi.com/1424-8220/24/9/2899}.

\bibitem[{Mainetti(2023)}]{mainetti_acoustic_2023}
Mainetti, L. (2023) Acoustic {Identification} of {Wood}-{Boring} {Insects} with
  {TinyML}.
\newblock \urlprefix\url{https://hdl.handle.net/10589/208626}.

\bibitem[{Mainwaring et~al.(2002)Mainwaring, Culler, Polastre, Szewczyk and
  Anderson}]{mainwaring_wireless_2002}
Mainwaring, A., Culler, D., Polastre, J., Szewczyk, R. and Anderson, J. (2002)
  Wireless sensor networks for habitat monitoring.
\newblock In \textit{Proceedings of the 1st {ACM} international workshop on
  {Wireless} sensor networks and applications}, {WSNA} '02, 88--97. New York,
  NY, USA: Association for Computing Machinery.
\newblock \urlprefix\url{https://doi.org/10.1145/570738.570751}.

\bibitem[{Mart{\'i}nez~Balvanera et~al.(2025)Mart{\'i}nez~Balvanera, Mac~Aodha,
  Weldy, Pringle, Browning and Jones}]{martinez_balvanera_whombat_2025}
Mart{\'i}nez~Balvanera, S., Mac~Aodha, O., Weldy, M.~J., Pringle, H., Browning,
  E. and Jones, K.~E. (2025) Whombat: {An} open-source audio annotation tool
  for machine learning assisted bioacoustics.
\newblock \textit{Methods in Ecology and Evolution}, \textbf{16}, 19--28.
\newblock
  \urlprefix\url{https://onlinelibrary.wiley.com/doi/abs/10.1111/2041-210X.14468}.
\newblock \_eprint:
  https://besjournals.onlinelibrary.wiley.com/doi/pdf/10.1111/2041-210X.14468.

\bibitem[{Mathis and Mathis(2020)}]{mathis_deep_2020}
Mathis, M.~W. and Mathis, A. (2020) Deep learning tools for the measurement of
  animal behavior in neuroscience.
\newblock \textit{Current Opinion in Neurobiology}, \textbf{60}, 1--11.
\newblock
  \urlprefix\url{https://www.sciencedirect.com/science/article/pii/S0959438819301151}.

\bibitem[{Merenda et~al.(2020)Merenda, Porcaro and Iero}]{merenda_edge_2020}
Merenda, M., Porcaro, C. and Iero, D. (2020) Edge {Machine} {Learning} for
  {AI}-{Enabled} {IoT} {Devices}: {A} {Review}.
\newblock \textit{Sensors}, \textbf{20}, 2533.
\newblock \urlprefix\url{https://www.mdpi.com/1424-8220/20/9/2533}.

\bibitem[{Merri{\"e}nboer et~al.(2026)Merri{\"e}nboer, Dumoulin, Hamer,
  Harrell, Burns and Denton}]{merrienboer_perch_2026}
Merri{\"e}nboer, B.~v., Dumoulin, V., Hamer, J., Harrell, L., Burns, A. and
  Denton, T. (2026) Perch 2.0: {The} {Bittern} {Lesson} for {Bioacoustics}.
\newblock \urlprefix\url{http://arxiv.org/abs/2508.04665}.
\newblock ArXiv:2508.04665 [cs].

\bibitem[{Miao et~al.(2025)Miao, Zhang, Fabian, Hernandez~Celis, Beery, Li,
  Liu, Gupta, Nasir, Li, Holmberg, Palmer, Gaynor, Arbelaez, Wang, Dodhia and
  Ferres}]{miao_new_2025}
Miao, Z., Zhang, Y., Fabian, Z., Hernandez~Celis, A., Beery, S., Li, C., Liu,
  Z., Gupta, A., Nasir, M., Li, W., Holmberg, J., Palmer, M., Gaynor, K.,
  Arbelaez, P., Wang, P., Dodhia, R. and Ferres, J.~L. (2025) New frontiers in
  artificial intelligence for biodiversity research and conservation with
  multimodal language models.
\newblock \textit{Methods in Ecology and Evolution}, \textbf{n/a}.
\newblock
  \urlprefix\url{https://onlinelibrary.wiley.com/doi/abs/10.1111/2041-210X.70120}.
\newblock \_eprint:
  https://besjournals.onlinelibrary.wiley.com/doi/pdf/10.1111/2041-210X.70120.

\bibitem[{Millar et~al.(2024)Millar, Sethi, Haddadi and
  Madhavapeddy}]{millar_terracorder_2024}
Millar, J., Sethi, S., Haddadi, H. and Madhavapeddy, A. (2024) Terracorder:
  {Sense} {Long} and {Prosper}.
\newblock \urlprefix\url{http://arxiv.org/abs/2408.02407}.
\newblock ArXiv:2408.02407 [cs].

\bibitem[{Miquel et~al.(2023)Miquel, Latorre and
  Chamaill{\'e}-Jammes}]{miquel_energy-efficient_2023}
Miquel, J., Latorre, L. and Chamaill{\'e}-Jammes, S. (2023) Energy-{Efficient}
  {Audio} {Processing} at the {Edge} for {Biologging} {Applications}.
\newblock \textit{Journal of Low Power Electronics and Applications},
  \textbf{13}, 30.
\newblock \urlprefix\url{https://www.mdpi.com/2079-9268/13/2/30}.

\bibitem[{Monedero et~al.(2021)Monedero, Barbancho, M{\'a}rquez and
  Beltr{\'a}n}]{monedero_cyber-physical_2021}
Monedero, I., Barbancho, J., M{\'a}rquez, R. and Beltr{\'a}n, J.~F. (2021)
  Cyber-{Physical} {System} for {Environmental} {Monitoring} {Based} on {Deep}
  {Learning}.
\newblock \textit{Sensors}, \textbf{21}, 3655.
\newblock \urlprefix\url{https://www.mdpi.com/1424-8220/21/11/3655}.
\newblock Number: 11.

\bibitem[{Muramatsu et~al.(2025)Muramatsu, Shin, Deng, Markham and
  Patel}]{muramatsu_wildpose_2025}
Muramatsu, N., Shin, S., Deng, Q., Markham, A. and Patel, A. (2025) {WildPose}:
  a long-range {3D} wildlife motion capture system.
\newblock \textit{Journal of Experimental Biology}, \textbf{228}, JEB249987.
\newblock \urlprefix\url{https://doi.org/10.1242/jeb.249987}.

\bibitem[{O'Shea-Wheller et~al.(2024)O'Shea-Wheller, Corbett, Osborne, Recker
  and Kennedy}]{oshea-wheller_vespai_2024}
O'Shea-Wheller, T.~A., Corbett, A., Osborne, J.~L., Recker, M. and Kennedy,
  P.~J. (2024) {VespAI}: a deep learning-based system for the detection of
  invasive hornets.
\newblock \textit{Communications Biology}, \textbf{7}, 354.
\newblock \urlprefix\url{https://www.nature.com/articles/s42003-024-05979-z}.

\bibitem[{Otto(2018)}]{otto_adaptation_2018}
Otto, S.~P. (2018) Adaptation, speciation and extinction in the {Anthropocene}.
\newblock \textit{Proceedings of the Royal Society B: Biological Sciences},
  \textbf{285}, 20182047.
\newblock
  \urlprefix\url{https://royalsocietypublishing.org/doi/10.1098/rspb.2018.2047}.

\bibitem[{Pan and McElhannon(2018)}]{pan_future_2018}
Pan, J. and McElhannon, J. (2018) Future {Edge} {Cloud} and {Edge} {Computing}
  for {Internet} of {Things} {Applications}.
\newblock \textit{IEEE Internet of Things Journal}, \textbf{5}, 439--449.
\newblock \urlprefix\url{https://ieeexplore.ieee.org/document/8089336/}.

\bibitem[{Paoletti et~al.(2023)Paoletti, Rumes, Pierantonio, Panigada, Jan,
  Folegot, Schilling, Riviere, Carrier, Dumoulin, van Hamme,
  Marquis-Laisn{\'e}, Bruliard, Petitpierre and
  Demoor}]{paoletti_seadetect_2023}
Paoletti, S., Rumes, B., Pierantonio, N., Panigada, S., Jan, R., Folegot, T.,
  Schilling, A., Riviere, N., Carrier, V., Dumoulin, A., van Hamme, D.,
  Marquis-Laisn{\'e}, G., Bruliard, F.-A., Petitpierre, F. and Demoor, D.
  (2023) {SEADETECT}: developing an automated detection system to reduce
  whale-vessel collision risk.
\newblock \textit{Research Ideas and Outcomes}, \textbf{9}.
\newblock \urlprefix\url{https://hal.science/hal-04424217}.

\bibitem[{Pichler and Hartig(2022)}]{pichler_machine_2022}
Pichler, M. and Hartig, F. (2022) Machine learning and deep learning---{A}
  review for ecologists.
\newblock \textit{Methods in Ecology and Evolution}, \textbf{n/a}.
\newblock
  \urlprefix\url{https://onlinelibrary.wiley.com/doi/abs/10.1111/2041-210X.14061}.
\newblock \_eprint:
  https://onlinelibrary.wiley.com/doi/pdf/10.1111/2041-210X.14061.

\bibitem[{Pigot et~al.(2025)Pigot, Dee, Richardson, Cooper, Eisenhauer,
  Gregory, Lewis, Macgregor, Massimino, Maynard, Phillips, Rillo, Loreau and
  Haegeman}]{pigot_macroecological_2025}
Pigot, A.~L., Dee, L.~E., Richardson, A.~J., Cooper, D. L.~M., Eisenhauer, N.,
  Gregory, R.~D., Lewis, S.~L., Macgregor, C.~J., Massimino, D., Maynard,
  D.~S., Phillips, H. R.~P., Rillo, M., Loreau, M. and Haegeman, B. (2025)
  Macroecological rules predict how biomass scales with species richness in
  nature.
\newblock \textit{Science}, \textbf{387}, 1272--1276.
\newblock \urlprefix\url{https://www.science.org/doi/10.1126/science.adq3278}.

\bibitem[{Pocock et~al.(2015)Pocock, Newson, Henderson, Peyton, Sutherland,
  Noble, Ball, Beckmann, Biggs, Brereton, Bullock, Buckland, Edwards, Eaton,
  Harvey, Hill, Horlock, Hubble, Julian, Mackey, Mann, Marshall, Medlock,
  O'Mahony, Pacheco, Porter, Prentice, Procter, Roy, Southway, Shortall,
  Stewart, Wembridge, Wright and Roy}]{pocock_developing_2015}
Pocock, M. J.~O., Newson, S.~E., Henderson, I.~G., Peyton, J., Sutherland,
  W.~J., Noble, D.~G., Ball, S.~G., Beckmann, B.~C., Biggs, J., Brereton, T.,
  Bullock, D.~J., Buckland, S.~T., Edwards, M., Eaton, M.~A., Harvey, M.~C.,
  Hill, M.~O., Horlock, M., Hubble, D.~S., Julian, A.~M., Mackey, E.~C. et~al.
  (2015) Developing and enhancing biodiversity monitoring programmes: a
  collaborative assessment of priorities.
\newblock \textit{Journal of Applied Ecology}, \textbf{52}, 686--695.
\newblock
  \urlprefix\url{https://besjournals.pericles-prod.literatumonline.com/doi/10.1111/1365-2664.12423}.

\bibitem[{Pollock et~al.(2025)Pollock, Kitzes, Beery, Gaynor, Jarzyna,
  Mac~Aodha, Meyer, Rolnick, Taylor and Tuia}]{pollock_harnessing_2025}
Pollock, L.~J., Kitzes, J., Beery, S., Gaynor, K.~M., Jarzyna, M.~A.,
  Mac~Aodha, O., Meyer, B., Rolnick, D., Taylor, G.~W. and Tuia, D. (2025)
  Harnessing artificial intelligence to fill global shortfalls in biodiversity
  knowledge.
\newblock \textit{Nature Reviews Biodiversity}, 1--17.
\newblock \urlprefix\url{https://www.nature.com/articles/s44358-025-00022-3}.

\bibitem[{Porter et~al.(2012)Porter, Hanson and Lin}]{porter_staying_2012}
Porter, J.~H., Hanson, P.~C. and Lin, C.-C. (2012) Staying afloat in the sensor
  data deluge.
\newblock \textit{Trends in Ecology \& Evolution}, \textbf{27}, 121--129.
\newblock
  \urlprefix\url{https://www.sciencedirect.com/science/article/pii/S0169534711003326}.

\bibitem[{Prince et~al.(2019)Prince, Hill, Pi{\~n}a~Covarrubias, Doncaster,
  Snaddon and Rogers}]{prince_deploying_2019}
Prince, P., Hill, A., Pi{\~n}a~Covarrubias, E., Doncaster, P., Snaddon, J.~L.
  and Rogers, A. (2019) Deploying {Acoustic} {Detection} {Algorithms} on
  {Low}-{Cost}, {Open}-{Source} {Acoustic} {Sensors} for {Environmental}
  {Monitoring}.
\newblock \textit{Sensors}, \textbf{19}, 553.
\newblock \urlprefix\url{https://www.mdpi.com/1424-8220/19/3/553}.

\bibitem[{Pritchard et~al.(2022)Pritchard, Sauls, Oldekop, Kiwango and
  Brockington}]{pritchard_data_2022}
Pritchard, R., Sauls, L.~A., Oldekop, J.~A., Kiwango, W.~A. and Brockington, D.
  (2022) Data justice and biodiversity conservation.
\newblock \textit{Conservation Biology}, \textbf{36}, e13919.
\newblock
  \urlprefix\url{https://conbio.onlinelibrary.wiley.com/doi/10.1111/cobi.13919}.

\bibitem[{Reddi et~al.(2025)Reddi, {https://orcid.org/0000-0002-5259-7721} and
  {Search about this author}}]{reddi_generative_2025}
Reddi, V.~J., {https://orcid.org/0000-0002-5259-7721} and {Search about this
  author} (2025) Generative {AI} at the {Edge}: {Challenges} and
  {Opportunities}.
\newblock \textit{Queue}, \textbf{23}, 79--137.
\newblock \urlprefix\url{https://spawn-queue.acm.org/doi/abs/10.1145/3733702}.

\bibitem[{Reynolds et~al.(2025)Reynolds, Beery, Burgess, Burgman, Butchart,
  Cooke, Coomes, Danielsen, Minin, Dur{\'a}n, Gassert, Hinsley, Jaffer, Jones,
  Li, Aodha, Madhavapeddy, O'Donnell, Oxbury, Peck, Pettorelli, Rodr{\'i}guez,
  Shuckburgh, Strassburg, Yamashita, Miao and
  Sutherland}]{reynolds_potential_2025}
Reynolds, S.~A., Beery, S., Burgess, N., Burgman, M., Butchart, S. H.~M.,
  Cooke, S.~J., Coomes, D., Danielsen, F., Minin, E.~D., Dur{\'a}n, A.~P.,
  Gassert, F., Hinsley, A., Jaffer, S., Jones, J. P.~G., Li, B.~V., Aodha,
  O.~M., Madhavapeddy, A., O'Donnell, S. A.~L., Oxbury, W.~M., Peck, L. et~al.
  (2025) The potential for {AI} to revolutionize conservation: a horizon scan.
\newblock \textit{Trends in Ecology \& Evolution}, \textbf{40}, 191--207.
\newblock
  \urlprefix\url{https://www.cell.com/trends/ecology-evolution/abstract/S0169-5347(24)00286-6}.

\bibitem[{Rhinehart et~al.(2020)Rhinehart, Chronister, Devlin and
  Kitzes}]{rhinehart_acoustic_2020}
Rhinehart, T.~A., Chronister, L.~M., Devlin, T. and Kitzes, J. (2020) Acoustic
  localization of terrestrial wildlife: {Current} practices and future
  opportunities.
\newblock \textit{Ecology and Evolution}, \textbf{10}, 6794--6818.
\newblock
  \urlprefix\url{https://onlinelibrary.wiley.com/doi/abs/10.1002/ece3.6216}.
\newblock \_eprint: https://onlinelibrary.wiley.com/doi/pdf/10.1002/ece3.6216.

\bibitem[{Rigoudy et~al.(2023)Rigoudy, Dussert, Benyoub, Besnard, Birck, Boyer,
  Bollet, Bunz, Caussimont, Chetouane, Carriburu, Cornette, Delestrade,
  De~Backer, Dispan, Le~Barh, Duhayer, Elder, Fanjul, Fonderflick, Froustey,
  Garel, Gaudry, G{\'e}rard, Gimenez, Hemery, Hemon, Jullien, Knitter,
  Malafosse, Marginean, M{\'e}nard, Ouvrier, Pariset, Prunet, Rabault, Randon,
  Raulet, R{\'e}gnier, Ribi{\`e}re, Ricci, Ruette, Schneylin, Sentilles,
  Siefert, Smith, Terpereau, Touchet, Thuiller, Uzal, Vautrain, Vimal, Weber,
  Spataro, Miele and Chamaill{\'e}-Jammes}]{rigoudy_deepfaune_2023}
Rigoudy, N., Dussert, G., Benyoub, A., Besnard, A., Birck, C., Boyer, J.,
  Bollet, Y., Bunz, Y., Caussimont, G., Chetouane, E., Carriburu, J.~C.,
  Cornette, P., Delestrade, A., De~Backer, N., Dispan, L., Le~Barh, M.,
  Duhayer, J., Elder, J.-F., Fanjul, J.-B., Fonderflick, J. et~al. (2023) The
  {DeepFaune} initiative: a collaborative effort towards the automatic
  identification of {European} fauna in camera trap images.
\newblock \textit{European Journal of Wildlife Research}, \textbf{69}, 113.
\newblock \urlprefix\url{https://doi.org/10.1007/s10344-023-01742-7}.

\bibitem[{Sabbella et~al.(2022)Sabbella, Nair, Gumme, Yadav, Chakrabartty and
  Thakur}]{sabbella_always-tinyml_2022}
Sabbella, H., Nair, A., Gumme, V., Yadav, S., Chakrabartty, S. and Thakur, C.
  (2022) An {Always}-{On} {tinyML} {Acoustic} {Classifier} for {Ecological}
  {Applications}.
\newblock In \textit{2022 {IEEE} {International} {Symposium} on {Circuits} and
  {Systems} ({ISCAS})}, 2393--2396.
\newblock \urlprefix\url{https://ieeexplore.ieee.org/document/9937827/}.

\bibitem[{Sakschewski et~al.(2025)Sakschewski, Caesar, Andersen, Bechthold,
  Bergfeld, Beusen, Billing, Bodirsky, Botsyun, Dennis, Donges, Dou, Eriksson,
  Fetzer, Gerten, H{\"a}yh{\"a}, Hebden, Heckmann, Heilemann, Huiskamp, Jahnke,
  Kaiser, Kitzmann, Kr{\"o}nke, K{\"u}hnel, Laureanti, Li, Liu, Loriani,
  Ludescher, Mathesius, Norstr{\"o}m, Otto, Paolucci, Pokhotelov,
  Rafiezadeh~Shahi, Raju, Rostami, Schaphoff, Schmidt, Steinert, Stenzel,
  Virkki, Wendt-Potthoff, Wunderling, Rockstr{\"o}m, Kitzmann, Caesar,
  Sakschewski and Rockstr{\"o}m}]{sakschewski_planetary_2025}
Sakschewski, B., Caesar, L., Andersen, L., Bechthold, M., Bergfeld, L., Beusen,
  A., Billing, M., Bodirsky, B.~L., Botsyun, S., Dennis, D.~P., Donges, J.,
  Dou, X., Eriksson, A., Fetzer, I., Gerten, D., H{\"a}yh{\"a}, T., Hebden, S.,
  Heckmann, T., Heilemann, A., Huiskamp, W. et~al. (2025) Planetary {Health}
  {Check} 2025: {A} {Scientific} {Assessment} of the {State} of the {Planet}.
\newblock \textit{Tech. rep.}, Potsdam Institute for Climate Impact Research
  (PIK).
\newblock
  \urlprefix\url{https://publications.pik-potsdam.de/pubman/item/item_32589}.
\newblock Artwork Size: 144 pages, 22 MB Medium: application/pdf Publication
  Title: Planetary Boundaries Science (PBScience).

\bibitem[{Sanakoyeu et~al.(2020)Sanakoyeu, Khalidov, McCarthy, Vedaldi and
  Neverova}]{sanakoyeu_transferring_2020}
Sanakoyeu, A., Khalidov, V., McCarthy, M.~S., Vedaldi, A. and Neverova, N.
  (2020) Transferring {Dense} {Pose} to {Proximal} {Animal} {Classes}.
\newblock \urlprefix\url{http://arxiv.org/abs/2003.00080}.
\newblock ArXiv:2003.00080 [cs].

\bibitem[{Sandbrook et~al.(2021)Sandbrook, Clark, Toivonen, Simlai, O'Donnell,
  Cobbe and Adams}]{sandbrook_principles_2021}
Sandbrook, C., Clark, D., Toivonen, T., Simlai, T., O'Donnell, S., Cobbe, J.
  and Adams, W. (2021) Principles for the socially responsible use of
  conservation monitoring technology and data.
\newblock \textit{Conservation Science and Practice}, \textbf{3}, e374.
\newblock
  \urlprefix\url{https://onlinelibrary.wiley.com/doi/abs/10.1111/csp2.374}.
\newblock \_eprint:
  https://conbio.onlinelibrary.wiley.com/doi/pdf/10.1111/csp2.374.

\bibitem[{Sanguineti et~al.(2021)Sanguineti, Guidi, Kulikovskiy and
  Taiuti}]{sanguineti_real-time_2021}
Sanguineti, M., Guidi, C., Kulikovskiy, V. and Taiuti, M.~G. (2021) Real-{Time}
  {Continuous} {Acoustic} {Monitoring} of {Marine} {Mammals} in the
  {Mediterranean} {Sea}.
\newblock \textit{Journal of Marine Science and Engineering}, \textbf{9}, 1389.
\newblock \urlprefix\url{https://www.mdpi.com/2077-1312/9/12/1389}.

\bibitem[{Sethi et~al.(2018)Sethi, Ewers, Jones, Orme and
  Picinali}]{sethi_robust_2018}
Sethi, S.~S., Ewers, R.~M., Jones, N.~S., Orme, C. D.~L. and Picinali, L.
  (2018) Robust, real-time and autonomous monitoring of ecosystems with an
  open, low-cost, networked device.
\newblock \textit{Methods in Ecology and Evolution}, \textbf{9}, 2383--2387.
\newblock
  \urlprefix\url{https://onlinelibrary.wiley.com/doi/abs/10.1111/2041-210X.13089}.
\newblock \_eprint:
  https://besjournals.onlinelibrary.wiley.com/doi/pdf/10.1111/2041-210X.13089.

\bibitem[{Sethi et~al.(2020)Sethi, Ewers, Jones, Signorelli, Picinali and
  Orme}]{sethi_safe_2020}
Sethi, S.~S., Ewers, R.~M., Jones, N.~S., Signorelli, A., Picinali, L. and
  Orme, C. D.~L. (2020) {SAFE} {Acoustics}: {An} open-source, real-time
  eco-acoustic monitoring network in the tropical rainforests of {Borneo}.
\newblock \textit{Methods in Ecology and Evolution}, \textbf{11}, 1182--1185.
\newblock
  \urlprefix\url{https://onlinelibrary.wiley.com/doi/abs/10.1111/2041-210X.13438}.
\newblock \_eprint:
  https://onlinelibrary.wiley.com/doi/pdf/10.1111/2041-210X.13438.

\bibitem[{Shi et~al.(2016)Shi, Cao, Zhang, Li and Xu}]{shi_edge_2016}
Shi, W., Cao, J., Zhang, Q., Li, Y. and Xu, L. (2016) Edge {Computing}:
  {Vision} and {Challenges}.
\newblock \textit{IEEE Internet of Things Journal}, \textbf{3}, 637--646.
\newblock \urlprefix\url{https://ieeexplore.ieee.org/document/7488250/}.

\bibitem[{Showen et~al.(2018)Showen, Dunson, Woodman, Christopher, Lim and
  Wilson}]{showen_locating_2018}
Showen, R., Dunson, C., Woodman, G.~H., Christopher, S., Lim, T. and Wilson,
  S.~C. (2018) Locating fish bomb blasts in real-time using a networked
  acoustic system.
\newblock \textit{Marine Pollution Bulletin}, \textbf{128}, 496--507.
\newblock
  \urlprefix\url{https://www.sciencedirect.com/science/article/pii/S0025326X18300407}.

\bibitem[{Sittinger et~al.(2024)Sittinger, Uhler, Pink and
  Herz}]{sittinger_insect_2024}
Sittinger, M., Uhler, J., Pink, M. and Herz, A. (2024) Insect detect: {An}
  open-source {DIY} camera trap for automated insect monitoring.
\newblock \textit{PLOS ONE}, \textbf{19}, e0295474.
\newblock
  \urlprefix\url{https://journals.plos.org/plosone/article?id=10.1371/journal.pone.0295474}.

\bibitem[{Speaker et~al.(2022)Speaker, O'Donnell, Wittemyer, Bruyere, Loucks,
  Dancer, Carter, Fegraus, Palmer, Warren and Solomon}]{speaker_global_2022}
Speaker, T., O'Donnell, S., Wittemyer, G., Bruyere, B., Loucks, C., Dancer, A.,
  Carter, M., Fegraus, E., Palmer, J., Warren, E. and Solomon, J. (2022) A
  global community-sourced assessment of the state of conservation technology.
\newblock \textit{Conservation Biology}, \textbf{36}.
\newblock
  \urlprefix\url{https://www.scopus.com/inward/record.uri?eid=2-s2.0-85125146572&doi=10.1111%2fcobi.13871&partnerID=40&md5=c69d81e37d489961f0584cee58d0d1f3}.
\newblock Type: Article.

\bibitem[{St{\"a}hli et~al.(2022)St{\"a}hli, Ost and
  Studer}]{stahli_development_2022}
St{\"a}hli, O., Ost, T. and Studer, T. (2022) Development of an {AI}-based
  bioacoustic wolf monitoring system.
\newblock \textit{The International FLAIRS Conference Proceedings},
  \textbf{35}.
\newblock \urlprefix\url{https://journals.flvc.org/FLAIRS/article/view/130552}.

\bibitem[{Stephenson et~al.(2022)Stephenson, Londo{\~n}o-Murcia, Borges,
  Claassens, Frisch-Nwakanma, Ling, McMullan-Fisher, Meeuwig, Unter, Walls,
  Burfield, do~Carmo Vieira~Correa, Geller, Montenegro~Paredes, Mubalama,
  Ntiamoa-Baidu, Roesler, Rovero, Sharma, Wiwardhana, Yang and
  Fumagalli}]{stephenson_measuring_2022}
Stephenson, P.~J., Londo{\~n}o-Murcia, M.~C., Borges, P. A.~V., Claassens, L.,
  Frisch-Nwakanma, H., Ling, N., McMullan-Fisher, S., Meeuwig, J.~J., Unter, K.
  M.~M., Walls, J.~L., Burfield, I.~J., do~Carmo Vieira~Correa, D., Geller,
  G.~N., Montenegro~Paredes, I., Mubalama, L.~K., Ntiamoa-Baidu, Y., Roesler,
  I., Rovero, F., Sharma, Y.~P., Wiwardhana, N.~W. et~al. (2022) Measuring the
  {Impact} of {Conservation}: {The} {Growing} {Importance} of {Monitoring}
  {Fauna}, {Flora} and {Funga}.
\newblock \textit{Diversity}, \textbf{14}, 824.
\newblock \urlprefix\url{https://www.mdpi.com/1424-2818/14/10/824}.
\newblock Number: 10.

\bibitem[{Stowell(2022)}]{stowell_computational_2022}
Stowell, D. (2022) Computational bioacoustics with deep learning: a review and
  roadmap.
\newblock \textit{PeerJ}, \textbf{10}, e13152.
\newblock \urlprefix\url{https://peerj.com/articles/13152}.

\bibitem[{Sutherland et~al.(2026)Sutherland, Butchart, Clarke, Doar, Doran,
  Douglas, Field, Fleishman, Gaston, Herbert-Read, Hughes, Kaartokallio, Maggs,
  Palardy, Pearce-Higgins, Peck, Pettorelli, Schloss, Spalding, Timoshyna,
  Tubbs, Uehara, Watson, Wentworth, Wilson and
  Thornton}]{sutherland_horizon_2026}
Sutherland, W.~J., Butchart, S. H.~M., Clarke, S.~J., Doar, N.~R., Doran, H.,
  Douglas, I.~C., Field, D.~J., Fleishman, E., Gaston, K.~J., Herbert-Read,
  J.~E., Hughes, A.~C., Kaartokallio, H., Maggs, L., Palardy, J.~E.,
  Pearce-Higgins, J.~W., Peck, L.~S., Pettorelli, N., Schloss, I.~R., Spalding,
  M.~D., Timoshyna, A. et~al. (2026) A horizon scan of biological conservation
  issues for 2026.
\newblock \textit{Trends in Ecology \& Evolution}, \textbf{41}, 91--101.
\newblock
  \urlprefix\url{https://www.cell.com/trends/ecology-evolution/abstract/S0169-5347(25)00301-5}.

\bibitem[{Teixeira et~al.(2023)Teixeira, Gomes, Oliveira, Fortes-Da-Silva,
  Soares and Raymundo-Pereira}]{teixeira_review_2023}
Teixeira, S.~C., Gomes, N.~O., Oliveira, T. V.~d., Fortes-Da-Silva, P., Soares,
  N. d. F.~F. and Raymundo-Pereira, P.~A. (2023) Review and {Perspectives} of
  sustainable, biodegradable, eco-friendly and flexible electronic devices and
  ({Bio})sensors.
\newblock \textit{Biosensors and Bioelectronics: X}, \textbf{14}, 100371.
\newblock
  \urlprefix\url{https://www.sciencedirect.com/science/article/pii/S2590137023000687}.

\bibitem[{Theocharides et~al.(2025)Theocharides, Verhelst, Reddy and
  Gousev}]{theocharides_tinymlefficient_2025}
Theocharides, T., Verhelst, M., Reddy, V.~J. and Gousev, E. (2025)
  {TinyML}---{From} {Efficient} {Edge} {Inference} to {On}-{Device}
  {Intelligence}.
\newblock \textit{IEEE Design \& Test}, \textbf{42}, 5--7.
\newblock
  \urlprefix\url{https://ieeexplore.ieee.org/abstract/document/11131547}.

\bibitem[{Trevathan et~al.(2025)Trevathan, Tan, Xing, Holzner, Kerlin, Zhou and
  Castley}]{trevathan_computer_2025}
Trevathan, J., Tan, W.~L., Xing, W., Holzner, D., Kerlin, D., Zhou, J. and
  Castley, G. (2025) A computer vision enhanced {IoT} system for koala
  monitoring and recognition.
\newblock \textit{Internet of Things}, \textbf{29}, 101474.
\newblock
  \urlprefix\url{https://www.sciencedirect.com/science/article/pii/S2542660524004153}.

\bibitem[{Tuia et~al.(2026)Tuia, Beery, Costelloe, Oliver and
  Lecomte}]{tuia_towards_2026}
Tuia, D., Beery, S., Costelloe, B.~R., Oliver, R.~Y. and Lecomte, N. (2026)
  Towards `digital ecology': {Advances} in integrating artificial intelligence
  from data generation to ecological insight.
\newblock \textit{Methods in Ecology and Evolution}, \textbf{17}, 222--227.
\newblock
  \urlprefix\url{https://onlinelibrary.wiley.com/doi/abs/10.1111/2041-210x.70243}.
\newblock \_eprint:
  https://besjournals.onlinelibrary.wiley.com/doi/pdf/10.1111/2041-210x.70243.

\bibitem[{Tuia et~al.(2022)Tuia, Kellenberger, Beery, Costelloe, Zuffi, Risse,
  Mathis, Mathis, van Langevelde, Burghardt, Kays, Klinck, Wikelski, Couzin,
  van Horn, Crofoot, Stewart and Berger-Wolf}]{tuia_perspectives_2022}
Tuia, D., Kellenberger, B., Beery, S., Costelloe, B.~R., Zuffi, S., Risse, B.,
  Mathis, A., Mathis, M.~W., van Langevelde, F., Burghardt, T., Kays, R.,
  Klinck, H., Wikelski, M., Couzin, I.~D., van Horn, G., Crofoot, M.~C.,
  Stewart, C.~V. and Berger-Wolf, T. (2022) Perspectives in machine learning
  for wildlife conservation.
\newblock \textit{Nature Communications}, \textbf{13}, 792.
\newblock \urlprefix\url{https://www.nature.com/articles/s41467-022-27980-y}.
\newblock Number: 1.

\bibitem[{Verma and Kumar(2024)}]{verma_aviear_2024}
Verma, R. and Kumar, S. (2024) {AviEar}: {An} {IoT}-{Based} {Low}-{Power}
  {Solution} for {Acoustic} {Monitoring} of {Avian} {Species}.
\newblock \textit{IEEE Sensors Journal}, \textbf{24}, 42088--42102.
\newblock \urlprefix\url{https://ieeexplore.ieee.org/document/10742275/}.

\bibitem[{Vuilliomenet et~al.(2025)Vuilliomenet, Mart{\'i}nez~Balvanera,
  Mac~Aodha, Jones and Wilson}]{vuilliomenet_acoupi_2025}
Vuilliomenet, A., Mart{\'i}nez~Balvanera, S., Mac~Aodha, O., Jones, K.~E. and
  Wilson, D. (2025) acoupi: {An} open-source {Python} framework for deploying
  bioacoustic {AI} models on edge devices.
\newblock \textit{Methods in Ecology and Evolution}, \textbf{n/a}.
\newblock
  \urlprefix\url{https://onlinelibrary.wiley.com/doi/abs/10.1111/2041-210x.70208}.
\newblock \_eprint:
  https://besjournals.onlinelibrary.wiley.com/doi/pdf/10.1111/2041-210x.70208.

\bibitem[{W{\"a}gele et~al.(2022)W{\"a}gele, Bodesheim, Bourlat, Denzler,
  Diepenbroek, Fonseca, Frommolt, Geiger, Gemeinholzer, Gl{\"o}ckner, Haucke,
  Kirse, K{\"o}lpin, Kostadinov, K{\"u}hl, Kurth, Lasseck, Liedke, Losch,
  M{\"u}ller, Petrovskaya, Piotrowski, Radig, Scherber, Schoppmann, Schulz,
  Steinhage, Tschan, Vautz, Velotto, Weigend and
  Wildermann}]{wagele_towards_2022}
W{\"a}gele, J.~W., Bodesheim, P., Bourlat, S.~J., Denzler, J., Diepenbroek, M.,
  Fonseca, V., Frommolt, K.-H., Geiger, M.~F., Gemeinholzer, B., Gl{\"o}ckner,
  F.~O., Haucke, T., Kirse, A., K{\"o}lpin, A., Kostadinov, I., K{\"u}hl,
  H.~S., Kurth, F., Lasseck, M., Liedke, S., Losch, F., M{\"u}ller, S. et~al.
  (2022) Towards a multisensor station for automated biodiversity monitoring.
\newblock \textit{Basic and Applied Ecology}, \textbf{59}, 105--138.
\newblock
  \urlprefix\url{https://www.sciencedirect.com/science/article/pii/S1439179122000032}.

\bibitem[{Wang et~al.(2025)Wang, Tang, Guo, Meng, Wang, Wang and
  Jia}]{wang_empowering_2025}
Wang, X., Tang, Z., Guo, J., Meng, T., Wang, C., Wang, T. and Jia, W. (2025)
  Empowering {Edge} {Intelligence}: {A} {Comprehensive} {Survey} on
  {On}-{Device} {AI} {Models}.
\newblock \textit{ACM Comput. Surv.}, \textbf{57}, 228:1--228:39.
\newblock \urlprefix\url{https://dl.acm.org/doi/10.1145/3724420}.

\bibitem[{Warden and Situnayake(2020)}]{warden_tinyml_2020}
Warden, P. and Situnayake, D. (2020) \textit{{TinyML}: {Machine} {Learning}
  with {TensorFlow} {Lite} on {Arduino} and {Ultra}-{Low}-{Power}
  {Microcontrollers}}.
\newblock Sebastopol, CA: O'Reilly Media Inc.

\bibitem[{Whytock et~al.(2023)Whytock, Suijten, van Deursen,
  {\'S}wie{\.z}ewski, Mermiaghe, Madamba, Mouckoumou, Zwerts, Pambo, Bahaa-el
  din, Brittain, Cardoso, Henschel, Lehmann, Momboua, Makaga, Orbell, White,
  Iponga and Abernethy}]{whytock_real-time_2023}
Whytock, R.~C., Suijten, T., van Deursen, T., {\'S}wie{\.z}ewski, J.,
  Mermiaghe, H., Madamba, N., Mouckoumou, N., Zwerts, J.~A., Pambo, A. F.~K.,
  Bahaa-el din, L., Brittain, S., Cardoso, A.~W., Henschel, P., Lehmann, D.,
  Momboua, B.~R., Makaga, L., Orbell, C., White, L. J.~T., Iponga, D.~M. and
  Abernethy, K.~A. (2023) Real-time alerts from {AI}-enabled camera traps using
  the {Iridium} satellite network: {A} case-study in {Gabon}, {Central}
  {Africa}.
\newblock \textit{Methods in Ecology and Evolution}, \textbf{14}, 867--874.
\newblock
  \urlprefix\url{https://onlinelibrary.wiley.com/doi/abs/10.1111/2041-210X.14036}.
\newblock \_eprint:
  https://besjournals.onlinelibrary.wiley.com/doi/pdf/10.1111/2041-210X.14036.

\bibitem[{Wild et~al.(2023)Wild, van Schalkwyk, Viljoen, Heine, Richter,
  Vorneweg, Koblitz, Dechmann, Rogers, Partecke, Linek, Volkmer, Gregersen,
  Havm{\o}ller, Morelle, Daim, Wiesner, Wolter, Fiedler, Kays, Ezenwa, Meboldt
  and Wikelski}]{wild_multi-species_2023}
Wild, T.~A., van Schalkwyk, L., Viljoen, P., Heine, G., Richter, N., Vorneweg,
  B., Koblitz, J.~C., Dechmann, D. K.~N., Rogers, W., Partecke, J., Linek, N.,
  Volkmer, T., Gregersen, T., Havm{\o}ller, R.~W., Morelle, K., Daim, A.,
  Wiesner, M., Wolter, K., Fiedler, W., Kays, R. et~al. (2023) A multi-species
  evaluation of digital wildlife monitoring using the {Sigfox} {IoT} network.
\newblock \textit{Animal Biotelemetry}, \textbf{11}, 13.
\newblock \urlprefix\url{https://doi.org/10.1186/s40317-023-00326-1}.

\bibitem[{{WILDLABS}(2024)}]{wildlabs_2024_2024}
{WILDLABS} (2024) 2024 {Annual} {Report}.
\newblock \textit{Annual {Report}}, WILDLABS.NET.
\newblock
  \urlprefix\url{https://wildlabs.net/sites/default/files/2025-07/WILDLABS%202024%20Annual%20Report.pdf}.

\bibitem[{Wilkinson et~al.(2016)Wilkinson, Dumontier, Aalbersberg, Appleton,
  Axton, Baak, Blomberg, Boiten, da~Silva~Santos, Bourne, Bouwman, Brookes,
  Clark, Crosas, Dillo, Dumon, Edmunds, Evelo, Finkers, Gonzalez-Beltran, Gray,
  Groth, Goble, Grethe, Heringa, 't~Hoen, Hooft, Kuhn, Kok, Kok, Lusher,
  Martone, Mons, Packer, Persson, Rocca-Serra, Roos, van Schaik, Sansone,
  Schultes, Sengstag, Slater, Strawn, Swertz, Thompson, van~der Lei, van
  Mulligen, Velterop, Waagmeester, Wittenburg, Wolstencroft, Zhao and
  Mons}]{wilkinson_fair_2016}
Wilkinson, M.~D., Dumontier, M., Aalbersberg, I.~J., Appleton, G., Axton, M.,
  Baak, A., Blomberg, N., Boiten, J.-W., da~Silva~Santos, L.~B., Bourne, P.~E.,
  Bouwman, J., Brookes, A.~J., Clark, T., Crosas, M., Dillo, I., Dumon, O.,
  Edmunds, S., Evelo, C.~T., Finkers, R., Gonzalez-Beltran, A. et~al. (2016)
  The {FAIR} {Guiding} {Principles} for scientific data management and
  stewardship.
\newblock \textit{Scientific Data}, \textbf{3}, 160018.
\newblock \urlprefix\url{https://www.nature.com/articles/sdata201618}.

\bibitem[{Williams et~al.(2025)Williams, van Merri{\"e}nboer, Dumoulin, Hamer,
  Fleishman, McKown, Munger, Rice, Lillis, White, Hobbs, Razak, Curnick, Jones
  and Denton}]{williams_using_2025}
Williams, B., van Merri{\"e}nboer, B., Dumoulin, V., Hamer, J., Fleishman,
  A.~B., McKown, M., Munger, J., Rice, A.~N., Lillis, A., White, C., Hobbs, C.,
  Razak, T., Curnick, D., Jones, K.~E. and Denton, T. (2025) Using tropical
  reef, bird and unrelated sounds for superior transfer learning in marine
  bioacoustics.
\newblock \textit{Philosophical Transactions of the Royal Society B: Biological
  Sciences}, \textbf{380}, 20240280.
\newblock \urlprefix\url{https://doi.org/10.1098/rstb.2024.0280}.

\bibitem[{Wood et~al.(2024)Wood, G{\"u}nther, Rex, Hofstadter, Reers, Kahl,
  Peery and Klinck}]{wood_real-time_2024}
Wood, C.~M., G{\"u}nther, F., Rex, A., Hofstadter, D.~F., Reers, H., Kahl, S.,
  Peery, M.~Z. and Klinck, H. (2024) Real-time acoustic monitoring facilitates
  the proactive management of biological invasions.
\newblock \textit{Biological Invasions}.
\newblock \urlprefix\url{https://link.springer.com/10.1007/s10530-024-03426-y}.

\bibitem[{{XPrize Rainforest}(2024)}]{xprize_rainforest_xprize_2024}
{XPrize Rainforest} (2024) {XPrize} {Rainforest} {Solution} {Catalogue}: {The}
  innovative {Technologies} revolutionizing tropical rainforest conservation.
\newblock \textit{Tech. rep.}, XPrize Rainforest, alana.
\newblock
  \urlprefix\url{https://assets-us-01.kc-usercontent.com/9bc15d1f-8a5c-007d-b507-e3496e85af86/12d20e04-b5f2-452d-b694-91ae1a219aa1/XPRF%20SOLUTION%20SHOWCASE.pdf}.

\bibitem[{Yu et~al.(2024)Yu, Amador, Cribellier, Klaassen, de~Knegt, Naguib,
  Nijland, Nowak, Prins, Snijders, Tyson and Muijres}]{yu_edge_2024}
Yu, H., Amador, G.~J., Cribellier, A., Klaassen, M., de~Knegt, H.~J., Naguib,
  M., Nijland, R., Nowak, L., Prins, H.~H., Snijders, L., Tyson, C. and
  Muijres, F.~T. (2024) Edge computing in wildlife behavior and ecology.
\newblock \textit{Trends in Ecology \& Evolution}, \textbf{39}, 128--130.
\newblock
  \urlprefix\url{https://www.sciencedirect.com/science/article/pii/S0169534723003312}.

\bibitem[{Zakaria et~al.(2020)Zakaria, Nordin and Ruslan}]{zakaria_iot_2020}
Zakaria, M.~S., Nordin, R. and Ruslan, M.~S. (2020) {IoT} {Infrastructure} for
  {Environmental} {Monitoring}: {A} {Case} at {Tasik} {Chini} {Biosphere}
  {Reserve}.
\newblock In \textit{2020 {IEEE} {Conference} on e-{Learning}, e-{Management}
  and e-{Services} ({IC3e})}, 43--48.
\newblock Journal Abbreviation: 2020 IEEE Conference on e-Learning,
  e-Management and e-Services (IC3e).

\bibitem[{Zeuss et~al.(2024)Zeuss, Bald, Gottwald, Becker, Bellafkir, Bendix,
  Bengel, Beumer, Brandl, Br{\"a}ndle, Dahlke, Farwig, Freisleben, Friess,
  Heidrich, Heuer, H{\"o}chst, Holzmann, Lampe, Leberecht, Lindner, Masello,
  Mielke~M{\"o}glich, M{\"u}hling, M{\"u}ller, Noskov, Opgenoorth, Peter,
  Quillfeldt, R{\"o}sner, Royaut{\'e}, Mestre-Runge, Schabo, Schneider, Seeger,
  Shayle, Steinmetz, Tafo, Vogelbacher, W{\"o}llauer, Younis, Zobel and
  Nauss}]{zeuss_nature_2024}
Zeuss, D., Bald, L., Gottwald, J., Becker, M., Bellafkir, H., Bendix, J.,
  Bengel, P., Beumer, L.~T., Brandl, R., Br{\"a}ndle, M., Dahlke, S., Farwig,
  N., Freisleben, B., Friess, N., Heidrich, L., Heuer, S., H{\"o}chst, J.,
  Holzmann, H., Lampe, P., Leberecht, M. et~al. (2024) Nature 4.0: {A}
  networked sensor system for integrated biodiversity monitoring.
\newblock \textit{Global Change Biology}, \textbf{30}, e17056.
\newblock
  \urlprefix\url{https://onlinelibrary.wiley.com/doi/abs/10.1111/gcb.17056}.
\newblock \_eprint: https://onlinelibrary.wiley.com/doi/pdf/10.1111/gcb.17056.

\bibitem[{Zhao et~al.(2021)Zhao, Kam, Kymissis and
  Culligan}]{zhao_lorawan-based_2021}
Zhao, H., Kam, K.~A., Kymissis, I. and Culligan, P.~J. (2021) A
  {LoRaWAN}-{Based} {Environmental} {Sensor} {System} for {Urban} {Tree}
  {Health} {Monitoring}.
\newblock In \textit{2021 {IEEE} {Sensors}}, 1--4. Sydney, Australia: IEEE.
\newblock \urlprefix\url{https://ieeexplore.ieee.org/document/9639788/}.

\bibitem[{Zhou et~al.(2019)Zhou, Chen, Li, Zeng, Luo and
  Zhang}]{zhou_edge_2019}
Zhou, Z., Chen, X., Li, E., Zeng, L., Luo, K. and Zhang, J. (2019) Edge
  {Intelligence}: {Paving} the {Last} {Mile} of {Artificial} {Intelligence}
  {With} {Edge} {Computing}.
\newblock \textit{Proceedings of the IEEE}, \textbf{107}, 1738--1762.
\newblock \urlprefix\url{https://ieeexplore.ieee.org/document/8736011/}.

\bibitem[{Zikulnig et~al.(2025)Zikulnig, Carrara and
  Kosel}]{zikulnig_life_2025}
Zikulnig, J., Carrara, S. and Kosel, J. (2025) A life cycle assessment approach
  to minimize environmental impact for sustainable printed sensors.
\newblock \textit{Scientific Reports}, \textbf{15}, 10866.
\newblock \urlprefix\url{https://www.nature.com/articles/s41598-025-95682-8}.

\bibitem[{Zuffi et~al.(2019)Zuffi, Kanazawa, Berger-Wolf and
  Black}]{zuffi_three-d_2019}
Zuffi, S., Kanazawa, A., Berger-Wolf, T. and Black, M.~J. (2019) Three-{D}
  {Safari}: {Learning} to {Estimate} {Zebra} {Pose}, {Shape}, and {Texture}
  from {Images} "{In} the {Wild}".
\newblock \urlprefix\url{http://arxiv.org/abs/1908.07201}.
\newblock ArXiv:1908.07201 [cs].

\end{thebibliography}
    
    \clearpage
    \appendix{}
\section{Appendix}

\subsection{Review Methods}
\label{appendix:review_methods}
\textbf{Search Strategy}\\
Between October and November 2025, a structured literature search was conducted across five databases: ACM Digital Library, arXiv, bioRxiv, IEEE Xplore, and Web of Science. 
All retrieved records were exported in BibTeX format and imported into Zotero, where title/abstract screening, eligibility assessment, and final inclusion were managed. 
To ensure transparency and reproducibility, only English-language, open-access journal articles and conference papers were retained for further analysis.
Our review does not restrict its scope to a specific ecological level of organisation (e.g. individuals, populations, communities, or ecosystems). 
Instead, we adopt a methodological focus, encompassing all ecological applications that engage with edge computing or edge artificial intelligence, defined here as systems that either (i) execute AI models on embedded or resource-constrained hardware platforms at or near the point of data collection, or (ii) transmit ecological data via wireless connectivity for remote or cloud-based inference.
This broad definition aim to capture the continuum of edge-to-cloud architectures employ in biodiversity monitoring.

The below boolean query was adapted to the indexing conventions of each database and applied to titles and abstracts, or equivalent metadata fields. 

\begin{wileyboxTitled}[]
("biodiversity" OR "ecology" or "wildlife")  
AND
("edge computing" OR "edgeAI" OR "edge AI" OR "edge artificial intelligence" OR "embedded artificial intelligence" OR "embedded AI" OR "embeddedAI" 
OR "tiny ML" OR "tinyML" OR "tiny machine learning"
OR "IoT" OR "internet of things" OR "cyber-physical system"
OR "WSN" OR "wireless sensor network" OR "sensor network") 
\end{wileyboxTitled}

\textbf{Publications Screening}\\
An initial screening was performed based on titles and abstracts. Remaining papers were then assessed by reviewing methods and results sections, confirming the presence of an implemented edge system and a relevant contribution to edge or distributed sensing.
We excluded articles meeting any of the following criteria:

 \begin{itemize}
     \item The study focused on IoT applications in manufacturing, industrial automation, smart homes, or governance, without an ecological or biodiversity monitoring component.
     \item The study proposed theoretical or conceptual frameworks for IoT, networking, security, or governance without implementing or evaluating a sensing or edge-AI prototype.
     \item The paper outlines a theoretical framework and empirical guidelines for the deployment of an edge AI system but does not develop a prototype.
     \item The study developed novel ML or DL techniques for modelling ecological patterns (e.g. species distributions, population dynamics, behaviour), but makes no reference to wireless sensor networks, edge-based data processing, or field deployment architecture.
     \item The study deploys stand-alone sensing devices or develops new biologging hardware for field data collection but did not include wireless connectivity or on-device processing.
     \item The study introduced ML algorithms to analyse ecological or environmental datasets, without addressing their deployment, constraints, or performance on edge devices.
 \end{itemize}
 
\textbf{Limitations}\\
Although the search query captured a broad range of ecological sensing systems, this review intentionally narrows its analytical scope. 
Articles centred on environmental sensing networks where excluded where they capture solely abiotic parameters (e.g., temperature, humidity, particulate matter) without organism-level inference.
Readers are referred to reviews of environmental sensing and IoT-based environmental monitoring.
Similarly, studies employing environmental DNA (eDNA) were excluded, as most current eDNA workflows require laboratory-based DNA extraction and computationally intensive DNA sequence matchings against large loud-based genetic databases genomic databases.
While, research is advancing towards mobile, autonomous eDNA sequencing systems using nanopore technology \citep{hammad_nanopore-aware_2025}, and on-device genomic inference using custom designed hardware \citep{magierowski_sequencing_2025}, these solutions remains at an early stage, lacking sufficient field validation for comparative synthesis. 
Finally, robotic, aerial, and satellite-based remote sensing were excluded, as these systems rely predominantly on cloud-based processing.



\clearpage

\subsection{Hardware Overview}
\label{appendix:hardware_table}
\newcolumntype{L}{>{\RaggedRight\arraybackslash}X}

\begin{table}[!htbp]
\caption{Overview of popular edge devices.}
\begin{threeparttable}
\small
\setlength\tabcolsep{5pt} 
\setlength{\arrayrulewidth}{0.1pt} 

\begin{tabularx}{\textwidth}{
    >{\hsize=1.35\hsize}L     
    >{\hsize=1.3\hsize}L     
    >{\hsize=1.35\hsize}L     
    >{\hsize=0.5\hsize}L     
    >{\hsize=1.05\hsize}L     
    >{\hsize=0.75\hsize}L     
    >{\hsize=1.0\hsize}L     
    >{\hsize=0.7\hsize}L     
    >{\hsize=1.0\hsize}L     
}
\headrow
\thead{Device Name} & \thead{Type}  & \thead{CPU, GPU} & \thead{NPU} & \thead{Clock Freq.} & 
\thead{RAM} & \thead{Memory} & \thead{ISA} & 
\thead{Power} \\ \hline

AudioMoth & MCU EFM32 Gecko & 1-core CPU Cortex-M3 & --- & 32 MHz & 16kB & Flash 128MB & Arm v7-M & 180 $\mu$A/Hz D.Sleep (0.9 $\mu$A) \\

Arduino Nano 33 BLE & MCU nRF52850 & 1-core CPU Cortex-M4 & --- & 64 MHz & 256kB & Flash 1MB & Arm v7E-M & --- \\

SparkFun Edge Dev Apollo3 Blue & MCU ambiq Apollo3 Blue & 1-core CPU Cortex-M4 & --- & 48 MHz / 96 MHz & 384kB & Flash 1MB & Arm v7E-M & {<}1 mW / PMU D.Sleep (1 $\mu$A) \\

Raspberry Pi Pico & MCU RP2040 & 2-core CPU Cortex-M0+ & --- & 133 MHz & 264kB & Flash 2MB & Arm v6-M & 2.4 W --- 3.2 W  \\

B-U585I-IOT02A Discovery kit & MCU STM32U585AI & 1-core CPU Cortex-M33 & --- & 160 MHz & 786kB & Flash 2MB & Arm v8-M & --- \\

Arduino Portenta H7 & MCU STM32H747XI & 2-core CPU Cortex-M4, Cortex-M7 & --- & 240 MHz / 480 MHz & 8MB & Flash 16MB & Arm v7E-M & 120 mA D.Sleep 2.95$\mu$A \\

OpenMV Cam & MCU NXP i.MX RT1060 & 1-core CPU Cortex-M7 & --- & 600 MHz & 33MB & Flash 16MB & Arm v7E-M & 100 mA D.Sleep 30$\mu$A at 3.7V \\

Raspberry Pi 4B & SBC with SoC BCM2711 & 4-core CPU Cortex-A72 & --- & 1.8 GHz & 2-8GB & microSD & Arm v8-A & 540 mA  \\

Nvidia Jetson Nano Dev Kit & SoC & 4-core CPU Cortex-A57, 128-core Maxwell GPU & --- & 1.43 GHz & 4GB & microSD & Arm v8-A & --- \\

Google Coral Dev Board & SBC with SoC NXP i.MX 8M & 4-core CPU Cortex A-53, 1-core CPU Cortex-M4F & TPU & 1.5 GHz & 1GB & eMMC 8GB & Arm v8-A & --- \\

Raspberry Pi 5 & SBC with SoC BCM2712 & 4-core CPU Cortex-A76 & --- & 2.4 GHz & 2-16GB & microSD & Arm v8.2-A & --- \\


\hline
\end{tabularx}
\end{threeparttable}
\end{table}

\clearpage

\subsection{List of reviewed publications}
\label{appendix:list_of_publications}
\begin{enumerate}
    \item Trevathan J., Tan, W. L., Xing, W., Holzner, D., Kerlin, D., Zhou, J. and Castley, G. (2025).
    \textit{A computer vision enhanced IoT system for koala monitoring and recognition}.
    \url{https://doi.org/10.1016/j.iot.2024.101474}

    \item Zhao, H., Kam, K. A., Kymissis, I. and Culligan, P. J (2021)
    \textit{A LoRaWAN based environemntal sensor system for urban tree health monitoring}
    \url{https://doi.org/10.1109/SENSORS47087.2021.9639788}

    \item Vasconcelos, D., and Nunes, N. J. (2022)
    \textit{A Low-Cost Multi-Purpose IoT Sensor for Biologging and Soundscape Activities}
    \url{https://doi.org/10.3390/s22197100}

    \item Kimutai, G. and Förster, A. (2023)
    \textit{A low-cost TinyML model for Mosquito Detection in Resource-Constrained Environments}
    \url{https://doi.org/10.1145/3582515.3609514}

    \item Alippi, C., Ambrosini, R., Longoni, V., Cogliati, D. and Roveri, M. (2017)
    \textit{A lightweight and energy-efficient Internet-of-birds tracking system}
    \url{https://doi.org/10.1109/PERCOM.2017.7917862}

    \item Wild, T.A., van Schalkwyk, L., Viljoen, P. et al. (2023)
    \textit{A multi-species evaluation of digital wildlife monitoring using the Sigfox IoT network}
    \url{https://doi.org/10.1186/s40317-023-00326-1}

    \item Li, Y., Liu, J., Kusy, B., Marchant, R., Do, B., Merz,  T., Crosswell, J., Steven, A., Tychsen-Smith, L., Ahmedt-Aristizabal, D., Oorloff, J., Moghadam, P., Babcock, R., Malpani, M. and Oerlemans, A. (2022) 
    \textit{A Real-time Edge-AI System for Reef Surveys}
    \url{https://doi.org/10.48550/arXiv.2208.00598}
    
    \item Gazis, A. and Katsiri, E. (2020)
    \textit{A wireless sensor network for underground passages: Remote sensing and wildlife monitoring}
    \url{https://doi.org/10.1002/eng2.12170}

    \item Mainetti, L. (2023)
    \textit{Acoustic Identification of Wood-Boring Insects with TinyML}
    \url{https://hdl.handle.net/10589/208626}

    \item Vuilliomenet A., Martínez Balvanera S., Mac Aodha O.,  Jones K.E., Wilson D. (2025)
    \textit{acoupi: An open-source Python framework for deploying bioacoustic AI models on edge devices}
    \url{https://doi.org/10.1111/2041-210x.70208}

    \item De Simone, A., Barbisan, L., Turvani, G. and Riente, F. (2024)
    \textit{Advancing Beekeeping: IoT and TinyML for Queen Bee Monitoring Using Audio Signals}
    \url{https://doi.org/10.1109/TIM.2024.3449981}

    \item Ratsakatika, T. (2025)
    \textit{AI-based alert system for bears and wild boars in Romania’s Carpathian Mountains}
    \url{https://github.com/ratsakatika/camera-traps}

    \item Caillau, A. (2024)
    \textit{AI for Bears Challenge: Empowering developers while fostering conservation}
    \url{https://github.com/earthtoolsmaker/bear-conservation }

    \item Ibraheam, M., Li, K. F., Gebali, F. (2023)
    \textit{An Accurate and Fast Animal Species Detection System for Embedded Devices}
    \url{https://doi.org/10.1109/ACCESS.2023.3252499}

    \item Sabbella, H., Nair, A., Gumme, V., Yadav, S., Chakrabartty, S., Thakur, C. (2022)
    \textit{An Always-On tinyML Acoustic Classifier for Ecological Applications}
    \url{https://doi.org/10.1109/ISCAS48785.2022.9937827}

    \item Luder, V., Schulthess, L., Cortesi, S., Davis, L. R. and Magno, M. (2025)
    \textit{AniTrack: A Power-Efficient, Time-Slotted and Robust UWB Localization System for Animal Tracking in a Controlled Setting}
    \url{https://doi.org/10.48550/arXiv.2506.00216}

    \item Fourcade, P., Cabanillas, E. and Brulais, S. (2024)
    \textit{Autonomous, Connected and AI-embedded Multi-Sensor System for Wildlife Monitoring}
    \url{https://doi.org/10.1109/SSI63222.2024.10740549}

    \item Ramasubramanian, C., Lokiah, S., Viswanath, Y. and Jamthe, S. (2022)
    \textit{Averting Human-Elephant Conflict using IoT and Machine Learning of Elephant Vocalizations}
    \url{https://doi.org/10.1109/WF-IOT54382.2022.10152220}

    \item Verma, R. and Kumar, S. (2024)
    \textit{AviEar: An IoT-Based Low-Power Solution for Acoustic Monitoring of Avian Species}
    \url{https://doi.org/10.1109/JSEN.2024.3487638}

    \item Mahbub, T., Bhagwagar, A., Chand, P., Zualkernan, I., Judas, J. and Dghaym, D. (2024)
    \textit{Bat2Web: A Framework for Real-Time Classification of Bat Species Echolocation Signals Using Audio Sensor Data}
    \url{https://doi.org/10.3390/s24092899}

    \item Kaur, M., Singh, K. and Kumar, S. (2024)
    \textit{Biodiversity Sensor: A Customized and Power Efficient Solution for Biodiversity Surveillance}
    \url{https://doi.org/10.1109/JSEN.2023.3328067}

    \item Contina, A., Abelson, E., Allison, B., Stokes, B., Sanchez, K. F., Hernandez, H. M., Kepple, A. M., Tran, Q., Kazen, I., Brown, K. A., Powell, J. H. and Keitt, T. H. (2024)
    \textit{BioSense: An automated sensing node for organismal and environmental biology}
    \url{https://doi.org/10.1016/j.ohx.2024.e00584}

    \item Höchst, J., Bellafkir, H., Lampe, P., Vogelbacher, M., Mühling, M., Schneider, D., Lindner, K., Rösner, S., Schabo, D. G., Farwig, N. and Freisleben, B. (2022)
    \textit{Bird@Edge: Bird Species Recognition at the Edge}
    \url{https://jonashoechst.de/assets/papers/hoechst2022birdedge.pdf}

    \item Disabato, S., Canonaco, G., Flikkema, P. G., Roveri, M. and Alippi, C. (2021)
    \textit{Birdsong Detection at the Edge with Deep Learning}
    \url{https://doi.org/10.1109/SMARTCOMP52413.2021.00022 }

    \item Das, N., Padhy, N., Dey, N., Mukherjee, A., Maiti, A. (2022)
    \textit{Building of an edge enabled drone network ecosystem for bird species identification}
    \url{https://doi.org/10.1016/j.ecoinf.2021.101540}

    \item Roe, P., Ferroudj, M., Towsey, M., Schwarzkopf, L. (2018)
    \textit{Catching Toad Calls in the Cloud: Commodity Edge Computing for Flexible Analysis of Big Sound Data}
    \url{https://doi.org/10.1109/eScience.2018.00022}

    \item Gauld, J., Atkinson, P.W., Silva, J.P. et al. (2023)
    \textit{Characterisation of a new lightweight LoRaWAN GPS bio-logger and deployment on griffon vultures Gyps fulvus}
    \url{https://doi.org/10.1186/s40317-023-00329-y}

    \item Monedero, I., Barbancho, J., Márquez, R. and Beltrán, J. F. (2021)
    \textit{Cyber-Physical System for Environmental Monitoring Based on Deep Learning}
    \url{https://doi.org/10.3390/s21113655}

    \item Duhart, C., Dublon, G., Mayton, B. and Paradiso, J. (2018)
    \textbf{Deep Learning Locally Trained Wildlife Sensing  in Real Acoustic Wetland Environment}
    \url{https://doi.org/10.1007/978-981-13-5758-9_1}

    \item De Luca, M., Loreti, P., Bracciale, L., Colosimo, G., et al. (2025)
    \textit{Design of a LoRa-Based Multisensor Device for the Internet of Animals}
    \url{https://doi.org/10.1109/JIOT.2025.3574494}

    \item Kubizňák, P., Hochachka, W. M., Osoba, V., Kotek, T., Kuchař, J., Klapetek, V., Hradcová, K., Růžička, J., Zárybnická, M. (2019)
    \textit{Designing network-connected systems for ecological research and education}
    \url{https://doi.org/10.1002/ecs2.2761}

    \item Sharma, S., Pansri, B., Timilsina, S., Gautam, B. P., Okada, Y., Watanabe, S., Kondo, S., Sato, K. (2024)
    \textit{Developing an Alert System for Agricultural Protection: Sika Deer Detection Using Raspberry Pi}
    \url{https://doi.org/10.3390/electronics13234852}

    \item Stähli, O., Ost, T., Studer, T. (2022)
	\textit{Development of an AI-based bioacoustic wolf monitoring system}
    \url{https://doi.org/10.32473/flairs.v35i.130552}

    \item Ayankoso, S., Wang, Z., Shi, D., Yang, W., Vikiru, A., Kamau, S., Muchiri, H. and Gu, F. (2024)
    \textit{Development of Long-Range, Low-Powered and Smart IoT Device for Detecting Illegal Logging in Forests}
    \url{https://doi.org/10.37965/jdmd.2024.550}

    \item Tydén, A. and Olsson, S. (2020)
    \textit{Edge Machine Learning for Animal Detection, Classification, and Tracking}
    \url{https://api.semanticscholar.org/CorpusID:263758169}

    \item Hing, K. K. and Behjati, M. (2025)
    \textit{Edge Intelligence for Wildlife Conservation: Real-Time Hornbill Call Classification Using TinyML}
    \url{https://doi.org/10.48550/arXiv.2504.12272}

    \item Bandara, N. S., Bandara, D. P., et al. (2025)
    \textit{Elemantra: An End-to-End Framework Empowered with Edge AI to Tackle Human-Elephant Conflict}
    \url{https://doi.org/10.1109/APSCON63569.2025.11144360}

    \item Ciapponi, S., Mannini, L., Scanferla, J., Anderle, M. and Farella, E. (2025)
    \textit{Enabling Multi-Species Bird Classification on Low-Power Bioacoustic Loggers}
    \url{https://doi.org/10.48550/arXiv.2509.20103}

    \item Brunelli, D., Albanese, A., d’Acunto, D. and Nardello, M. (2019)
    \textit{Energy Neutral Machine Learning Based IoT Device for Pest Detection in Precision Agriculture}
    \url{https://doi.org/10.1109/IOTM.0001.1900037}

    \item Li, K., Shan, M., Perez, S. B., Luo, K. and Worrall, S. (2024)
    \textit{Endangered Alert: A Field-Validated Self-Training Scheme for Detecting and Protecting Threatened Wildlife on Roads and Roadsides}
    \url{https://doi.org/10.48550/arXiv.2412.12222}

    \item Darras, K. F. A., Balle, M., Xu, W., Yan, Y., Zakka, V. G., Toledo-Hernández, M., Sheng, D., Lin, W., Zhang, B., Lan, Z., Fupeng, L. and Wanger, T. C. (2024)
    \textit{Eyes on nature: Embedded vision cameras for terrestrial biodiversity monitoring}
    \url{ https://doi.org/10.1111/2041-210X.14436}

    \item Delgado-Rajó, F.A., Travieso-Gonzalez, C.M. (2025)
    \textit{Flexible hybrid edge computing IoT architecture for low-cost bird songs detection system}
    \url{https://doi.org/10.1016/j.ecoinf.2025.103231}

    \item Ogami, R., Yamamoto, H., Kato, T. and Utsunomiya, E. (2019)
    \textit{Harmful Wildlife Detection System Utilizing Deep Learning for Radio Wave Sensing on Multiple Frequency Bands}
    \url{https://doi.org/10.1109/ICAIIC.2019.8668967}
        
    \item Sittinger, M., Uhler, J., Pink, M. and Herz, A. (2024)
    \textit{Insect detect: An open-source DIY camera trap for automated insect monitoring}
    \url{https://doi.org/10.1371/journal.pone.0295474}

    \item Chen, P., Fei, T., Du, Y., Yi, J., Li, Y. and Kupfer, J. A. (2025)
    \textit{Intelligent Bear Prevention System Based on Computer Vision: An Approach to Reduce Human-Bear Conflicts in the Tibetan Plateau Area, China}
    \url{https://doi.org/10.48550/arXiv.2503.23178}
    
    \item Knyva, M., Gailius, D., Balčiūnas, G., Pratašius, D., Kuzas, P., and Kukanauskaitė, A. (2023)
    \textit{IoT Sensor Network for wild-animal detection near roads}
    \url{https://doi.org/10.3390/s23218929}

    \item Zakaria, M. S., Nordin, R. and Ruslan, M. S. (2020)
    \textit{IoT Infrastructure for Environmental Monitoring: A Case at Tasik Chini Biosphere Reserve}
    \url{https://doi.org/10.1109/IC3e50159.2020.9288403}

    \item Vasconcelos, D. and Nunes, N. J., Ribeiro, M., Prandi, C. and Rogers A. (2019)
    \textit{LOCOMOBIS: a low-cost acoustic-based sensing system to monitor and classify mosquitoes}
    \url{https://doi.org/10.1109/CCNC.2019.8651767}

    \item Jose, T. and Mayan, J. A. (2025)
    \textit{LoRaWAN and artificial intelligence integrated smart acoustic sensor network for bird species identification and deterrence system for farm protection}
    \url{https://doi.org/10.1016/j.measurement.2025.118452}

    \item Andreadis, A., Giambene, G., and Zambon, R. (2021)
    \textit{Monitoring Illegal Tree Cutting through Ultra-Low-Power Smart IoT Devices}
    \url{https://doi.org/10.3390/s21227593}

    \item Campagnaro, F., Ghalkhani, M., Tumiati, R., Marin, F., Dal Grande, M., Pozzebon, A., De Battisti, D., Francescon, R. and Zorzi, M. (2025)
    \textit{Monitoring the Venice Lagoon: An IoT Cloud-Based Sensor Network Approach}
    \url{https://doi.org/10.1109/JOE.2024.3459483}

    \item Aira, J., Olivares, T., Delicado, F. M. and Vezzani, D. (2023)
    \textit{MosquIoT: A System based on IoT and machine learning for the monitoring of Aedes aegypti (Diptera: Culicidea)}
    \url{https://doi.org/10.1109/TIM.2023.3265119 }
    
    \item Rochester, E., Ma, J., Lee, B. and Ghaderi, M. (2019)
    \textit{Mountain Pine Beetle Monitoring with IoT}
    \url{https://doi.org/10.1109/WF-IoT.2019.8767291}

    \item Bonino, M., et al. (2025)
    \textit{Open-Source Solutions for Amphibian Monitoring: Adapting Autonomous Recording Devices (ARDs) and AI-Based Detection in Patagonia}
    \url{https://wildlabs.net/discussion/wildlabs-awards-2025-open-source-solutions-amphibian-monitoring-adapting-autonomous}

    \item Brüggemann, L., Schütz, B., and Aschenbruck, N. (2021)
    \textit{Ornithology meets the IoT: Automatic Bird Identification, Census, and Localization}
    \url{https://doi.org/10.1109/WF-IoT51360.2021.9595401 }

    \item Babu, V., Pineda, R. F., Bizan, M., Wojak, A., Wierzowiecki, S., Gervásio, J., Szklarz, J., Castriotta, L. A. and Carlo, A. D. (2024)
    \textit{Perovskite Solar Module Enabled IoT Asset Tracking for Wildlife Conservation}
    \url{https://doi.org/10.1109/JPHOTOV.2024.3355406}

    \item Acebrón, F., Rosas, E., Cano, J.-C., Manzoni, P., Cecilia, J. and Sebastian, E. (2025)
    \textit{Protecting Endangered Birds with Edge-AI: Real-Time Detection of Invasive Cats in Natural Parks}
    \url{https://doi.org/10.1109/IE64880.2025.11130121}

    \item Treskova, M., Aamer, H., Mack, H., Wahlberg, M., Airom, O., Dayabandara, S., Denkinger, C. M., Diener, E., Fransson, P., Heidecke, J., Ulusoy, I. and Rocklöv, J. (2025)
    \textit{Prototyping an internet-of-things-based bioacoustics system to support research and surveillance of avian-associated infectious diseases}
    \url{https://doi.org/10.1016/j.sbsr.2025.100817}

    \item Whytock, R. C., Suijten, T., van Deursen, T., Świeżewski, J., et al. (2022)
    \textit{Real-time alerts from AI-enabled camera traps using the Iridium satellite network: A case study in Gabon, Central Africa}
    \url{https://doi.org/10.1111/2041-210X.14036}

    \item Bjerge, K., Mann, H.M.R. and Høye, T.T. (2022)
    \textit{Real-time insect tracking and monitoring with computer vision and deep learning}
    \url{https://doi.org/10.1002/rse2.245}

    \item Sethi, S. S., Ewers, R. M., Jones, N. S., Signorelli, A., Picinali, L. and Orme, C. D. L. (2020)
    \textit{SAFE Acoustics: An open-source, real-time eco-acoustic monitoring network in the tropical rainforests of Borneo}
    \url{https://doi.org/10.1111/2041-210X.13438 }
    
    \item Uchiyama, K., Yamamoto, H., Utsunomiya, E. and Yoshihara K. (2022)
    \textit{Sensor Network System for Condition Detection of Harmful Animals by Step-by-step Interlocking of Various Sensors}
    \url{https://doi.org/10.1109/ICAIIC54071.2022.9722692 }

    \item Gallacher, S., Wilson, D., Fairbrass, A., Turmukhambetov, D., Firman, M., Kreitmayer, S., Mac Aodha, O., Brostow, G. and Jones, K. (2021)
    \textit{Shazam for bats: Internet of Things for continuous real-time biodiversity monitoring}
    \url{https://doi.org/10.1049/smc2.12016}

    \item Nguyen, Q. M., Pham, D. A., Pham, D. T., Le, A. D., H. Vo, N. Q. and Vo, H. B. (2023)
    \textit{SmartTrap: An On-Field Insect Monitoring System Empowered by Edge Computing Capabilities}
    \url{https://doi.org/10.1109/RIVF60135.2023.10471810 }

    \item Tulasi, D., Granados, A., Gunawardane, P., Kashyap, A., McDonald, Z. and Thulasidasan, S. (2023)
    \textit{Smart Camera Traps: Enabling Energy-Efficient Edge-AI for Remote Monitoring of Wildlife}
    \url{https://doi.org/10.1145/3615893.362876}

    \item Peter, J., Luder, V., Davis, L. R., Schulthess, L. and Magno, M. (2025)
    \textit{Smart Feeding Station: Non-Invasive, Automated IoT Monitoring of Goodman's Mouse Lemurs in a Semi-Natural Rainforest Habitat}
    \url{https://doi.org/10.48550/arXiv.2503.09238}

    \item Abdelouahid, R. A., Debauche, O., Mahmoudi, S., Marzak, A., Manneback, P. and Lebeau, F. (2020)
    \textit{Smart Nest Box: IoT Based Nest Monitoring In Artificial Cavities}
    \url{https://doi.org/10.1109/CommNet49926.2020.9199624}
    
    \item Ma, A. (2022)
    \textit{Smart Wildlife Sentinel (SWS): Preventing Wildlife-Vehicle Collisions and Monitoring Road Ecology with Embedded IoT Systems and Machine Learning}
    \url{https://doi.org/10.1109/URTC56832.2022.10002174}

    \item Mayton, B., Dublon, G., Russell, S., Lynch, E. F., Haddad, D. D., Ramasubramanian, V., Duhart, C., Davenport, G., Paradiso, J. A. (2017)
    \textit{The Networked Sensory Landscape: Capturing and Experiencing Ecological Change Across Scales}
    \url{https://doi.org/10.1162/PRES_a_00292}
    
    \item Lawrence, Z., Lal, A. and Shen, M. (2024)
    \textit{TinyML : Acoustic Burrowing Owl Vocalization Detection on STM32}
    \url{https://api.semanticscholar.org/CorpusID:280037754}

    \item Wägele, J. W., Bodesheim, P., Bourlat, S. J., Denzler, J., Diepenbroek, M., Fonseca, V., et al. (2022)
    \textit{Towards a multisensor station for automated biodiversity monitoring}
    \url{https://doi.org/10.1016/j.baae.2022.01.003}

    \item Zualkernan, I. A., Dhou, S., Judas, J., Sajun, A. R., Gomez, B. R., Hussain, L. A. and Sakhnini, D. (2020)
    \textit{Towards an IoT-based Deep Learning Architecture for Camera Trap Image Classification}
    \url{https://doi.org/10.1109/GCAIoT51063.2020.9345858}
    
    \item Bjerge, K., Karstoft, H. and Høye, T.T. (2024)
    \textit{Towards edge processing of images from insect camera traps}
    \url{https://doi.org/10.1002/rse2.70007}
    
    \item Arratia, B., Rosas, E., Prades, J., Peña-Haro, S., Cecilia, J. M. and Manzoni, P. (2025)
    \textit{Towards efficient stream monitoring: A systematic approach for model selection and continuous improvement in Tiny Machine Learning applications}
    \url{https://doi.org/10.1016/j.engappai.2025.112415}

    \item Gardiner, R. J., Rowlands, S. and Simmons, B. I. (2025)
    \textit{Towards scalable insect monitoring: Ultra-lightweight CNNs as on-device triggers for insect camera traps}
    \url{https://doi.org/10.1111/2041-210X.70098}
    
    \item Briceño, C. and Tarrillo, J. (2023)
    \textit{Trap camera prototype with remote monitoring system through LoRa technology}
    \url{https://doi.org/10.1109/INTERCON59652.2023.10326034}

    \item O’Shea-Wheller, T. A., Corbett, A., Osborne, J. L., Recker, M. and Kennedy, P. J. (2024)
    \textit{VespAI: a deep learning-based system for the detection of invasive hornets}
    \url{https://doi.org/10.1038/s42003-024-05979-z}

    \item Arshad, B., Barthelemy, J., Pilton, E. and Perez, P. (2020)
    \textit{Where is my Deer?-Wildlife Tracking And Counting via Edge Computing And Deep Learning}
    \url{https://doi.org/10.1109/SENSORS47125.2020.9278802}

    \item Elias, A. R., Golubovic, N., Krintz, C. and Wolski, R. (2017)
    \textit{Where's The Bear? Automating Wildlife Image Processing Using IoT and Edge Cloud Systems}
    \url{https://doi.org/10.1145/3054977.3054986}

    \item John, R. and Issac, J. J. (2025)
    \textit{Wildlife Monitoring Using YOLOv8 and Edge-AI: A Real-Time Approach}
    \url{https://doi.org/10.71329/IUPJTC/2025.17.3.20-32}

    \item Ross, R., Anderson, B., Bienvenu, B., Scicluna, E. L., and Robert, K. A. (2022)
    \textit{WildTrack: An IoT system for tracking passive-RFID Microchipped wildlife for ecology research}
    \url{https://doi.org/10.3390/automation3030022}

    \item Sheng, Z., Pfersich, S., Eldridge, A., Zhou, J., Tian, D., Leung, V. C. M. (2019)
    \textit{Wireless acoustic sensor networks and edge computing for rapid acoustic monitoring}
    \url{https://doi.org/10.1109/JAS.2019.1911324}

\end{enumerate}

\clearpage
    \clearpage
    
\end{document}